\documentclass[11pt]{article}

\usepackage{xr-hyper}
\pdfoutput=1
\usepackage{hyperref}
\hypersetup{
    colorlinks=true,
    linkcolor=black,
    citecolor=black,
    filecolor=black,
    urlcolor=black,
}

\RequirePackage{amssymb}
\RequirePackage{amsmath}
\RequirePackage{latexsym}
\RequirePackage{mathrsfs}

\DeclareMathOperator*{\var}{var}
\DeclareMathOperator*{\supp}{supp}

\newcommand{\R}{\ensuremath{\mathbb{R}}}
\newcommand{\Exp}{\ensuremath{\mathbb{E}}}
\newcommand{\Prob}{\ensuremath{\mathbb{P}}}

\renewcommand{\a}{\ensuremath{\mathbf{a}}}
\renewcommand{\b}{\ensuremath{\mathbf{b}}}
\newcommand{\s}{\ensuremath{\mathbf{s}}}
\newcommand{\x}{\ensuremath{\mathbf{x}}}

\newcommand{\X}{\ensuremath{\mathbf{X}}}
\newcommand{\Y}{\ensuremath{\mathbf{Y}}}
\newcommand{\W}{\ensuremath{\mathbf{W}}}
\renewcommand{\S}{\ensuremath{\mathbf{S}}}

\renewcommand{\P}{\ensuremath{\mathbf{P}}}
\newcommand{\A}{\ensuremath{\mathbf{A}}} 
\newcommand{\Q}{\ensuremath{\mathbf{Q}}}
\newcommand{\U}{\ensuremath{\mathbf{U}}}

\usepackage{bbm}
\newcommand{\indicator}{\ensuremath{\mathbbm{1}}}

\makeatletter
\newcommand*{\indep}{
  \mathbin{
    \mathpalette{\@indep}{}
  }
}
\newcommand*{\nindep}{
  \mathbin{
    \mathpalette{\@indep}{\not}
  }
}
\newcommand*{\@indep}[2]{
  \sbox0{$#1\perp\m@th$}
  \sbox2{$#1=$}
  \sbox4{$#1\vcenter{}$}
  \rlap{\copy0}
  \dimen@=\dimexpr\ht2-\ht4-.2pt\relax
  \kern\dimen@
  {#2}
  \kern\dimen@
  \copy0
} 
\makeatother

\usepackage{color}

\usepackage{wrapfig}
\usepackage{tikz-cd}
\usepackage{tikz}
\usepackage{pgfplots}

\RequirePackage{amsthm}

\theoremstyle{definition}

\newtheorem{proposition}{Proposition}
\newtheorem{lemma}{Lemma}

\newtheorem{definition}{Definition}
\newtheorem{theorem}{Theorem}

\newtheorem{Aassumption}{Assumption}

\newtheorem{Bassumption}{Assumption}

\usepackage{setspace}
\usepackage[margin=1in]{geometry}

\usepackage[longnamesfirst]{natbib}
\usepackage{ulem}
\usepackage{wrapfig}
\usepackage{float}

\usepackage{epstopdf}
\usepackage{mathtools}

\usepackage{graphicx}
\usepackage{subcaption} 

\usepackage{multirow}
\usepackage{booktabs}

\usepackage{placeins}

\usepackage{tikz}
\usetikzlibrary{fit,shapes}

\newcommand{\cellnode}[2]{
  \tikz[remember picture,baseline=(#1.base)]
    \node[inner sep=1pt] (#1) {#2};
}

\title{
\textbf{Finite Population Identification and \\[0.15em] Design-Based Sensitivity Analysis}\footnote{First arXiv draft: April 19, 2025. This paper was presented at UC Santa Cruz, Notre Dame, Duke, the University of Virginia, the University of Chicago Booth School of Business, the University of Washington, Simon Fraser University, the University of British Columbia, Pennsylvania State University, the University of Toronto, the 2025 Duke Causation Workshop, the 2025 Triangle Econometrics Conference, University College London, the 2025 Oxford Econometrics Workshop, the 2025 Southern Economic Association Annual Meeting, UC Santa Barbara, and the Princeton Quantitative Social Science Colloquium. We thank audiences at those seminars and conferences, as well as Chuck Manski, John Pepper, Alex Poirier, Michael Pollmann, Muyang Ren, Jonathan Roth, and Zeyang Yu for helpful comments and conversations. We thank Daria Soboleva for excellent research assistance. Masten thanks the National Science Foundation for research support under Grant 1943138.}
} 

\author{Brendan Kline\thanks{
    Department of Economics, University of Texas at Austin,
    \texttt{brendan.kline@utexas.edu}}
\qquad
Matthew A. Masten\footnote{Department of Economics, Duke University,
        \texttt{matt.masten@duke.edu}}
}
\date{May 8, 2026}

\begin{document}

\maketitle
\begin{abstract}
We develop a new approach for quantifying uncertainty in finite populations, by using design distributions to calibrate sensitivity parameters in finite population identified sets. This yields uncertainty intervals that can be interpreted as identified sets, robust Bayesian credible sets, or uniform frequentist design-based confidence sets. We focus on quantifying uncertainty about the average treatment effect, where our approach (1) yields design-based confidence intervals which allow for heterogeneous treatment effects without using asymptotics, (2) provides a new motivation for examining covariate balance, and (3) gives a new formal analysis of the role of randomization. We illustrate our approach in three empirical applications.
\end{abstract}



\onehalfspacing
\normalsize

\newpage
\section{Introduction}\label{sec:intro}

 Many empirical papers use variation across a small number of units to learn about causal effects. For example, in many settings this variation is across units like states or industries. How does identification of causal effects work in such settings, and how should empirical researchers quantify uncertainty? Our paper gives new answers to these questions, focusing on the context of randomized experiments with a finite number of units.

Traditionally, the $N$ units are assumed to be sampled from a hypothetical infinite ``super'' population, but recently there has been a resurgence in the ``design-based inference'' literature (reviewed in section \ref{sec:litReview}) which studies explicitly finite populations.\footnote{The ``design-based'' versus ``super-population'' distinction and jargon come from the survey sampling literature (e.g., \citealt{SarndalEtAl1978}), which is how we use them here. In economics, some researchers now use the term ``design-based'' in a related but not equivalent sense, to refer to specific identification strategies (e.g. \citealt{Card2022}). We focus on randomized experiments in this paper and hence both meanings of ``design-based'' apply.} With very few exceptions (e.g., \citealt{ManskiPepper2018}, \citealt{BorusyakHullJaravel2024}, and \citealt{RambachanRoth2025}), most papers that study finite populations do not discuss identification and instead focus solely on proving frequentist properties like consistency or confidence interval validity. In contrast, there is a vast and rich literature on identification in infinite populations\footnote{See \cite{ChristensenConnault2023}, \cite{Obradovic2024}, \cite{Spini2024}, \cite{Deaner2025}, \cite{Sloczynski2025}, \cite{BlandholEtAl2025}, and \cite{CaetanoEtAl2026} for a few recent examples, among many others; for comprehensive surveys see \cite{Matzkin2007}, \cite{Manski2009}, \cite{BontempsMagnac2017}, \cite{Lewbel2019}, \cite{ChesherRosen2020}, \cite{Molinari2020}, and \cite{KlineTamer2023}.}; no corresponding literature on identification in finite populations exists. Our first main contribution, therefore, is to develop a formal theory of identification in finite populations.

The literature on identification in finite populations is so sparse that we are not aware of any source which formally defines identification in an \emph{explicitly} finite population setting (though some references come close). Nonetheless, it is quite common and ``conventional'' in the prior literature (following the term used by \citealt{HeckmanVytlacil2007}, footnote 6, page 4786) to informally state that a parameter is point identified if there is a known mapping from the distribution of the data induced by randomization (of treatment assignment, for example) into the parameter. We state this prior definition formally, and then argue that it has two major problems.

First, we show that this prior conventional definition relies on hypothetical data that is impossible to observe, since it involves averaging over counterfactual worlds that are incompatible with each other. As a consequence, unlike in infinite populations, we prove that this prior conventional definition in finite populations implies \emph{the values of every potential outcome for every unit are point identified in any setting where the probability of each treatment value is positive} (a type of overlap restriction), including non-experimental settings where treatment assignment depends almost arbitrarily on potential outcomes (this finding builds on ideas in \citealt{Imbens2018} and \citealt{ChenRothSpiess2026}, as we discuss in section \ref{sec:tradDef}). For example, according to the prior conventional definition, assumptions like unconfoundedness or instrument exogeneity are unnecessary to obtain point identification of treatment effects, so long as an overlap restriction holds. Therefore, we argue that the prior conventional definition is not adequate for assessing how informative the actual data one has are about treatment effects.

Second, the prior conventional definition of identification, as well as the broader literature on finite population design-based inference, \emph{pre-supposes} that there is some kind of known a priori randomization. For example, that randomization may involve treatment assignment, an instrument assignment, or sampling, depending on the setting. These may be plausible in randomized experiments or designed surveys, but in many observational settings it is not clear that the variable of interest (like treatment assignment) is randomly chosen, let alone that the distribution it was drawn from is known to the researcher (a point discussed in \citealt{ManskiPepper2018}, pages 234--235). Consequently, the prior conventional definition of identification in finite populations does not apply to such cases.

To address these two problems, we develop a different, formal definition of identification for a parameter in a finite population setting. Our definition is analogous to the usual definition of identification in infinite population settings (e.g., \citealt{Manski2003}, \citealt{Lewbel2019}, \citealt{Molinari2020}): The \emph{finite population identified set} for a parameter (such as the average treatment effect, ATE) is the set of parameter values consistent with (a) the finite matrix of data the researcher actually has and (b) any assumptions that they make.

Our definition directly addresses the two problems with the prior conventional definition described above: First, our definition does not require any knowledge about how treatment is assigned. Hence it can be applied in a wider variety of settings. Second, it uses actually observed data, rather than hypothetical data that is impossible to observe. Relatedly, it does not require any kind of asymptotics or hypothetical infinite populations. Consequently, any parameter which cannot be written as a known function of the actually observed data will typically be partially identified. The size of the identified set, however, will generally vary with the assumptions and data available, as in the usual infinite population definition of identified sets.

As a leading example of our identification analysis, we show that the average treatment effect in a randomized experiment is generally partially identified in finite populations.\footnote{For a recent survey of infinite population approaches to randomized experiments, see \cite{BaiShaikhTabordMeehan2025}.} This conclusion arises from a subtle distinction between the finite and infinite population settings: Randomization guarantees that the treatment and control groups are exactly balanced in an infinite population, but not in finite populations. Since randomization cannot guarantee any level of balance between the treatment and control groups in a finite population, it does not have any \emph{identifying} power. 

Nonetheless, we show how to use our finite population identified sets combined with assumptions on treatment assignment to obtain a new way to quantify uncertainty in finite populations, which we call \emph{design-based sensitivity analysis}. This is our second main contribution. 

Overall, our method yields uncertainty intervals that can be interpreted as (1) identified sets, (2) robust Bayesian credible sets, or (3) uniform frequentist design-based confidence sets. This approach is non-asymptotic, which allows it to be reliably applied in small populations. And it allows researchers to flexibly impose additional assumptions on unobserved values, transparently quantifying the impact of these assumptions on conclusions. We summarize our approach in section \ref{sec:introPractitionerGuideRCTs}.

In particular, focusing on their frequentist interpretation, our uncertainty intervals avoid several limitations of the confidence intervals that are currently available in the literature: (i) They do not rely on large-$N$ asymptotics, which can perform poorly if $N$ is small (e.g., $N=10$) and (ii) they allow for heterogeneous treatment effects. They are also valid uniformly over the set of all possible matrices of potential outcomes consistent with the assumptions (an important feature in partially identified settings; see \citealt{CanayShaikh2017}). Furthermore, unlike the confidence intervals in the literature, our uncertainty intervals have both Bayesian and identification-based interpretations. This allows empirical researchers to choose their most preferred interpretation. This interpretational flexibility is one benefit obtained from a quantification of uncertainty that begins with identification. In particular, our uncertainty intervals are not based on hypothesis test-inversion, but are instead based on ideas from partial identification.

As mentioned above, we develop this approach in the context of randomized experiments with a binary treatment, where the goal is to quantify the uncertainty in the average treatment effect due to missing potential outcomes. We extend the analysis in three main directions: In section \ref{sec:covariates} we give a new motivation for examining covariate balance. Specifically, by making an assumption that explicitly links covariates to potential outcomes, we show that observed imbalances in covariates have identifying power for the unobserved \emph{realized} imbalance in potential outcomes, which allows researchers to obtain tighter bounds on ATE. In section \ref{sec:IV} we study instrumental variable based analysis under one-sided noncompliance. And in section \ref{sec:sampling} we discuss uncertainty due to non-sampled units. 

\subsection{Our Approach to Quantifying Uncertainty in Experiments}\label{sec:introPractitionerGuideRCTs}

\nocite{Manski1990}

Here we sketch our approach in the context of causal inference. See appendix \ref{sec:sampleSurveyGuide} for a parallel exposition in the context of survey sampling. Consider the population of six units in table \ref{table:examplePopulation} (left). Here $Y_i(1)$ and $Y_i(0)$ denote potential outcomes and $X_i \in \{0,1\}$ denotes realized treatment. The true ATE is 0.55, the difference between the average of $Y_i(1)$ over $i=1,\ldots,6$ and the average of $Y_i(0)$ over $i=1,\ldots,6$. The data (table \ref{table:examplePopulation} right) only reveals half of the potential outcomes, however. Suppose we know that all outcomes must be between 0 and 1. Then without further assumptions, all we can say \emph{for sure} about ATE is that it is in the set
\begin{align}\label{eq:exampleNoAssumpsBounds}
	&\left[ \frac{(0.6+0.9+0.8)+(0+0+0)}{6} - \frac{(0.1+0+0.2) + (1+1+1)}{6}, \right. \\
	&\quad \left. \frac{(0.6+0.9+0.8)+(1+1+1)}{6} - \frac{(0.1+0+0.2) + (0+0+0)}{6} \right]
	= \left[ - \frac{1}{6}, \frac{5}{6} \right]. \notag
\end{align}
This is an example of what we call the \emph{finite population identified set} for ATE; in this case, it is the set of ATE values consistent with the observed data and known bounds on outcomes, but no other assumptions are imposed (see section \ref{sec:ourDef} for a formal definition). It is obtained by filling in the unobserved values in table \ref{table:examplePopulation} (right) with values in $[0,1]$ to either maximize or minimize the corresponding ATE value (analogous to Manski's 1990 derivations). 

\begin{table}[t]
  \centering
  \begin{subtable}[t]{0.4\textwidth}
  \vspace{0pt}
    \centering
    \begin{tabular}{c|ccc}
      $i$ & $Y_i(0)$ & $Y_i(1)$ & $X_i$ \\%
      \hline
      1 & 0.1 & 0.8 & 0 \\%
      2 & 0 & 0.5 & 0 \\%
      3 & 0.2 & 0.7 & 0 \\%
      4 & 0.4 & 0.6 & 1 \\%
      5 & 0.1 & 0.9 & 1 \\%
      6 & 0.2 & 0.8 & 1 \\%
    \end{tabular}
  \end{subtable}
  \begin{subtable}[t]{0.4\textwidth}
  \vspace{-10pt}
    \centering
    \begin{tikzpicture}[remember picture,
                        baseline=(current bounding box.north)]
      \node (mytable) {%
        \begin{tabular}{c|ccc}
          $i$ & $Y_i(0)$ & $Y_i(1)$ & $X_i$ \\
          \hline
          1 & \cellnode{r1c2}{0.1} & \cellnode{r1c3}{?} & 0 \\
          2 & \cellnode{r2c2}{0} & \cellnode{r2c3}{?} & 0 \\
          3 & \cellnode{r3c2}{0.2} & \cellnode{r3c3}{?} & 0 \\[0.5em]
          4 & \cellnode{r4c2}{?} & \cellnode{r4c3}{0.6} & 1 \\
          5 & \cellnode{r5c2}{?} & \cellnode{r5c3}{0.9} & 1 \\
          6 & \cellnode{r6c2}{?} & \cellnode{r6c3}{0.8} & 1 \\
        \end{tabular}
      };
      \node[fit=(r1c2)(r2c2)(r3c2),draw,inner sep=2.5pt]{};
      \node[fit=(r4c2)(r5c2)(r6c2),draw,dashed,inner sep=4pt]{};
      \node[fit=(r4c3)(r5c3)(r6c3),draw,shape=rectangle,
            rounded corners=8pt,inner xsep=3pt,inner ysep=3pt]{};
      \node[fit=(r1c3)(r2c3)(r3c3),draw,dashed,shape=rectangle,
            rounded corners=8pt,inner xsep=4pt,inner ysep=4pt]{};
    \end{tikzpicture}
  \end{subtable}
  \caption{Left: Example population, $N=6$. Right: The observed data from that population.}
  \label{table:examplePopulation}
\end{table}

Suppose we also knew that treatment was randomly assigned. In infinite populations, this assumption is routinely interpreted to imply $\Exp[Y(x) \mid X=1] = \Exp[Y(x) \mid X=0]$ for $x \in \{0,1\}$, a post-randomization exact balance condition which leads to point identification of ATE (see appendix \ref{sec:popDistributions} for a discussion of this notation in the context of finite populations). Unfortunately, as is well known (e.g., \citealt{Altman1985}, \citealt{GreenlandRobins1986}, and \citealt{Greenland1990}), in \emph{finite} populations randomization does not \emph{guarantee} any kind of post-randomization balance in potential outcomes, even if it makes such balance ``likely''. Consequently, ATE is not point identified in finite populations, even when treatment is randomly assigned; its identified set continues to be equation \eqref{eq:exampleNoAssumpsBounds}. We formalize this in theorem \ref{thm:randomizationUseless}.

What explains the difference in the identification status of ATE under randomly assigned treatment between the infinite and finite population settings? We first emphasize what is \emph{not} different about the two settings: In both settings, sampling uncertainty has been assumed away---data on all units in the population are observed. Moreover, in both settings we only observe data from \emph{a single realization of random assignment of treatment} for each unit in the population. In particular, classical infinite population identification analyses (e.g., \citealt{HeckmanVytlacil2007}) always assume that we only observe one potential outcome per unit, because of the fundamental problem of causal inference---they can either get treatment or not. Analogously, our finite population identification analysis is based on data like table \ref{table:examplePopulation} right. We discuss this similarity in more detail at the end of section \ref{sec:tradDef}.

This discussion shows that the only difference in the identification analysis between the two settings is population size: Exact balance obtains for infinitely many units but not finitely many. We discuss that distinction further in section \ref{sec:randomizationIdentification}.

Although random assignment of treatment does not guarantee exact balance in finite populations, it is ``likely'' to deliver ``some'' balance. We formalize this idea in two steps, starting with a non-probabilistic analysis and then moving to a probabilistic analysis. Specifically, suppose we were willing to assume that (a) the average of the unobserved values in the dashed box is not farther than $K$ from the observed mean $(0.1+0+0.2)/3 = 0.1$ in the solid box and (b) the average of the unobserved values in the dashed oval is not farther than $K$ from the observed mean $(0.6+0.9+0.8)/3 = 0.77$ in the solid oval (a finite population version of an assumption proposed in \citealt{Manski2003}). We call this the \emph{$K$-approximate mean balance} assumption and derive the identified set for ATE under this assumption in theorem \ref{thm:KapproxiBalanceATEset}, which we denote by $\Theta_I(K)$. For sufficiently small $K$, we can conclude that ATE is in a strictly smaller set than equation \eqref{eq:exampleNoAssumpsBounds}, where the size of this set depends on $K$, the maximal magnitude of imbalance between the treatment and control groups. Note, in particular, that ATE is point identified if $K=0$ (exact balance), as in infinite population analyses, although exact balance is implausible for small finite $N$, as discussed earlier.

\begin{figure}[t]
  \centering

  \begin{subfigure}[c]{0.45\textwidth}
    \centering
\includegraphics[width=\linewidth]{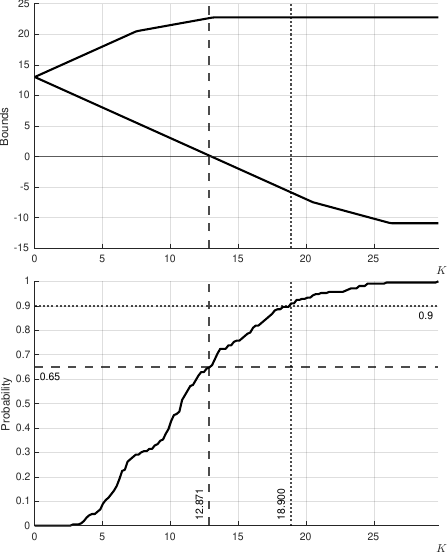}
\caption{Top: $\Theta_I(K)$, bounds on ATE as a function of the maximum magnitude of ex post imbalance, $K$. Bottom: $\underline{p}(K)$. \label{fig:GneezyDesignBased}}
  \end{subfigure}
  \hfill
  \begin{subfigure}[c]{0.45\textwidth}
    \centering
\includegraphics[width=\linewidth]{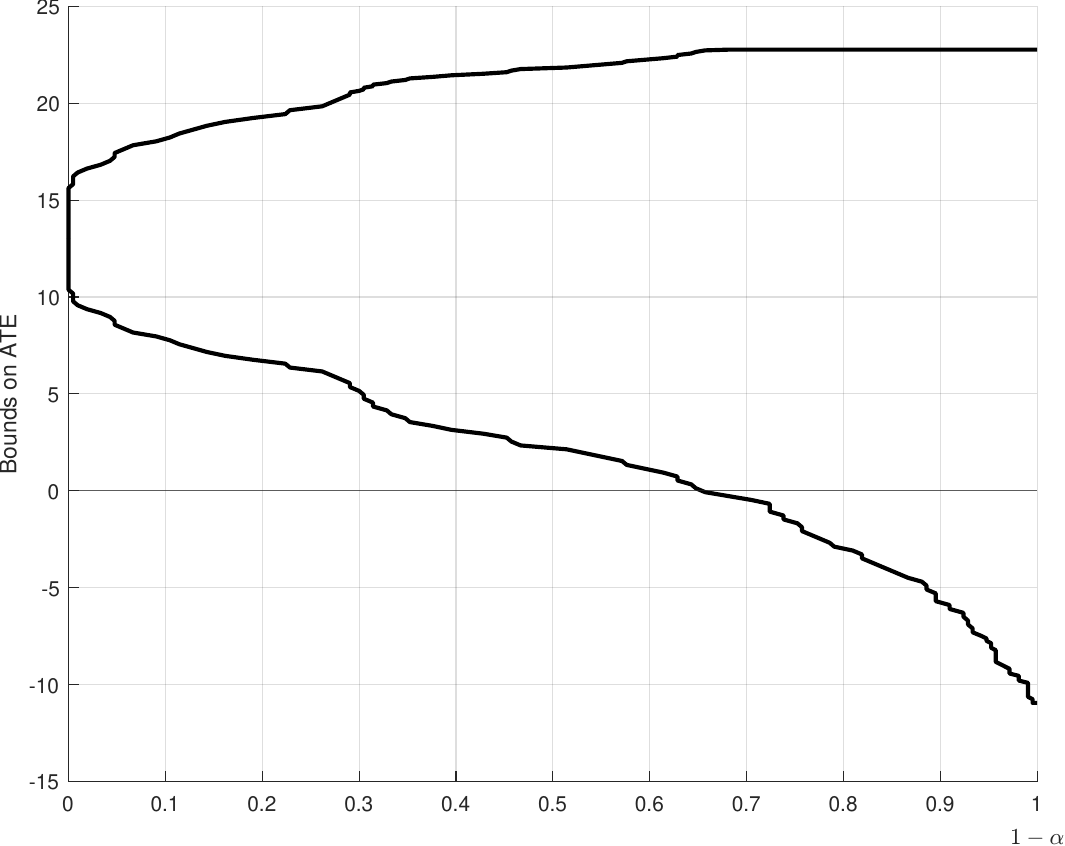}
\caption{$\Theta_I(K(\alpha))$, bounds on the ATE as a function of $1-\alpha$. \label{fig:GneezyDesignBased2}}
  \end{subfigure}

  \caption{Example output for a design-based sensitivity analysis, $N=10$. See section \ref{sec:Gneezy} for data details.}
  \label{fig:Gneezy}
\end{figure}

The top plot in figure \ref{fig:GneezyDesignBased} shows an example of these identified sets $\Theta_I(K)$ as a function of $K$ (using real data; see section \ref{sec:Gneezy}). By examining how these sets change with $K$, we can examine the sensitivity of conclusions about ATE to assumptions about the magnitude of the \emph{realized} differences in average potential outcomes across the treatment and control group. These sets provide a non-probabilistic approach to quantifying uncertainty about ATE.

Researchers may be unsure how to select plausible values of $K$, however. In order for $\Theta_I(K)$ to contain the true ATE, $K$ must be weakly larger than the \emph{true}, but unknown magnitude of imbalance between the treatment and control groups, which we denote by $K^\text{true,max}$ (e.g., it is $\max \{0.13, 0.1\}$ in table \ref{table:examplePopulation}). Consequently, we next use random assignment of treatment to probabilistically quantify uncertainty in $K^\text{true,max}$. We use this to calibrate values of $K$, which will lead to a probabilistic quantification of uncertainty in ATE.

Specifically, suppose momentarily that the true values of all potential outcomes were known. Then, since we know the treatment assignment distribution, we can compute the \emph{ex ante} probability that the difference in potential outcome means across the treatment and control groups is at most $K$. This can be done for any $K$. This yields a distribution of possible values of the ex post imbalance $K^\text{true,max}$. Since we do not actually know all potential outcomes, however, this distribution itself is partially identified. Nonetheless, we can find the smallest possible value of this probability, across all logically possible completions of table \ref{table:examplePopulation} (right). We do this using modern computational methods, described in section \ref{sec:computation}. Denote this smallest probability by $\underline{p}(K)$. We can now plot this function $\underline{p}(\cdot)$, which converts $K$-values into ex ante design-probabilities. The bottom plot in figure \ref{fig:GneezyDesignBased} shows an example.

Finally, leading to our main empirical recommendation, we combine the two plots in figure \ref{fig:GneezyDesignBased} as follows: For any desired ex ante probability of balance $1-\alpha \in (0,1)$, use the bottom plot to find the smallest magnitude of imbalance that occurs with at least $100(1-\alpha)$\% ex ante probability; denote this value by $K(\alpha)$. For example, if we set $1-\alpha = 0.9$ then the figure shows us that $K=18.9$ satisfies $\underline{p}(K) = 1-\alpha$ (see the dotted lines). Then use the top plot to read off bounds on ATE for this value of $K$, as shown in the dotted lines. In this case the bounds are about $[-6, 22]$.

Our main construction is a set of \emph{uncertainty intervals}, such as those shown in figure \ref{fig:GneezyDesignBased2}, which plots the bounds $\Theta_I(K(\alpha))$ for all values of $1-\alpha$.\footnote{Note: For $1-\alpha$ close to zero but strictly positive, the bounds in figure \ref{fig:GneezyDesignBased2} are non-singleton. This arises since exact balance is impossible in the worst case completions of table \ref{table:examplePopulation}; this is why the plot of $\underline{p}$ in figure \ref{fig:GneezyDesignBased} is flat at zero up to a strictly positive value of $K$. We discuss this point further in section \ref{sec:randomizationIdentification}.} We show that these uncertainty intervals have three interpretations:
\begin{enumerate}
\item First, $\Theta_I(K(\alpha))$ is the finite population identified set for ATE, under the \emph{deterministic} assumption that the maximum magnitude of ex post imbalance is at most $K(\alpha)$.

\item Second, $\Theta_I(K(\alpha))$ is a robust empirical Bayesian $1-\alpha$ credible set for ATE. That is, $\Theta_I(K(\alpha))$ contains the true ATE with posterior probability at least $1-\alpha$ for any prior over the vector of all potential outcomes (such as the 12 potential outcomes in table \ref{table:examplePopulation} left) that satisfies our bounded outcome assumption. Essentially, we show that even the ``most pessimistic'' Bayesian has a posterior distribution over the true, but unknown, magnitude of imbalance in potential outcomes $K^\text{true,max}$ that is no more pessimistic than the distribution $\underline{p}$ that we constructed above (and we show that $\underline{p}$ is in fact a valid cdf).

\item Third, $\Theta_I(K(\alpha))$ is a uniformly valid design-based $1-\alpha$ confidence interval for ATE. That is, across repeated re-assignments of treatment, these bounds will contain the true ATE at least $100(1-\alpha)$\% of the time. Uniformity here means that this property holds for all possible vectors of potential outcomes that satisfy our bounded outcome assumption (such as the 12 potential outcomes in table \ref{table:examplePopulation} left).
\end{enumerate}
We discuss each of these interpretations further in section \ref{sec:interpretation}. Overall, we recommend that researchers report plots like figure \ref{fig:GneezyDesignBased2}, which can then be interpreted any of these three ways.

Besides reporting interval quantifications of uncertainty, researchers also often report summary statistics that are meant to measure the ``strength of evidence'' in favor of a hypothesis. Our results also allow for such analyses. For example, we may ask: How much evidence is there in the data for the conclusion that ATE is nonnegative? The horizontal intercept in figure \ref{fig:GneezyDesignBased2}, called a \emph{breakdown point}, provides a simple answer. In figure \ref{fig:GneezyDesignBased}, we first compute the horizontal intercept in the top plot, which is the largest magnitude of imbalance between mean potential outcomes that can be allowed while still allowing us to conclude that ATE is nonnegative. We then plug it into $\underline{p}$ to convert it to design probabilities (see the dashed lines). Here we obtain $\underline{p}(12.871) = 0.65$. This tells us that, given the data, and regardless of what the true unknown potential outcomes actually are, there was an ex ante probability of at least 65\% that the treatment and control groups would be sufficiently balanced to ensure that the identified set for ATE only contains non-negative numbers. Or, using the Bayesian interpretation of our analysis, this number says there is at least a 65\% posterior probability that ATE is non-negative.

We also show how to obtain tighter bounds by making further assumptions about unobserved potential outcomes, which illustrates the flexibility of our approach. For example, in many experiments researchers observe covariates for each unit. It is often reasonable to expect that these covariates are predictive of potential outcomes. If researchers are willing to place a lower bound on this predictive power, as measured by the R-squared from the regression of potential outcomes onto covariates, then they can obtain tighter uncertainty intervals for ATE. We illustrate this analysis in figure \ref{fig:BloomDesignBased2} in section \ref{sec:Bloom}.

Beyond this basic setup, we discuss a variety of extensions and complementary results throughout the paper, including (1) instrumental variables and noncompliance, (2) random sampling of units, in addition to randomization, (3) other restrictions on the unobserved potential outcomes, like having an a priori bounded variance, (4) identification of parameters besides the average treatment effect, and (5) using other measures of balance beyond means.

\subsection{Related Literature}\label{sec:litReview}

Our paper relates to a variety of literatures, but for brevity we only discuss the most closely related work, rather than attempt a comprehensive survey. \cite{ManskiPepper2018} derive identified sets in an explicitly finite population setting. Like them, we also consider the identifying power of bounded variation type assumptions. A main difference is that they focus on a setting with observational data whereas we primarily consider randomized experiments. Our focus on experiments allows us to use randomization itself to calibrate the magnitude of the bounded-variation sensitivity parameter $K$. Like us, \cite{GreenlandRobins1986} also describe the problem of drawing conclusions in finite populations as an identification problem which must be solved by making assumptions about unobserved values of variables. They formally show how exact balance assumptions yield point identification. They then informally explain that ``if we randomize...when both samples are large, random differences will in probability be small'' (page 415). Many other papers make similar informal remarks, like \citet[page 590]{Lindley1980} and \citet[page 480]{RoyallPfeffermann1982}. Relatedly, the distinction between ex ante and ex post balance is well known; \cite{Greenland1990} gives a particularly clear discussion, where he considers a randomized experiment with $N=2$. \cite{GreenlandRobins2009} more recently discuss this distinction, and conclude that ``randomization (or more generally, ignorability) does \textbf{not} impose ``no [ex post] confounding''...rather, it provides...a randomization-based (``objective'') derivation of a prior...that applies after allocation as well as before, and becomes more narrowly centered around zero as the sample size increases. This is a key post-allocation benefit of randomization'' (page 5, emphasis in the original). That paper as well as the rest of this prior literature, however, is largely verbal discussion. One of our main contributions is to build on those observations to provide a new, detailed, and formal approach to identification and quantifying uncertainty in finite populations.

The standard approach to uncertainty quantification in finite populations is to construct frequentist design-based confidences sets. For example, chapter 5 of \cite{ImbensRubin2015} discusses confidence sets which are valid for finite $N$ but are based on homogeneous treatment effect assumptions, while chapter 6 discusses confidence sets which allow for heterogeneous treatment effects but which are based on asymptotics. Our uncertainty intervals also have a confidence set interpretation, and so fit within this standard approach. Importantly, however, our intervals also have two additional non-frequentist interpretations, as either identified sets or Bayesian credible sets. We are not aware of any other intervals in the literature which have these three distinct interpretations. Moreover, viewed purely as confidence sets for ATE, our intervals have valid coverage for any fixed $N$, allow for heterogeneous treatment effects, and do not require outcomes to be binary. We are only aware of one feasible alternative design-based confidence interval with these properties, which we call the Hoeffding CI, as discussed in \cite{AronowEtAl2025} and \cite{Ding2025} (who build on the binary outcome analysis of \citealt{RobinsRitov1997}). In our empirical applications this interval is always at least 3.5 times wider than ours, and can be up to 20 times wider, depending on the nominal coverage probability (see appendix \ref{sec:AlternativeEmpiricalCIs}). Since both intervals have valid coverage, this suggests that the Hoeffding CI is substantially less powerful than our interval. We discuss these issues as well as further related literature in appendix \ref{sec:simulations}.

Finally, many recent papers on causal inference study explicitly finite populations, including \cite{LiDing2017}, \cite{LiDingMiratrix2017}, \cite{AronowSamii2017}, \cite{Ding2017}, \cite{AtheyEcklesImbens2018}, \cite{KangPeckKeele2018}, \cite{AbadieAtheyImbensWooldridge2020, AbadieAtheyImbensWooldridge2023}, \cite{HongLeungLi2020}, \cite{WuDing2021}, \cite{EcklesEtAl2020}, \cite{ImbensMenzel2021}, \cite{ZhaoDing2021}, \cite{Savje2021}, \cite{BojinovRambachanShephard2021}, \cite{Xu2021}, \cite{XuWooldridge2022}, \cite{AtheyImbens2022}, \cite{RothSantAnna2023}, \cite{Pollmann2023}, \cite{Wooldridge2023}, \cite{StartzSteigerwald2023}, \cite{Sancibrian2024}, \cite{BorusyakHullJaravel2024}, \cite{RambachanRoth2025}, and \cite{BaiEtAl2025}, among many others. Also see \cite{ImbensRubin2015} and \cite{Ding2024} for book level surveys. Empirical researchers have also recently argued in favor of these methods, leading to renewed applied interest as well; see \cite{Young2019}. The theoretical literature largely follows a template laid out by Neyman (1923) that jumps straight to deriving various frequentist properties, without doing any identification analysis. Two exceptions are \cite{BorusyakHullJaravel2024} and \cite{RambachanRoth2025} which both discuss identification, based on the conventional definition, and do asymptotics-based inference (section 4.3 of \citealt{BorusyakHullJaravel2024} also discuss exact inference under a constant effects assumption). Our finding that all potential outcomes are point identified under the conventional definition builds on ideas in \cite{Imbens2018} and \cite{ChenRothSpiess2026}, as we discuss in section \ref{sec:tradDef}. Rather than further studying identification, \cite{ChenRothSpiess2026} derive various frequentist testing and Bayesian updating implications of the conventional definition of identification, concluding that this conventional definition ``translates to little, if any, practical learning'' (page 14).

\nocite{Neyman1923}

\section{Finite Population Identification}\label{sec:finitePopIdent}

We begin by describing our approach to identification: We define the finite population identified set in section \ref{sec:ourDef}. We then use this definition to derive identified sets for finite population average treatment effects under various assumptions in section \ref{section:ATEidentifiedSets}, and we discuss the role of randomization for identification in section \ref{sec:randomizationIdentification}. We explain our concerns with the prior definition of identification in section \ref{sec:tradDef}.

\subsection{The Finite Population Identified Set}\label{sec:ourDef}

Consider the standard potential outcomes model with a binary treatment. There are four components required to conduct an identification analysis:

\smallskip

\textbf{1. The population:} Suppose there are $N$ units in our population. Let $\mathcal{I} \coloneqq \{1,\ldots,N \}$ be the set of indices for these units. Each unit $i \in \mathcal{I}$ is associated with the vector of numbers $(Y_i(1), Y_i(0), X_i, W_i)$, where $Y_i(1)$ and $Y_i(0)$ are potential outcomes, $X_i \in \{0,1\}$ is a realized binary treatment, and $W_i$ is a $d_W$-vector of covariates. The \emph{population} is the $N \times (3+d_W)$ matrix of all of these numbers. Let $\P^\text{true} \coloneqq (\Y(1),\Y(0),\X,\W)$ denote this population matrix. We use $\P$ to denote alternative possible values of this population matrix. Table \ref{table:examplePopulation} shows an example population.

\smallskip

\textbf{2. Assumptions:} We do not observe all elements of $\P^\text{true}$ because of the fundamental problem of causal inference. We typically make assumptions about its missing elements, the missing potential outcomes. Formalize these assumptions as the restriction that $\P^\text{true} \in \mathcal{P}$ where $\mathcal{P}$ is a known set of $N \times (3+d_W)$ matrices. We give examples of $\mathcal{P}$ in section \ref{section:ATEidentifiedSets}.

\smallskip

\textbf{3. Parameters:} Parameters of interest are functionals of the population matrix. For example, the average treatment effect is defined as $\text{ATE} \coloneqq \frac{1}{N} \sum_{i=1}^N \big( Y_i(1) - Y_i(0) \big)$. Similarly, the average effect of treatment on the treated is $\text{ATT} \coloneqq \Big( \sum_{i=1}^N \big( Y_i(1) - Y_i(0) \big) \allowbreak \indicator(X_i = 1) \Big)  / \sum_{i=1}^N \indicator(X_i = 1)$, assuming the denominator is nonzero. In general, let $\theta(\P)$ be a functional defined on the set of logically possible values of the population matrix. Let $\Theta$ denote the set of logically possible values of this parameter. Let $\theta^\text{true} \coloneqq \theta(\P^\text{true})$ denote its true value.

\smallskip

\textbf{4. The data:} Finally, we must describe the data that is observed to the econometrician. To allow for sampling, let $\S = (S_1,\ldots,S_N)$ be a vector of indicators where $S_i = 1$ if unit $i$ appears in the dataset, and zero otherwise. Let $\Y = (Y_1,\ldots,Y_N)$ where $Y_i \coloneqq Y_i(1) X_i + Y_i(0) (1-X_i)$ for all $i \in \mathcal{I}$. Then $\P^\text{data}$ is the matrix $(\Y, \X, \W)$ with row $i$ empty if $S_i = 0$. Formalize this construction via the function $\text{MakeData}(\P,\S)$, so that $\P^\text{data} = \text{MakeData}(\P^\text{true}, \S)$.\footnote{We define the population at the time period after which treatment has been assigned but before units are sampled. This is standard in the identification literature (e.g., \citealt{Manski2003}, ch.\ 7). One could modify the notation to further distinguish population values of potential outcomes and covariates from realized treatment assignment, as none of our concepts or procedures depend on this particular notational choice, but at the cost of additional notation throughout, especially since we mainly focus on the case where all units are sampled.}

\smallskip

We are now ready to define the identified set.

\begin{definition}\label{def:identifiedSet}
The \emph{identified set} for $\theta$ is
\[
	\Theta_I \coloneqq \{ \theta \in \Theta : \theta = \theta(\P) \text{ for some } \P \in \mathcal{P} \text{ such that } \text{MakeData}(\P,\S) = \P^\text{data} \}.
\]
\end{definition}

\nocite{Koopmans1949}

This definition follows the usual description of the identified set as ``the set of parameters...that are consistent with the model and the data'' (\citealt{Tamer2010}, page 184), but with close attention to what is meant by ``data'': In a finite population, the ``data'' is defined by a finite dimensional matrix rather than a joint distribution of random variables (as in def. 3.1 on page 5324 of \citealt{Matzkin2007}, for example; also see our appendix \ref{sec:popDistributions} for a related discussion). Our definition also allows for data on some units to be missing (i.e., not sampled, so that $S_i=0$); we discuss this further in section \ref{sec:sampling}.

$\Theta_I$ has two key features: (1) It can always be computed with the data the analyst actually has at hand, and (2) like the usual definition of the identified set in an infinite population, it is \emph{guaranteed} to contain the true parameter of interest, so long as the model is not false. We will revisit these properties below.

\subsection{Identified Sets for ATE}\label{section:ATEidentifiedSets}

To simplify the exposition, we generally assume that all units in the population are observed ($S_i=1$ for all $i \in \mathcal{I}$). We show that the extension to incorporate sampling is straightforward in section \ref{sec:sampling}. We also focus on the identification of the average treatment effect for brevity. We briefly discuss other parameters in section \ref{sec:beyondATE}. 

It is well known that bounds on mean based parameters are usually infinite without some kind of restriction on the values they can take. Hence we maintain the following assumption throughout the paper, which is standard in the literature on partial identification.

\begin{Aassumption}[Bounded outcomes]\label{assump:boundedSupport1}
There are known values $-\infty < y_\text{min} < y_\text{max} < \infty$ such that $Y_i(x) \in [y_\text{min},y_\text{max}]$ for all $i \in \mathcal{I}$, for each $x \in \{0,1\}$.
\end{Aassumption}

In some applications the values of $y_\text{min}$ and $y_\text{max}$ can be set to their logical values, such as 0 to 100 for test scores. In other settings, like when outcomes are wages, these values are sensitivity parameters that reflect our beliefs about the smallest and largest possible values of potential outcomes in the population under consideration. We discuss several variations on this assumption in section \ref{sec:boundedOutcomesVariations}, including an assumption that does not require selecting a priori bounds on outcomes.

\ref{assump:boundedSupport1} alone has identifying power for ATE, via a finite population version of the Manski \citeyearpar{Manski1990} bounds. Those bounds are a special case of theorem \ref{thm:KapproxiBalanceATEset} below, which uses the following assumption.

\begin{Aassumption}[$K$-approximate mean balance]\label{assump:KapproxBalance}
There is a known $K \geq 0$ such that $K^\text{true}(x) \leq K$ for $x \in \{0,1\}$, where
\begin{equation}\label{eq:Ktrue}
	K^\text{true}(x) \coloneqq \left| \frac{1}{N_1} \sum_{i=1}^N Y_i(x) \indicator(X_i = 1) - \frac{1}{N_0} \sum_{i=1}^N Y_i(x) \indicator(X_i = 0) \right|
\end{equation}
and $N_x \coloneqq \sum_{i=1}^N \indicator(X_i = x)$ is the number of units who receive treatment $x$, with $N_1, N_0 > 0$.
\end{Aassumption}

This assumption was proposed by \citet[page 149]{Manski2003}, who called it approximate mean independence. He noted that if $K = 0$ then it is equivalent to mean independence of potential outcomes from realized treatment. \cite{ManskiPepper2018} call this a \emph{bounded variation} type assumption. \ref{assump:KapproxBalance} could be generalized to allow a different $K$ for each potential outcome; we use a common $K$ for simplicity. Let $\overline{\Y(x)} \coloneqq \frac{1}{N} \sum_{i=1}^N Y_i(x)$ and $\overline{Y}_x \coloneqq \frac{1}{N_x} \sum_{i=1}^N Y_i \indicator(X_i = x)$. With a minor variation on Manski's \citeyearpar{Manski1990} derivations, we obtain the following characterization of the finite population identified set for ATE under the $K$-approximate mean balance assumption.

\begin{theorem}\label{thm:KapproxiBalanceATEset}
Suppose \ref{assump:boundedSupport1} and \ref{assump:KapproxBalance} hold, and $\P^\text{data}$ is known. Then, for each $x \in \{0,1\}$, the identified set for $\overline{\Y(x)}$ is $[\text{LB}_K(x), \text{UB}_K(x)]$ where
\begin{align*}
	\text{LB}_K(x) &\coloneqq \overline{Y}_x \frac{N_x}{N}  + \max \{ y_\text{min}, \overline{Y}_x - K \} \frac{N_{1-x}}{N}, \text{ and} \\
	\text{UB}_K(x) &\coloneqq \overline{Y}_x \frac{N_x}{N} + \min \{ y_\text{max}, \overline{Y}_x + K \} \frac{N_{1-x}}{N}.
\end{align*}
Moreover, the identified set for ATE is $\Theta_I(K) \coloneqq [\text{LB}_K(1) - \text{UB}_K(0), \; \text{UB}_K(1) - \text{LB}_K(0)]$.
\end{theorem}

The magnitude of uncertainty about ATE depends on the choice of $K$. At one extreme, for sufficiently large values ($K \geq \overline{K} \coloneqq \max \{ \overline{K}_1, \overline{K}_0 \}$ where $\overline{K}_x \coloneqq \max \{ y_\text{max} - \overline{Y}_x, \; \overline{Y}_x - y_\text{min} \}$), $\Theta_I(K)$ gives what \cite{Manski2003} calls a ``domain of consensus'', a set of ATE values that are consistent with the observed data and \ref{assump:boundedSupport1}, but otherwise no restriction at all on the relationship between realized treatment and potential outcomes. Denote this set by $\Theta_I(\infty)$. At the other extreme, $K=0$, mean independence holds. In this case, ATE is point identified and equals the observed difference in means, $\overline{Y}_1 - \overline{Y}_0$. In general, smaller $K$'s lead to smaller identified sets. So how should researchers assess the credibility of a choice of $K$?

\subsection{Randomization and Identification in Finite Populations}\label{sec:randomizationIdentification}

In practice, researchers often motivate assumptions that two groups are in some sense comparable by appealing to random assignment. So consider the following common formalization of random assignment in finite populations.

\begin{Aassumption}[Random assignment]\label{assump:uniformRandomization}
The size of the treatment and control groups, $N_1$ and $N_0$, are fixed a priori, with $0 < N_1 < N$. $\X = (X_1,\ldots,X_N)$ is a single realization from the known probability distribution $\Prob_\text{design}$ on $\{ 0,1\}^N$ defined by
\[
	\Prob_\text{design}(X_1^\text{new} = x_1,\ldots,X_N^\text{new} = x_N) = 1 \Big/ {N \choose N_1}
\]
for all $(x_1,\ldots,x_N) \in \{0,1\}^N$ with $\sum_{i=1}^N x_i = N_1$, and equal to 0 otherwise.
\end{Aassumption}

In this assumption we use the notation $\X^\text{new} = (X_1^\text{new},\ldots,X_N^\text{new})$ to denote the random vector with distribution $\Prob_\text{design}$. $\X = (X_1,\ldots,X_N)$ is a single, non-random realization of this random variable. Call $\Prob_\text{design}$ the \emph{design distribution} of treatment. This particular choice of design distribution is often called \emph{uniform randomization}. It is a standard formalization of randomization in the design-based inference literature; for example, see \citet[section 4.4]{ImbensRubin2015}. We conjecture that most of our results extend to other design distributions commonly used in randomized experiments (as surveyed in \citealt{RosenbergerLachin2015}, for example), but we focus on uniform randomization for brevity.

As has long been recognized, randomization in finite populations does not guarantee any level of \emph{ex post} balance (e.g., \citealt{Greenland1990}). The identified set, by definition, must contain the true parameter value whenever the model assumptions are true. Consequently, for any finite population, randomization has no \emph{identifying} power. The following theorem formalizes this result.

\begin{theorem}\label{thm:randomizationUseless}
Suppose \ref{assump:boundedSupport1} and \ref{assump:uniformRandomization} hold and $\P^\text{data}$ is known. Then the identified set for ATE is $\Theta_I(\infty)$.
\end{theorem}

Put differently, after treatment has been assigned, we cannot rule out the possibility that potential outcomes are substantially imbalanced across the treatment and control groups, even if it was unlikely a priori. Thus the assumption that treatment was randomly assigned does not rule out any values of the unknown potential outcomes. Hence it does not shrink the domain-of-consensus bounds in finite populations. Nonetheless, in section \ref{sec:sensitivityAnalysis} we will reinterpret randomization as a procedure that \emph{affects our beliefs about ex post balance}. Specifically, we will use randomization to assess the plausibility of a specific choice of the sensitivity parameter $K$. This will let us use the identification result in theorem \ref{thm:KapproxiBalanceATEset} to perform a sensitivity analysis motivated by random assignment of treatment. That analysis will yield uncertainty intervals that have interpretations as identified sets, robust Bayesian credible sets, or uniform frequentist confidence sets.

\subsubsection*{Exact Balance in Infinite versus Finite Populations}

In infinite population identification analysis, random assignment is typically formalized as the assumption that realized treatment is statistically independent from potential outcomes, $X \indep (Y(1),Y(0))$. This assumption implies mean independence, which in a finite population is equivalent to $K=0$. But in a finite population, $K=0$ is an \emph{exact} balance assumption. It requires that $\frac{1}{N_1} \sum_{i=1}^N Y_i(x) \indicator(X_i = 1) = \frac{1}{N_0} \sum_{i=1}^N Y_i(x) \indicator(X_i = 0)$ for $x \in \{0,1\}$. In fact, for many values of the vectors $\Y(x) = (Y_1(x),\ldots, Y_N(x))$, it is impossible for exact balance to hold regardless of the values of realized treatment $\X = (X_1,\ldots,X_N)$, a point noted by \citet[page 415]{GreenlandRobins1986}. Hence statistical independence is not an appropriate formalization of random assignment in finite populations. 

This qualitative distinction between the finite and infinite setting explains why theorem \ref{thm:randomizationUseless} comes to a different conclusion about identification of ATE from traditional infinite population identification analyses: In the infinite population setting, exact balance is assumed to hold and hence ATE is point identified. But in the finite population setting, exact balance is not guaranteed even under random assignment, and is sometimes even logically impossible, and hence ATE is partially identified. That said, because any level of approximate balance is more likely when $N$ is large, there is a connection between these two results, which we study in section \ref{sec:valueOfRandomization}. Also see section \ref{sec:beyondATE}, where we discuss various measures of balance.

\subsection{The Prior Definition of Finite Population Identification}\label{sec:tradDef}

Now that we have explained our approach to identification, we turn to examine the prior definition. First, what \emph{is} the prior definition? We are not aware of any books or papers that explicitly and formally define identification for finite populations. For example, \cite{BorusyakHullJaravel2024} discuss identification but do not provide a formal definition, and as representative of the rest of the literature, their first formal result is a consistency proof, not an identification theorem. The book by \cite{Tille2020} is an extensive and detailed survey of sampling from finite populations, but does not discuss identification. The textbook by \cite{Ding2024} discusses causal inference in finite populations in detail, starting from chapter 2, but a definition of identification does not appear until chapter 10, only after shifting focus from randomized experiments to observational data. That definition (10.1 on page 128) states that ``a parameter $\theta$ is nonparametrically identifiable if it can be written as a function of the distribution of the observed data without any parametric model assumptions''. Similar definitions appear in other references, going back to the earliest sources (such as \citealt{Koopmans1949}, \citealt{KoopmansReiersol1950}, \citealt{Hurwicz1950}, \citealt{Rothenberg1971}). Verbal definitions like these are ambiguous about the meaning of ``the distribution of the observed data'' (\citealt{Ding2024}) or the ``joint distribution of the observations'' (\citealt{Koopmans1949}, page 125). There are different ways to formalize such statements and they lead to substantially different definitions of identification. As we argue below, this issue is particularly important in the finite population setting.

Our definition \ref{def:identifiedSet} is one such formalization, where by ``data'' we mean the matrix $\P^\text{data}$ that the researcher actually has at hand. In this case the ``distribution'' of this data can be formalized as a specific discrete distribution (see appendix \ref{sec:popDistributions} for details). However, based on the informal discussions in the literature, and our personal communication with researchers in this field, ``data'' is conventionally interpreted as the \emph{design distribution} of the observed data, rather than the dataset that is actually observed. This interpretation leads to the following definition of point identification for finite populations.

\begin{definition}[Conventional]\label{def:conventionalIdentification}
Let $(\Y(1),\Y(0),\W)$ be the $N \times (2+ d_W)$ matrix of potential outcomes and covariate values for all units in the population $\mathcal{I} = \{1,\ldots,N\}$. Let $\theta(\cdot)$ be a functional defined on the set of logically possible values of this matrix. Let $\theta^\text{true} \coloneqq \theta((\Y(1),\Y(0),\W))$ denote its true value. For each $i \in \mathcal{I}$, let $X_i^\text{new} \in \{0,1\}$ and $S_i^\text{new} \in \{0,1\}$ be random variables. Let $\X^\text{new} \coloneqq (X_1^\text{new},\ldots,X_N^\text{new})$ and $\S^\text{new} \coloneqq (S_1^\text{new}, \ldots, S_N^\text{new})$. Let $\Prob_\text{design}$ denote the joint distribution of $(\X^\text{new},\S^\text{new})$; it is called the design distribution. Let $Y_i^\text{new} \coloneqq Y_i(X_i^\text{new})$. Let $\mathcal{S} = \{ i \in \mathcal{I} : S_i^\text{new} = 1\}$ denote the random set of sampled indices. Let $\mathcal{D} \coloneqq \{ (i, Y_i^\text{new}, X_i^\text{new}, W_i) \in \mathcal{I} \times \R^{2+d_W} : i \in \mathcal{S} \}$. Say $\theta^\text{true}$ is \emph{point identified} if there is a known mapping from the probability distribution of the random set $\mathcal{D}$ to $\theta^\text{true}$.
\end{definition}

We believe definition \ref{def:conventionalIdentification} accurately formalizes the conventional definition of finite population identification in the prior literature. For example, \citet[footnote 6, page 4786]{HeckmanVytlacil2007} say ``identification in small samples requires establishing the sampling distribution of estimators, and adopting bias as the criterion for identifiability'' (see our proposition \ref{prop:designMomentConditions} below) and that ``this approach is conventional in classical statistics,'' although they do not give a formal definition which we could compare against. This definition also resembles the definition of ``sampling identification'' in \citet[definition 2.1 on page 5]{FlorensSimoni2021}, although that paper is not about design-based inference or finite populations. 

Definition \ref{def:conventionalIdentification} has the following seemingly innocuous and familiar implication.

\begin{proposition}\label{prop:designMomentConditions}
Consider the setting of definition \ref{def:conventionalIdentification}. Suppose there is a known function $m$ such that $\theta^\text{true}$ is the unique solution to $\Exp_\text{design}[ m(\mathcal{D}, \theta)] = 0$. Then $\theta^\text{true}$ is point identified.
\end{proposition}

This result says that a parameter is point identified if it is the unique solution to a set of design-based moment conditions. A special case arises when there is an $m$ is such that we can write $\theta^\text{true} = \Exp_\text{design}[m(\mathcal{D})]$. In this case the function $m(\mathcal{D})$ is usually called a design-unbiased estimator of $\theta^\text{true}$. Moment-based reasoning like in proposition \ref{prop:designMomentConditions} is used in the identification analysis of \citet[page 88, first paragraph of section 3.1]{BorusyakHullJaravel2024} and \citet[page 13, first displayed equation in second column]{RambachanRoth2025}. Proposition \ref{prop:designMomentConditions}, however, implies the following result.

\begin{proposition}\label{prop:weirdWeird}
Consider the setting of definition \ref{def:conventionalIdentification}. Suppose $\Prob_\text{design}(S_i^\text{new}=1) = 1$ for all $i \in \mathcal{I}$. Then, according to definition \ref{def:conventionalIdentification}, the following hold.
\begin{enumerate}
\item $p_i \coloneqq \Prob_\text{design}(X_i^\text{new} = 1)$ is point identified for all $i \in \mathcal{I}$.

\item $Y_i(1)$ is point identified for each $i \in \mathcal{I}$ with $p_i > 0$. $Y_i(0)$ is point identified for each $i \in \mathcal{I}$ with $p_i < 1$.
\end{enumerate}
\end{proposition}

\begin{proof}[Proof of proposition \ref{prop:weirdWeird}]
Part 1 follows from $p_i = \Exp_\text{design}(X_i^\text{new})$ and proposition \ref{prop:designMomentConditions}. For part 2, define
\[
	\widehat{Y_i(1)} \coloneqq \frac{X_i^\text{new} Y_i^\text{new}}{p_i}
	\qquad \text{and} \qquad
	\widehat{Y_i(0)} \coloneqq \frac{(1-X_i^\text{new}) Y_i^\text{new}}{1-p_i}.
\]
Then
\[
	\Exp_\text{design}[\widehat{Y_i(1)}]
	= \frac{1}{p_i} \Exp_\text{design}[ X_i^\text{new} Y_i(X_i^\text{new}) ] \\
	= \frac{1}{p_i} \Exp_\text{design}[ X_i^\text{new} ] Y_i(1) \\
	= Y_i(1).
\]
Similarly, $\Exp_\text{design}[\widehat{Y_i(0)}] = Y_i(0)$. Apply proposition \ref{prop:designMomentConditions}.\footnote{An alternative proof defines $\theta \coloneqq Y_i(1)$, sets $m(\mathcal{D},\theta) \coloneqq Y_i^\text{new} X_i^\text{new} - \theta X_i^\text{new}$, and then notes that $\theta$ solves $\Exp_\text{design}[m(\mathcal{D},\theta)] = 0$.}
\end{proof}

We assumed all units are sampled in proposition \ref{prop:weirdWeird} for simplicity, but the result can be generalized to allow for sampling. Proposition \ref{prop:weirdWeird} combines ideas from \citet[footnote 3]{Imbens2018} and \cite{ChenRothSpiess2026}. Specifically, using a different proof strategy and restricting attention to the binary outcome case, \cite{ChenRothSpiess2026} showed that the proportion of units who benefit from treatment is point identified, according to the conventional definition \ref{def:conventionalIdentification} of identification in finite populations. \cite{Imbens2018} considered a randomized experiment with $N=1$ and $\Prob_\text{design}(X_1^\text{new}=1) = 0.5$ and showed that there exists a design-unbiased estimator of that units' unit-level treatment effect, $Y_1(1) - Y_1(0)$. Our proposition \ref{prop:weirdWeird} builds on these ideas to show that all potential outcomes are point identified, according to definition \ref{def:conventionalIdentification}, as long as the probability of treatment is strictly between zero and one for all units.\footnote{Following the original design-based literature in survey sampling (e.g., page 19 of \citealt{CasselSarndalWretman1977} or page 10 of \citealt{Thompson1997}) and more recent design-based literature in economics such as \citealt{AbadieAtheyImbensWooldridge2020} (who work with data matrices where the row position encodes the unit identifier), in proposition \ref{prop:weirdWeird} we assume that unit identifiers are observed in the data set. For example, if the population consists of all U.S. states, then the data consists of each state's name $i$ and its corresponding realized treatment $X_i$ and outcome $Y_i$. Chen et al's \citeyearpar{ChenRothSpiess2026} proposition 3.1 shows that the proportion of units who benefit from treatment is point identified according to the conventional definition \ref{def:conventionalIdentification} even if only \emph{anonymized} data is observed, meaning that the data is a list of $(Y_i,X_i)$ values, but it is not known which units these values belong to. Our proposition \ref{prop:weirdWeird} does not apply to this kind of anonymized data. However, suppose that the anonymized data includes covariates $W_i$. Then as long as the covariates are rich enough to uniquely identify units---which will typically be the case as long as there is a single ``continuous'' covariate, for example---the conclusion of proposition \ref{prop:weirdWeird} will continue to hold, even with anonymized data. This follows because the covariates effectively become unit identifiers.}

Proposition \ref{prop:weirdWeird} applies to randomized experiments, which typically guarantee $p_i \in (0,1)$ by design (e.g., \ref{assump:uniformRandomization} holds), but it also applies to many observational settings as well, because this result does \emph{not} assume that $p_i$ is known for all $i \in \mathcal{I}$. For example, unconfoundedness restricts the treatment assignment probabilities $p_i$ to be functions of observed covariates (in which case it is called the propensity score). Proposition \ref{prop:weirdWeird} implies that such a priori restrictions on treatment probabilities are unnecessary to point identify potential outcomes, according to definition \ref{def:conventionalIdentification}; it could even be that the unit-level treatment assignment probabilities $p_i = f(Y_i(1),Y_i(0))$ are functions of potential outcomes themselves; all that matters is that the ex ante probability of treatment is not degenerate on 0 or 1. Put differently, unconfoundedness is unnecessary according to the conventional definition of identification; overlap ($p_i \in (0,1)$ for all $i \in \mathcal{I}$) is sufficient for point identification of all unit-level treatment effects. Similarly, the conventional definition implies that all unit-level treatment effects are point identified for compliers in an instrumental variables setting, even if the instrument is endogenous (see appendix \ref{sec:conventionalAppendix} for details).

Essentially all of these conclusions implied by proposition \ref{prop:weirdWeird} differ from the conclusions of classical infinite population identification theory. The fact that a parameter can be partially identified under infinite population theory but point identified according the conventional definition was first pointed out by \cite{ChenRothSpiess2026}, who focused on the proportion of units who benefit from treatment with a binary outcome.\footnote{They use this finding to motivate a careful study of the implications of the conventional definition of identification for Bayesian updating and various frequentist concepts, rather than exploring different definitions of identification as we do.} In a randomized experiment, that parameter is generally partially identified in infinite population theory (\citealt{Makarov1982}, \citealt{Manski1997}, \citealt{FanPark2010}), and yet the conventional definition \ref{def:conventionalIdentification} implies that it is point identified. As they note, in the binary treatment and binary instrument setting of \cite{ImbensAngrist1994}, this implies that the proportion of compliers is point identified even without a no-defiers assumption, another inconsistency with infinite population theory. Our result goes further to show that the compliance type (i.e., complier, defier, always taker, or never taker) is point identified for every unit, according to the conventional definition of identification (see appendix \ref{sec:conventionalAppendix}), a further inconsistency with infinite population theory. 

\subsubsection*{Discussion: Averaging over Units or over Counterfactual Worlds}

The different conclusions implied by different definitions of identification are a direct consequence of how ``data'' is defined. The conventional definition \ref{def:conventionalIdentification} uses data that is not even in principle available to real-world researchers. That follows since means like $\Exp_\text{design}$ average over hypothetical, counterfactual worlds that are mutually incompatible with each other---such as a world where unit $i$ is treated and a world where unit $i$ is not treated. By allowing identification to depend on information from both such worlds, this definition allows one to conclude that both $Y_i(1)$ and $Y_i(0)$ are point identified. This conclusion suggests that there is, in fact, no fundamental problem of causal inference (\citealt{Holland1986}).

In stark contrast, both our definition \ref{def:identifiedSet} and the standard definition of infinite population identification do \emph{not} assume one can observe ``data'' from such mutually incompatible worlds. This holds for our definition \ref{def:identifiedSet} since the identified set only depends on $\P^\text{data}$, the matrix of data that is actually observed by real empirical researchers. The standard infinite population definition is also not based on data from incompatible worlds, and hence does not lead to results like proposition \ref{prop:weirdWeird}. Specifically, consider a joint distribution of random variables $(Y,X)$, representing the population ``data'' of realized outcomes and treatments. As explained in \citet[section 2.2]{HeckmanVytlacil2007}, for example, each realization $(Y(\omega), X(\omega))$ from this distribution represents a single unit $\omega$ in some set of units $\Omega$ (they assume $\Omega = [0,1]$) who is either treated ($X(\omega)=1)$ or not ($X(\omega)=0$). Consequently, even though the population is infinite, each unit is only treated once. The ``data'' available to researchers comes from a \emph{single} assignment of each of the infinitely many units to be treated or not. Concretely, infinite population expectations like $\Exp \left[ \frac{Y X}{\Prob(X=1 \mid W)} \right]$, as used in a selection-on-observables analysis (e.g., \citealt{ImbensRubin2015}, section 12.4), do not involve averaging worlds where some units are treated with other worlds where those \emph{same} units are \emph{not} treated. Rather, such averages are over different \emph{units} in the infinite population instead of over different \emph{counterfactual worlds} involving different treatment assignments, as in $\Exp_\text{design}$.

In summary, these issues with the conventional definition motivate our development of a different theory of identification, based on definition \ref{def:identifiedSet}.

\section{Design-Based Sensitivity Analysis}\label{sec:sensitivityAnalysis}

Thus far we have defined and studied identification in finite populations. We also derived the identified set for ATE, $\Theta_I(K)$, under the $K$-approximate mean balance assumption. That set provides a non-stochastic quantification of uncertainty about ATE, a mapping from a class of deterministic assumptions about balance into values of ATE consistent with the observed data and the assumptions. Researchers may be unsure about which specific values of $K$ are plausible, however. In this section, we show how to use a known design distribution of treatment to probabilistically quantify uncertainty about $K$, which can then be combined with our identified sets to probabilistically quantify uncertainty about ATE.

\subsection{A Design-Based Approach to Calibrating $K$}\label{sec:designBasedMainSection}

The identified set $\Theta_I(K)$ as a function of the sensitivity parameter $K$, as in the top plot of figure \ref{fig:GneezyDesignBased}, shows how our conclusions can vary from point identification of ATE (under exact balance, $K=0$) to partial identification under approximate balance ($K > 0$). Like any sensitivity analysis, however, there is an important question: Which values of $K$ are most plausible? In this section, we provide an objective approach to \emph{calibrating} this sensitivity parameter, based on an assumption that treatment was randomly assigned. Specifically, define
\begin{multline*}
	p(K, \Y(1), \Y(0)) \coloneqq \\
	\Prob_\text{design} \left( \left| \frac{1}{N_1} \sum_{i=1}^N Y_i(x) \indicator(X_i^\text{new} = 1) - \frac{1}{N_0} \sum_{i=1}^N Y_i(x) \indicator(X_i^\text{new} = 0) \right| \leq K \text{ for $x \in \{0,1\}$} \right).
\end{multline*}
This is the design-probability that the $K$-approximate mean balance assumption holds, when the true potential outcomes are $\Y(1)$ and $\Y(0)$. Although $\Prob_\text{design}$ is known, $p(K,\Y(1),\Y(0))$ is unknown since it depends on unknown values of the potential outcomes.

We can, however, think of $p(K,\Y(1),\Y(0))$ itself as a parameter---it is a functional of the population matrix of potential outcomes, holding $K$ fixed. We will show that this parameter is itself partially identified, and we can derive its identified set, in the same sense as we developed in section \ref{sec:finitePopIdent}. Specifically, we obtain bounds on this probability by using two constraints: (1) Half of all potential outcomes are known (namely, those associated with the observed treatments) and (2) Outcomes are bounded (\ref{assump:boundedSupport1}).

We focus on the lower bound on $p(K,\Y(1),\Y(0))$, since this tells us the smallest design probability such that $K$-approximate mean balance is guaranteed. So let
\begin{equation}\label{eq:defOfunderlinep}
	\underline{p}(K) \coloneqq \inf_{
	\substack{
	\Y(1), \Y(0) \in [y_\text{min},y_\text{max}]^N \\
	\Y = \Y(1) \times \X + \Y(0) \times (\bold{1} - \X) 
	}
	} \; p\big( K, \Y(1), \Y(0) \big)
\end{equation}
where $\times$ denotes component-wise multiplication and $\bold{1}$ is an $N$-vector of ones. $\underline{p}$ is a known function. It depends on the realized data $(\Y,\X)$ and the design distribution $\Prob_\text{design}$. We discuss how to feasibly compute this function in section \ref{sec:computation}. For now we focus on its interpretation and use. Note that, by construction, $\underline{p}(K) \leq p(K,\Y(1),\Y(0))$ for any $\Y(1),\Y(0)$ satisfying the constraints in \eqref{eq:defOfunderlinep}; in particular, this holds for the true population values of potential outcomes.

We use $\underline{p}(K)$ to \emph{calibrate plausible values of the sensitivity parameter $K$ in the identification results of section \ref{sec:finitePopIdent}}. Specifically, we suggest performing what we call a \emph{design-based sensitivity analysis}:
\begin{enumerate}
\item First, plot $\Theta_I(K)$, the identified set for ATE, as a function of $K$. Below this, plot the function $\underline{p}(K)$. Figure \ref{fig:GneezyDesignBased} gives an example of this paired plot. The top graph shows the sequence of finite population identified sets alone. The second plot then converts its horizontal axis values of $K$ into design-probabilities.

\item Focus on particular values of $K$. There are several reasonable ways to do this:
\begin{enumerate}
\item Define the breakdown point $K^\text{bp} \coloneqq \sup \{ K \geq 0 : 0 \notin \Theta_I(K) \}$, the largest relaxation of exact balance such that zero is not in the identified set. Researchers can compute $\underline{p}(K^\text{bp})$ to interpret this value. In figure \ref{fig:GneezyDesignBased} it is 0.65. Thus there was at least a 65\% ex ante probability that potential outcomes would be sufficiently balanced that we can conclude that ATE is positive.

\item Next, notice that $\underline{p}(K)$ decreases as $K$ decreases, because smaller $K$ implies a stronger balance assumption that is therefore less likely to hold. Let $\alpha \in (0,1)$. Define $K(\alpha) \coloneqq \inf \{ K \geq 0 : \underline{p}(K) \geq 1-\alpha \}$, the closest we can get to exact balance while still ensuring that approximate balance holds with design-probability at least $1-\alpha$. Researchers can then present the set $\Theta_I \big( K(\alpha) \big)$ as a function of $1-\alpha$. Figure \ref{fig:GneezyDesignBased2} gives an example.

\end{enumerate}
\end{enumerate}

\subsection{Interpreting $\Theta_I(K(\alpha))$}\label{sec:interpretation}

Our main recommendation is that researchers report the set $\Theta_I(K(\alpha))$ as a function of $1-\alpha$, as in figure \ref{fig:GneezyDesignBased2}. This set can then be interpreted three different ways. The first interpretation is non-probabilistic while the second two are probabilistic.

First, $\Theta_I(K(\alpha))$ can be interpreted as the finite population identified set for ATE under different deterministic assumptions about the true, but unknown, magnitude of ex post imbalance, $K^\text{true,max} \coloneqq \max \{ K^\text{true}(1), \allowbreak K^\text{true}(0) \}$ (recall that $K^\text{true}(x)$ is defined in eq.\ \eqref{eq:Ktrue}); namely, under the assumption that $K^\text{true,max}$ is at most $K(\alpha)$. This follows immediately from the identification analysis of section \ref{sec:finitePopIdent}. However, since researchers may not know how large $K^\text{true,max}$ is likely to be, our second and third interpretations view $[0, K(\alpha)]$ as an uncertainty interval for $K^\text{true,max}$, using either a robust Bayesian or uniform frequentist interpretation. Using our identification analysis in section \ref{sec:finitePopIdent}, these uncertainty intervals over $K^\text{true,max}$ then translate into uncertainty intervals over ATE, yielding $\Theta_I(K(\alpha))$, as we discuss next.

Second, viewed as a function of $\alpha$, $\Theta_I(K(\alpha))$ can be interpreted as a robust empirical Bayesian $1-\alpha$ credible set for ATE. To see this, consider the classical, ``fully'' Bayesian approach (e.g., \citealt{Rubin1978} or section 8.4 of \citealt{ImbensRubin2015}). This approach starts with a prior on the entire matrix $(\Y(1),\Y(0))$. The random assignment of treatment assumption \ref{assump:uniformRandomization} then implies corresponding beliefs over the magnitude of imbalance between the treatment and control groups. This is sufficient to obtain a distribution over $K^\text{true,max}$. Knowing that distribution alone would allow us construct a Bayesian credible set for ATE, by using our identified sets $\Theta_I(K)$ from theorem \ref{thm:KapproxiBalanceATEset} and treating $K$ as the uncertain quantity.

However, as usual with Bayesian analyses, it is unclear what specific prior over $(\Y(1),\Y(0))$ should be used. To address this, we first show that $\underline{p}$ is a valid cdf on $\R_+$:

\begin{proposition}\label{prop:underlinePisAcdf}
Suppose \ref{assump:boundedSupport1} and \ref{assump:uniformRandomization} hold. Then for any realization $\X$, $\underline{p}$ is monotonic, $\underline{p}$ is right continuous, $\lim_{K \rightarrow 0} \; \underline{p}(K) = 0$, and $\lim_{K \rightarrow \infty} \; \underline{p}(K) = 1$.
\end{proposition}

From this result and the definition of $\underline{p}$, we can show that $\underline{p}$ is a ``worst case'' or ``robust'' distribution over $K^\text{true,max}$, in the sense that every prior on $(\Y(1),\Y(0))$ yields a distribution over $K^\text{true,max}$ that is ``more optimistic'' about balance than the distribution $\underline{p}$. In other words, $[0,K(\alpha)]$ is a $1-\alpha$ credible set for $K^\text{true,max}$, based on the distribution $\underline{p}$, and this set is also a $1-\alpha$ credible set based on any prior distribution over $(\Y(1),\Y(0))$ satisfying \ref{assump:boundedSupport1}.

This distribution $\underline{p}$ over $K^\text{true,max}$ induces a distribution over the sets $\Theta_I(K)$. Since $\Theta_I(K)$ contains the true ATE so long as $K \leq K^\text{true,max}$ and the model is not falsified, this implies that $\Theta_I(K(\alpha))$ is a robust empirical Bayesian $1-\alpha$ credible set for ATE: It is a $1-\alpha$ credible set for ATE for any prior on $(\Y(1),\Y(0))$ that is consistent with our baseline assumption on potential outcomes, \ref{assump:boundedSupport1}. We call it an ``empirical'' Bayes credible set because the distribution $\underline{p}$ depends on the observed data $(\Y,\X)$. We give additional details and discussions of these results in appendix \ref{sec:moreBayesian}.

This Bayesian interpretation is based on obtaining a distribution over the sets $\Theta_I(K)$, rather than obtaining a posterior distribution for ATE itself. This distinction is well known from the previous literature on Bayesian inference in partially identified settings. One consequence is that, like that previous literature (e.g., \citealt{poirier1998revising}, \citealt{moon2012bayesian}, \citealt{kline2016bayesian}, \citealt{giacomini2021robust}), we cannot make probabilistic statements about where ATE is \emph{within} the set $\Theta_I(K(\alpha))$. This analysis nonetheless allows us to make some probabilistic statements about ATE directly. For example, in figure \ref{fig:Gneezy}, we can say that there is at least a 90\% posterior probability that ATE is in $\Theta_I(K(0.1)) = [-6,22]$, and that there is at least a 65\% posterior probability that ATE is non-negative, regardless of what one's prior on $(\Y(1),\Y(0))$ is.

Third, $\Theta_I(K(\alpha))$ has a frequentist interpretation as a uniform design-based confidence interval. To make the frequentist thought experiment explicit, write $\mathcal{C}(\Y(1) \times \X + \Y(0) \times (\bold{1}-\X),\X) = \Theta_I(K(\alpha))$ to emphasize that both the identified set $\Theta_I(K)$ and $K(\alpha)$ depend on the vector of realized treatments $\X$ and the potential outcome vectors $\Y(1)$ and $\Y(0)$. Let $\theta\big( \Y(1), \Y(0) \big) \coloneqq \frac{1}{N} \sum_{i=1}^N Y_i(1) - Y_i(0)$ denote the average treatment effect functional.

\begin{theorem}\label{thm:ValidCoverage}
Suppose \ref{assump:boundedSupport1} and \ref{assump:uniformRandomization} hold. Then
\[
\resizebox{0.98\linewidth}{!}{$
	\displaystyle \inf_{\Y(1), \Y(0) \in [y_\text{min},y_\text{max}]^N}
	\;
	\Prob_\text{design}\Big( \mathcal{C}(\Y(1) \times \X^\text{new} + \Y(0) \times (\bold{1}-\X^\text{new}),\X^\text{new}) \ni \theta \big( \Y(1),\Y(0) \big) \Big) \geq 1-\alpha.
$}
\]
\end{theorem}

As usual in frequentist design-based finite population analyses, $\X^\text{new}$ is the only random quantity under consideration. Theorem \ref{thm:ValidCoverage} shows that $\Theta_I(K(\alpha))$ is a $100 (1-\alpha)$\% design-based confidence interval. That is, across repeated random assignments of treatment, $\Theta_I(K(\alpha))$ will contain the true parameter value with design-probability at least $1-\alpha$.

Viewed as a confidence interval, $\Theta_I(K(\alpha))$ has several key features that distinguish it from alternatives in the literature: First, it is valid for any fixed $N$; that is, it does not rely on an assumption that a large-$N$ asymptotic approximation holds. Second, it does not require strong assumptions like treatment effect homogeneity. And third, it is uniformly valid over any $(\Y(1),\Y(0))$ matrix satisfying \ref{assump:boundedSupport1}. Uniform validity is well known to be an important property in partially identified settings (e.g, \citealt{CanayShaikh2017}). In appendix \ref{sec:simulations} we discuss all of these features in more detail, give further discussion of the literature on design-based confidence intervals for ATE, and use simulations to study coverage probabilities.

These features of $\Theta_I(K(\alpha))$ all follow from its construction as a finite population identified set for ATE with sensitivity parameter $K = K(\alpha)$ calibrated to be large enough that it will be larger than $K^\text{true,max}$ often enough, across repeated random assignments of treatment. That is, our approach of first studying identification of ATE under non-probabilistic assumptions helps focus attention on $K^\text{true,max}$ as a key unknown parameter, which led us to probabilistically quantify uncertainty on ATE by first probabilistically quantifying uncertainty in $K^\text{true,max}$. This approach helps us overcome the downsides of the traditional methods for constructing design-based confidence intervals based on inverting hypothesis tests for sharp nulls or using large-$N$ asymptotics.

\subsection{The Value of Randomization}\label{sec:valueOfRandomization}

In section \ref{sec:randomizationIdentification} we showed that, in any finite population, random assignment of treatment does not have any identifying power because it does not guarantee any particular level of ex-post balance. However, random assignment of treatment does impact our \emph{beliefs} about balance. This led to our construction of robust Bayesian credible sets above. Our next result shows that our set $\Theta_I(K(\alpha))$ yields tight conclusions about ATE in large populations.

\begin{theorem}\label{thm:randomizationGood}
Consider a sequence of finite populations, $\{ (Y_i(0)_N, Y_i(1)_N) : i =1,\ldots,N \}$. Assume that for each $x \in \{0,1\}$ there is a constant $\mu(x) \in \R$ such that $\frac{1}{N} \sum_{i=1}^N Y_i(x)_N \rightarrow \mu(x)$ as $N \rightarrow \infty$. Suppose \ref{assump:boundedSupport1} holds and that the bounds $y_\text{min}$ and $y_\text{max}$ do not depend on $N$. Assume $N_1 / N \rightarrow \rho$ for some constant $\rho \in (0,1)$. Suppose that for each $N$, a vector of treatment assignments $\X_N$ is drawn from the random vector $\X_N^\text{new}$ that satisfies \ref{assump:uniformRandomization}. Then:
\hfill
\begin{enumerate}
\item If $\mu(1) - \mu(0) \neq 0$, $\underline{p}(K^\text{bp}) \xrightarrow{p} 1$ as $N \rightarrow \infty$.

\item Let $\text{ATE}_N$ denote the finite population ATE. For any $\alpha \in (0,1)$,
\[
	\sup_{\theta \in \Theta_I(K(\alpha))} | \theta - \text{ATE}_N | \xrightarrow{p} 0
	\qquad \text{as $N \rightarrow \infty$.}
\]
\end{enumerate}
\end{theorem}

Theorem \ref{thm:randomizationGood} is one way to formally show that randomly assigning treatment helps with learning about causal effects, in an explicitly finite population setting. Essentially, unobserved potential outcomes are ``more likely'' to be ``more balanced'' across the treatment and control groups in larger populations than small ones.

The first part of theorem \ref{thm:randomizationGood} concerns identification of the sign of ATE. It shows that, in large enough finite populations, the assumption of $K^\text{bp}$-approximate mean balance will indeed hold with probability approaching $1$. By definition of the breakdown point, this means that the ``naive'' conclusion about the sign of ATE (based on the exact balance assumption of $K = 0$) will be robust, in the sense that it is almost guaranteed that the amount of \emph{imbalance} required to overturn that conclusion will not occur.

The second part of theorem \ref{thm:randomizationGood} concerns the behavior of the credible sets and confidences sets constructed above, $\Theta_I(K(\alpha))$. It is a consistency result: It shows that this set converges to the singleton true value of ATE in larger finite populations. This implies, for example, that if you invert $\Theta_I(K(\alpha))$ to construct a design-based test of the hypothesis $H_0: \text{ATE}_N = \theta_0$ then this test is consistent; it will reject false nulls with high design-probability when $N$ is large enough.

\subsection{Measuring the Strength of Evidence: $\underline{p}(K^\text{bp})$ as an Alternative to $p$-values}\label{sec:strengthOfEvidence}

$p$-values are commonly interpreted as quantitative measures of the evidence against the null hypothesis, with small values interpreted as ``strong evidence'' against the null and larger values interpreted as weaker evidence against it. This is a controversial interpretation (e.g., see the entire 2019 special issue of \emph{The American Statistician} on ``a world beyond $p < 0.05$''). In light of this debate about $p$-values, we suggest that our measure $\underline{p}(K^\text{bp})$ can be viewed as an alternative summary statistic that has a well justified interpretation as a quantitative measure of how much evidence the data provides for a specific conclusion. In particular, because our analysis has a Bayesian interpretation, it allows us to make certain probabilistic statements about the true value of the parameter. For example, when the naive difference in means estimate is positive, $\underline{p}(K^\text{bp})$ is a lower bound on the posterior probability that the true ATE is nonnegative.

\subsection{Computing Bounds on the Design-Probability of $K$-Approximate Balance}\label{sec:computation}

Our approach requires computing the function $\underline{p}(K)$, which involves solving an optimization problem over $N$ variables. In appendix \ref{sec:MILPappendix}, we show that $\underline{p}(K)$ is the optimized value in a mixed integer linear programming (MILP) problem. Consequently, standard software for solving these problems can be used. However, as we discuss in section \ref{sec:numericalIllustration}, the MILP solver can often be very slow. So we also recommend that users try alternative solvers to compute $\underline{p}$. We have found that genetic algorithms (\citealt{KochenderferWheeler2019}, pages 148--156) work exceptionally well in this setting, delivering results that are very close to the solution from MILP but in a small fraction of the time. See section \ref{sec:numericalIllustration} for a discussion of the numerical evidence.

A second issue concerns computation of the objective function, which involves a summation over the set of all possible treatment assignments. For small values like $N=10$ this is feasible but it can become infeasible for moderate values like $N=100$. This issue is well known in design-based inference; e.g., sec.\ 5.8 in \cite{ImbensRubin2015}. Like them, we address this issue by sampling from the set of all treatment assignments and using this to approximate the objective function. In section \ref{sec:numericalIllustration} we show that our results are quite insensitive to the choice of sample size.

\section{Numerical Illustration}\label{sec:numericalIllustration}

Next we use simulated data to illustrate the analysis of section \ref{sec:sensitivityAnalysis}. We generated a nested sequence of five populations, with $N \in \{ 10, 20, 40, 100, 400 \}$. These populations have $[y_\text{min}, y_\text{max}]=[0,1]$ and heterogeneous treatment effects such that $\text{ATE}_N$ converges to 0.25 as $N$ gets large. We assigned treatment via \ref{assump:uniformRandomization} such that $N_1 = N/2$ for all $N$ while ensuring the populations are nested. In this illustration the assignment happens only once. Appendix \ref{sec:numericalAppendix} gives further details on how we generated the data.

\begin{figure}[t]
  \centering

  \begin{subfigure}[c]{0.5\textwidth}
    \centering
    \includegraphics[width=\linewidth]{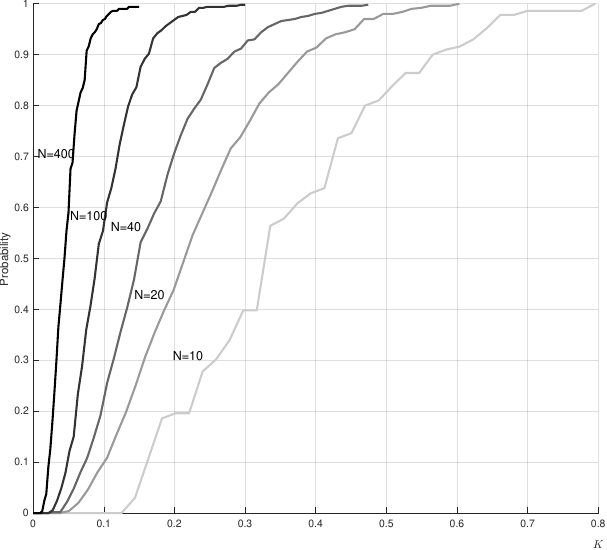}
    \caption{Convergence of $\underline{p}$ to a step function at zero as population size $N$ increases. \label{fig:covergence_of_p_as_N_increases}}
  \end{subfigure}
  \hfill
  \begin{subfigure}[c]{0.45\textwidth}
    \centering
       \includegraphics[width=\linewidth]{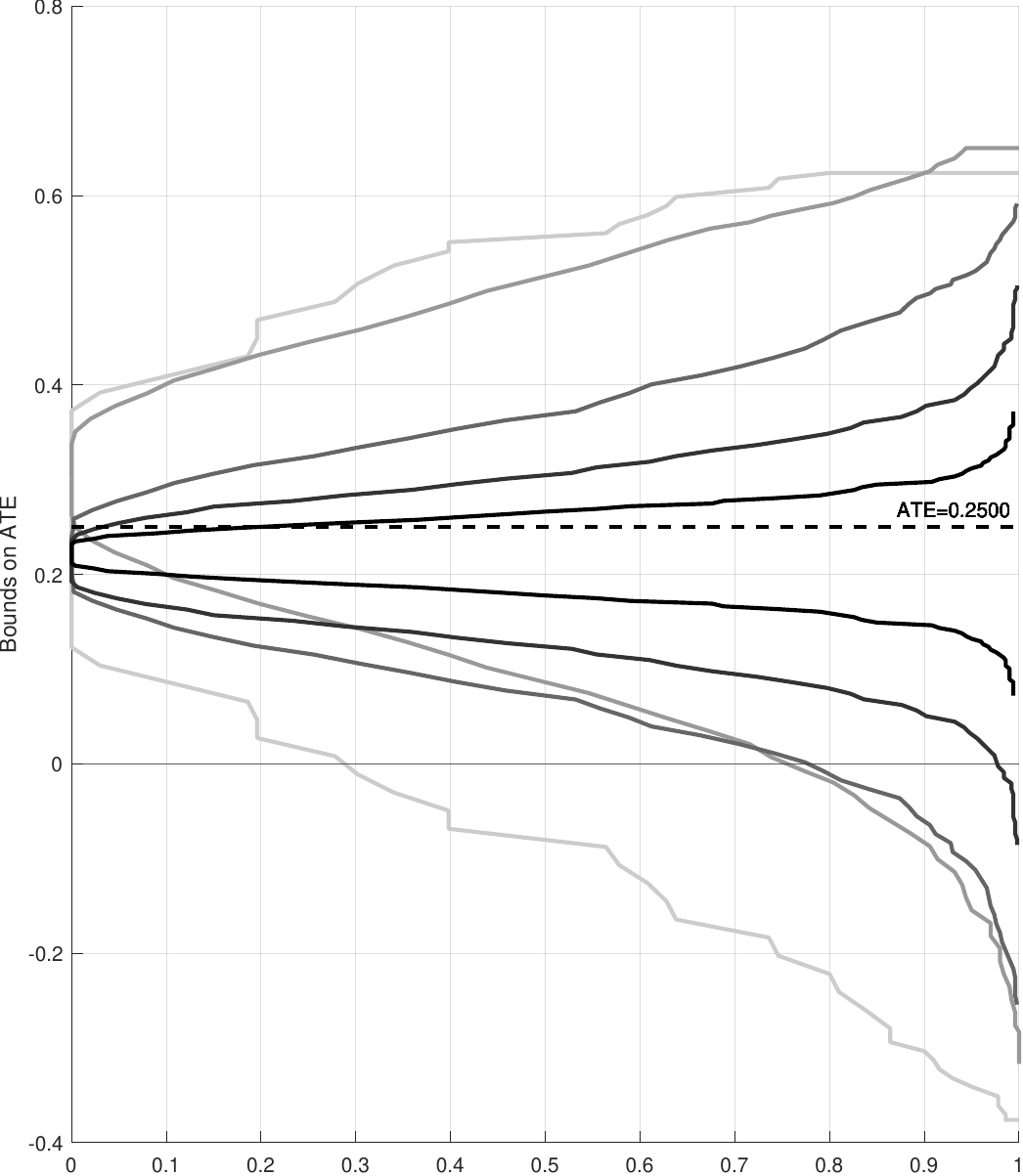}
    \caption{Convergence of the bounds on ATE, $\Theta_I(K(\alpha))$, as population size $N$ increases. The lightest gray line is $N =10$ while the darkest line is $N=400$.}
    \label{fig:bounds_ATE_as_N_increases_combined} 
  \end{subfigure}

  \caption{Illustration of the convergence results of theorem \ref{thm:randomizationGood}.}
  \label{fig:mainNumerical}
\end{figure}

First we demonstrate the convergence results of theorem \ref{thm:randomizationGood}. Its first part shows that $\underline{p}(K^\text{bp}) \xrightarrow{p} 1$ as $N \rightarrow \infty$. In the proof, we showed that $\underline{p}(\cdot)$ converges to a step function at zero. Figure \ref{fig:covergence_of_p_as_N_increases} demonstrates this convergence, which implies convergence of $\underline{p}(K^\text{bp})$ to 1 so long as the limiting breakdown point is nonzero. The second part of theorem \ref{thm:randomizationGood} shows that, for any $\alpha \in (0,1)$, the distance between the set $\Theta_I(K(\alpha))$ and ATE converges to zero as $N \rightarrow \infty$. Figure \ref{fig:bounds_ATE_as_N_increases_combined} plots $\Theta_I(K(\alpha))$ as a function of $1-\alpha$, for the five different values of $N$. The bounds are centered at the naive difference-in-means estimand $\overline{Y}_1 - \overline{Y}_0$, which varies with $N$ but converges to 0.25 by construction. Again, consistent with the theory, we see that for any $\alpha \in (0,1)$ our bounds shrink as $N$ gets larger. This plot also shows the first part of theorem \ref{thm:randomizationGood}, since $\underline{p}(K^\text{bp})$ is the point at which the bounds intersect the horizontal axis at zero. We see that this point converges to 1 as $N$ gets large. Indeed, for $N=100$ our bounds are strictly positive for almost all values of $1-\alpha$.

\begin{figure}[t]
    \centering
    \begin{minipage}{0.45\textwidth}
        \includegraphics[width=0.85\linewidth]{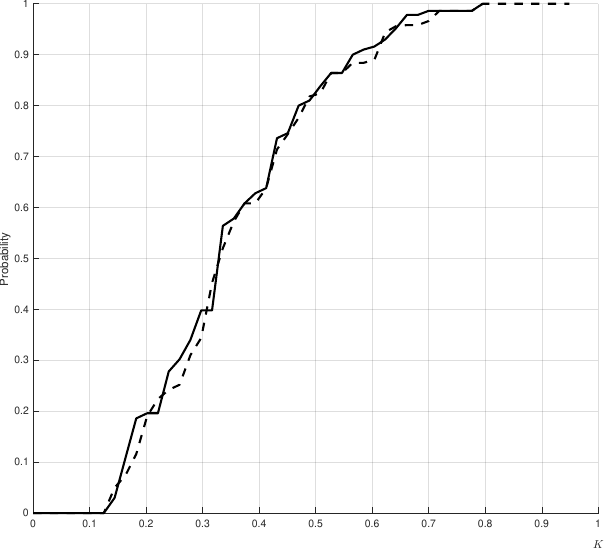}
    \end{minipage}
    \begin{minipage}{0.45\textwidth}
        \includegraphics[width=\linewidth]{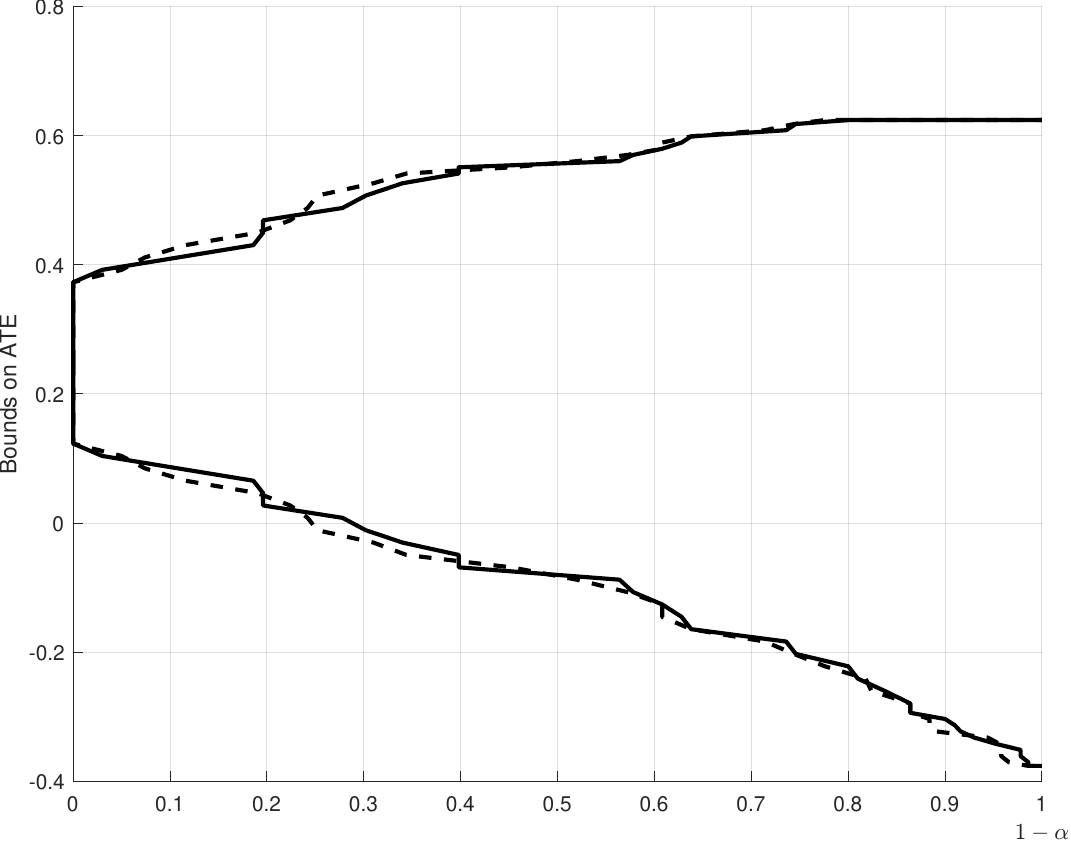}
    \end{minipage}
    \caption{Assessing accuracy of the genetic algorithm: $\underline{p}$ vs $K$ (left plot), $\Theta_I(K(\alpha))$ vs $1-\alpha$ (right plot) for GA (solid line) and MILP (dashed line), $N=10$.}
    \label{fig:MILP_GA_N=10}
\end{figure}

These results used a genetic algorithm (GA) to solve the optimization problem \eqref{eq:defOfunderlinep}. Next we verify that this algorithm is obtaining accurate results by comparing its output with output from the mixed-integer linear programming (MILP) approach that we discussed in section \ref{sec:computation}. For $N=10$, the left plot in figure \ref{fig:MILP_GA_N=10} shows the function $\underline{p}$ obtained using the genetic algorithm as a solid line, and the same function obtained using mixed-integer linear programming as a dashed line. The two lines are very similar, showing that the genetic algorithm is able to closely match the output of the MILP approach, despite being substantially faster. Specifically, for this plot the genetic algorithm took about 4 minutes, whereas MILP took about 34 \emph{hours}. Similarly, the right plot in figure \ref{fig:MILP_GA_N=10} shows $\Theta_I(K(\alpha))$ as a function of $1-\alpha$, as obtained by both algorithms. Again, the genetic algorithm is able to closely match the output from MILP. Finally, appendix figure \ref{fig:robustness_batch_size} shows that our results are robust to the choice of the number of treatment assignments used to approximate the objective function.

\section{The Role of Balance in Covariates}\label{sec:covariates}

We next illustrate the flexibility of our approach to finite population causal inference by extending it to incorporate covariates (sometimes called ``attributes'') in this section and noncompliance in section \ref{sec:IV}.

For each unit $i \in \mathcal{I}$, let $W_i$ denote a vector of covariate values. Let $\W = (W_1,\ldots,W_N)$ denote the collection of covariate values for all units in the population. In practice, researchers commonly examine the observed magnitude of balance in covariates across the treatment and control groups. In this section we give a new perspective on this kind of covariate balance analysis: Observed imbalances in covariates can be used to help identify the unobserved imbalance in potential outcomes. Informally, this additional information arises when the covariates are predictive of potential outcomes. For covariates to have identifying power, however, we must make an explicit assumption on this relationship. There are many different formal assumptions one could consider. For brevity we focus on a particularly simple one here, but it would be useful to explore variations in future work. We illustrate this analysis empirically in section \ref{sec:Bloom}.

Let $\Q = (q(W_1)', \ldots, q(W_N)')'$ be a $N \times \text{dim}(q(W_i))$ matrix of known transformations of the observed covariates. Assume $\Q' \Q$ is invertible. Let $\beta(x) \coloneqq (\Q' \Q)^{-1} \Q' \Y(x)$ denote the population regression coefficient from OLS of potential outcomes onto the transformed covariates. Let $U_i \coloneqq Y_i(x) - q(W_i)' \beta(x)$ denote the regression residual for unit $i$, with $\U \coloneqq (U_1,\ldots, U_N)$. Let $R_{Y(x) \sim q(W)}^2 \coloneqq 1 - \var(\U) / \var[\Y(x)]$, where for any vector $\A \coloneqq (A_1,\ldots,A_N)$ we let $\overline{\A} \coloneqq \frac{1}{N} \sum_{i=1}^N A_i$ denote its mean and $\var(\A) \coloneqq \frac{1}{N} \sum_{i=1}^N (A_i - \overline{\A})^2$ denote its variance. Assume $q(W_i)$ includes a constant, so that $\overline{\U} = 0$ by construction, since these are OLS residuals. Not all potential outcomes are observed. Hence $\beta(x)$, $(U_1,\ldots,U_N)$, and $R_{Y(x) \sim q(W)}^2$ are not point identified. Instead, we make assumptions about them, as follows.

\begin{Aassumption}[Predictive covariates]\label{assump:approxCovariateLinearityALT}
For each $x \in \{0,1\}$, $R_{Y(x) \sim q(W)}^2 \geq \lambda$ for a known $\lambda \in [0,1]$.
\end{Aassumption}

$\lambda$ is a sensitivity parameter that controls the minimal predictive power of the covariates relative to unobserved variables that determine variation in potential outcomes. Larger values of $\lambda$ require a closer connection between potential outcomes and covariates. Since some potential outcomes are known, not all values of the sensitivity parameter $\lambda$ are consistent with the data. The largest value of $\lambda$ that is consistent with the data and all of the maintained assumptions is called the falsification point (\citealt{MastenPoirier2021}). Any $\lambda$ values below this point will lead to a non-falsified model, and therefore a non-empty identified set for ATE. Note that we could use different $\lambda$ values for each potential outcome; we use a common value for simplicity.

\ref{assump:approxCovariateLinearityALT} has two implications: First, it has direct identifying power for potential outcomes $\Y(1)$ and $\Y(0)$, since it imposes a constraint the unobserved values of potential outcomes. Second, it affects $\underline{p}(K)$ because that constraint also shrinks the constraint set for the optimization in equation \eqref{eq:defOfunderlinep}. We consider each of these implications next.

\subsection{The Identifying Power of Covariates}\label{sec:covariatesIdentification}

In section \ref{section:ATEidentifiedSets} we derived an explicit expression for the finite population identified set for ATE. In this section we provide a numerical procedure for computing the identified set for ATE under the additional assumption of predictive covariates (\ref{assump:approxCovariateLinearityALT}). Specifically, the upper bound on ATE solves
\[
	\max_{\Y(1),\Y(0) \in [y_\text{min},y_\text{max}]^{2N}} \; \frac{1}{N} \sum_{i=1}^N \big( Y_i(1) - Y_i(0) \big)
\]
subject to (i) the data constraints that $Y_i(X_i) = Y_i$ for all $i=1,\ldots,N$, (ii) the $K$-approximate means balance constraint
\[
	-K \leq \frac{1}{N_1} \sum_{i=1}^N Y_i(x) \indicator(X_i = 1) - \frac{1}{N_0} \sum_{i=1}^N Y_i(x) \indicator(X_i = 0) \leq K,
\]
and (iii) the predictive covariates constraint $R_{Y(x) \sim q(W)}^2 \geq \lambda$, which is equivalent to
\begin{equation}\label{eq:approxLinearityInW}
	\textstyle \sum_{i=1}^N \big( Y_i(x) - q(W_i)'\beta(x) \big) ^2 \leq N (1-\lambda) \var[\Y(x)]
\end{equation}
where $\beta(x) = (\Q' \Q)^{-1} \Q' \Y(x)$. The lower bound can be obtained by computing the minimum rather than the maximum. The constraint \eqref{eq:approxLinearityInW} is smooth in the unknowns and the objective function and other constraints are linear, so any gradient-based nonlinear solver can be used to solve this program.

\subsection{The Impact of Covariates on Design Probabilities of Balance}

Because the predictive covariates assumption restricts the feasible values of unobserved potential outcomes, it also affects the design probability that the $K$-approximate mean balance assumption \ref{assump:KapproxBalance} is guaranteed to hold. Specifically, we modify the definition of $\underline{p}$ to also impose the constraint in equation \eqref{eq:approxLinearityInW}:
\begin{equation}
	\underline{p}^\text{mod}(K,\lambda) \coloneqq
	\inf_{
	\substack{
	\Y(1), \Y(0) \in [y_\text{min},y_\text{max}]^N \\
	\Y = \Y(1) \times \X + \Y(0) \times (\bold{1} - \X)  \\
	\text{s.t.\ equation \eqref{eq:approxLinearityInW} holds for $\lambda$}
	}
	} \; p\big( K, \Y(1), \Y(0) \big).
\end{equation}
This additional constraint does not meaningfully affect the computational time. Because adding constraints weakly reduces the set of feasible values, $\underline{p}^\text{mod}(K,\lambda)$ is weakly increasing in $\lambda$ for any fixed $K$.

\subsection{Interpretation and Discussion}

To build intuition for the identifying power of covariates, consider the case with a binary covariate $W_i$ and focus on the treated potential outcomes $(Y_1(1),\ldots,Y_N(1))$. Let $q(W_i) = (1,W_i)'$, and denote the corresponding components of $\beta(x)$ by $\beta_0(x)$ and $\beta_1(x)$. For any vector $(A_1,\ldots,A_N)$ let $\Exp(A \mid X=x) \coloneqq \left( \sum_{i=1}^N A_i \indicator(X_i=x) \right) / \sum_{i=1}^N \indicator(X_i=x)$. By the definition of the residuals $U_i(1)$, $\Exp[Y(1) \mid X=x] = \beta_0(1) + \beta_1(1) \Exp(W \mid X=x) + \Exp[U(1) \mid X=x]$ for $x \in \{0,1\}$ and hence
\begin{align*}
	&\Exp[Y(1) \mid X=0] - \Exp[Y(1) \mid X=1] \\
	&= \beta_1(1) \big( \Exp(W \mid X=0) - \Exp(W \mid X=1) \big) + \big( \Exp[U(1) \mid X=0] - \Exp[U(1) \mid X=1] \big).
\end{align*}
This equation shows that the magnitude of imbalance in potential outcomes depends directly on the magnitude of imbalance in covariates. It also depends on the magnitude of imbalance in the residuals, which is controlled by $\lambda$. In particular, for $\delta \coloneqq \sqrt{N (1-\lambda) \var(Y)}$, $\Exp[U(1) \mid X=x] \in [-\delta,\delta]$ for each $x \in \{0,1\}$. Hence
\[
	\big| \Exp[U(1) \mid X=0] - \Exp[U(1) \mid X=1] \big| \leq 2 \delta.
\]
This bound combined with the observed imbalance in covariates directly constrain the imbalance in potential outcomes, which is the source of identifying power of covariates. A similar analysis applies to balance in $Y(0)$. Note that this discussion is for intuition only; section \ref{sec:covariatesIdentification} describes how we obtain identified sets in general.

Random assignment of treatment does not imply anything about the value of $R_{Y(x) \sim q(W)}^2$; it is a population level parameter that does not depend on how treatment is assigned. Consequently, we cannot use random assignment of treatment to calibrate $\lambda$. However, assumptions like \ref{assump:approxCovariateLinearityALT} are not necessary for drawing relatively tight conclusions about ATE in large finite populations, as in the second part of theorem \ref{thm:randomizationGood}. Rather, additional assumptions like \ref{assump:approxCovariateLinearityALT} can be useful to provide tighter bounds in \emph{smaller} finite populations.

We conclude this section with several remarks about the literature. First, covariate balance is sometimes examined to test whether treatment was actually randomized. Here we simply assume treatment was in fact randomized. Second, the traditional design-based inference framework primarily uses covariates to motivate different choices of test statistics, with the goal of increasing the power of hypothesis tests. For example, see \citet[section 5.9]{ImbensRubin2015} or \cite{ZhaoDing2021}. This approach, like ours, uses covariates to derive stronger conclusions about the parameter of interest. Mathematically, however, our approach uses covariates for identification and does not rely on hypothesis testing theory. Finally, covariates are also used to define the parameter of interest (e.g., \citealt{AbadieAtheyImbensWooldridge2020}). In principle this can be done in our approach too, but we leave this to future work (also see section \ref{sec:beyondATE}).

\section{Noncompliance}\label{sec:IV}

We have focused on the case where all units comply with their treatment assignment. In this section we extend the analysis to a finite population version of the \cite{ImbensAngrist1994} model (as in section 3 of \citealt{HongLeungLi2020}), allowing us to study settings with noncompliance. When certain exact balance conditions hold, we show that the Wald estimand point identifies a realized local average effect of treatment on the treated (LATT) parameter. We then study finite population identification under $K$-approximate mean balance type assumptions. In the special case of one-sided noncompliance, we show that the finite population identified set for the realized LATT has a simple form that is analogous to classical infinite population results. We then show how to use this result to do design-based sensitivity analysis. In all of this analysis we allow for heterogeneous treatment effects while still not relying on asymptotics (in contrast to some prior work on design-based instrumental variable analysis, such as \citealt{Rosenbaum1996}).\footnote{See \cite{ImbensRosenbaum2005}, \cite{BaiocchiSmallLorchRosenbaum2010}, \cite{KeeleSmallGrieve2017}, \cite{KangPeckKeele2018}, and \cite{RambachanRoth2025} for additional prior work on design-based instrumental variable analysis.}

\subsection{Setup}

As before, there is a finite population of units $i=1,\ldots,N$. For each unit $i$: Let $Y_i(1)$ and $Y_i(0)$ denote potential outcomes, $X_i(1)$ and $X_i(0)$ the binary potential treatments, $Z_i$ the realized value of a binary instrument (assigned treatment in the noncompliance setting), $X_i = X_i(Z_i)$ the realized treatment, and $Y_i = Y_i(X_i)$ the realized outcome. Note that we impose the exclusion restriction implicitly here and maintain it throughout this section. Without loss of generality, write $Y_i(x) = \beta_i \cdot x + U_i$ for $x \in \{0,1\}$, where we defined $U_i  \coloneqq Y_i(0)$ and $\beta_i \coloneqq Y_i(1) - Y_i(0)$. Let $N_1 = \sum_{i=1}^N \indicator(Z_i = 1)$ denote the number of units whose instrument value equals 1. We'll call this the instrument-on group. Let $N_0 = N - N_1$. We'll call the set of units with $Z_i = 0$ the instrument-off group. Let $\overline{Y}_{z=1} \coloneqq \frac{1}{N_1} \sum_{i : Z_i = 1} Y_i$ denote the average outcome in the instrument-on group. Define $\overline{Y}_{z=0}$, $\overline{X}_{z=1}$, $\overline{X}_{z=0}$, $\overline{U}_{z=1}$, and $\overline{U}_{z=0}$ similarly.

\subsection{Identification under Exact Balance}

Define the compliance type variable
\[
	T_i =
	\begin{cases}
		c &\text{if $X_i(1)=1, X_i(0) = 0$} \\
		a &\text{if $X_i(1)=1, X_i(0)=1$} \\
		n &\text{if $X_i(1)=0, X_i(0)=0$} \\
		d &\text{if $X_i(1)=0, X_i(0) =1$}.
	\end{cases}
\]
We maintain the following finite population version of the monotonicity / no defiers assumption in \cite{ImbensAngrist1994}.

\begin{Bassumption}[No defiers]\label{assump:noDefiers}
$T_i \neq d$ for all $i=1,\ldots,N$.
\end{Bassumption}

Similarly, we assume the following finite population version of relevance holds for the specific realization of treatment assignment that is observed.

\begin{Bassumption}[Relevance]\label{assump:relevance}
$\overline{X}_{z=1} \neq \overline{X}_{z=0}$.
\end{Bassumption}

Let $\overline{T}_1(a) \coloneqq \frac{\sum_{i=1}^N \indicator(Z_i=1) \indicator(T_i=a)}{\sum_{i=1}^N \indicator(Z_i=1)}$ denote the proportion of always takers in the instrument-on group. Likewise, let $\overline{T}_0(a)$ denote the proportion of always takers in the instrument-off group, and $\overline{T}_1(c)$ the proportion of compliers in the instrument-on group. Let $\overline{\beta}_1(a) = \left( \sum_{i : Z_i=1, T_i=a} \beta_i \right) / \allowbreak \sum_{i=1}^N \indicator(Z_i=1) \indicator(T_i=a)$ denote the average treatment effect among the instrument-on always takers. Define $\overline{\beta}_0(a)$ and $\overline{\beta}_1(c)$ similarly.

\begin{lemma}\label{lemma:firstIVeq}
Suppose \ref{assump:noDefiers} (no defiers) and \ref{assump:relevance} (relevance) hold. Then
\begin{equation}\label{eq:mainIVequation}
\resizebox{0.94\linewidth}{!}{$
	\frac{\overline{Y}_{z=1} - \overline{Y}_{z=0}}{\overline{X}_{z=1} - \overline{X}_{z=0}}
	= \frac{\overline{U}_{z=1} - \overline{U}_{z=0}}{\overline{X}_{z=1} - \overline{X}_{z=0} }
	+ \frac{\overline{T}_1(a) \overline{\beta}_1(a) - \overline{T}_0(a) \overline{\beta}_0(a)}{\overline{X}_{z=1} - \overline{X}_{z=0} } + \frac{\overline{T}_1(c)}{\big( \overline{T}_1(a) - \overline{T}_0(a) \big) + \overline{T}_1(c)} \; \text{\footnotesize $\overline{\beta}_1(c)$}.
$}
\end{equation}
\end{lemma}

The left hand side of equation \eqref{eq:mainIVequation} is the finite population version of the Wald estimand. So this lemma decomposes the Wald estimand into three pieces that each depend on the magnitude of various realized imbalances. Consider the following exact mean balance assumption.

\begin{Bassumption}[Exact balance]\label{assump:IVexactBalance}
(i) $\overline{U}_{z=1} = \overline{U}_{z=0}$, (ii) $\overline{T}_1(a) = \overline{T}_0(a)$, and (iii) $\overline{\beta}_1(a) = \overline{\beta}_0(a)$.
\end{Bassumption}

Part (i) says that $Y_i(0)$ is balanced across the instrument-on and instrument-off groups. This is an instrument exogeneity assumption, with respect to the potential outcomes. It is analogous to $Z \indep Y(0)$ in an infinite population analysis. Part (ii) says that the proportion of always takers is the same in the instrument-on and instrument-off groups. It is also an instrument exogeneity assumption. It is often called ``unconfounded types'', because it is about the relationship between the instrument and the potential treatment variables. It is analogous to $Z \indep (X(1), X(0))$ in an infinite population analysis. Finally, part (iii) says that the average treatment effect for always takers is the same in the instrument-on and instrument-off groups. In an infinite population analysis, this kind of mean balance condition would hold if $Z \indep (Y(0), Y(1), X(0), X(1))$.

\begin{proposition}\label{prop:IVpointIdent}
Suppose \ref{assump:noDefiers} (no defiers), \ref{assump:relevance} (relevance), and \ref{assump:IVexactBalance} (exact balance) hold. Then $\frac{\overline{Y}_{z=1} - \overline{Y}_{z=0}}{\overline{X}_{z=1} - \overline{X}_{z=0}} = \overline{\beta}_1(c)$.
\end{proposition}

Proposition \ref{prop:IVpointIdent} shows that $\overline{\beta}_1(c)$ is point identified in finite populations under exact balance. In particular, it equals the finite population Wald estimand. This result is a finite population version of the \cite{ImbensAngrist1994} result, with one slight difference: The point identified parameter $\overline{\beta}_1(c) \coloneqq \left( \sum_{i=1}^N \beta_i \cdot \indicator(T_i=c) \indicator(Z_i=1) \right) / \sum_{i=1}^N \indicator(T_i=c) \indicator(Z_i=1)$ is a realized local average effect of treatment on the treated (LATT)---it is the average unit-level causal effect among treated compliers (recall that $Z_i=X_i$ for compliers). In contrast, the population LATE is $\overline{\beta}(c) \coloneqq \left( \sum_{i=1}^N \beta_i \indicator(T_i=c) \right) / \sum_{i=1}^N \indicator(T_i=c)$. The LATT is an ex ante random parameter, since the set of units who will be treated, and thus who appear in the parameter's definition, depends on the realization of treatment assignment. This is analogous to Rosenbaum's \citeyearpar{Rosenbaum2001} finite population analysis of ATT, where he noted that the ATT parameter is also ex ante random (he called the ATT the ``attributable effect''). Another slight difference from the usual infinite population analysis is that when there is one-sided noncompliance (so there are no always takers), the ATT and LATT are the same, but they do not equal LATE because there can be non-treated compliers. Finally, note that we can decompose LATE as $\overline{\beta}(c) = p_1(c) \overline{\beta}_1(c) + p_0(c) \overline{\beta}_0(c)$ where $p_1(c) \coloneqq \left( \sum_{i=1}^N \indicator(T_i=c) \indicator(Z_i=1) \right) / \sum_{i=1}^N \indicator(T_i=c)$ is the proportion of units who are treated, among all compliers, and likewise for $p_0(c)$. So if we further assume that the average treatment effect for compliers in the instrument-on group is the same as for compliers in the instrument-off group---$\overline{\beta}_1(c) = \overline{\beta}_0(c)$---then the finite population Wald estimand equals LATE. This additional condition is analogous to part (iii) of \ref{assump:IVexactBalance}.

\subsection{Design-Based Sensitivity Analysis}

Proposition \ref{prop:IVpointIdent} shows that the Wald estimand equals LATT under an exact balance assumption. However, as discussed in section \ref{sec:finitePopIdent}, exact balance rarely holds in finite populations. Instead, we can apply the same ideas from earlier to perform a design-based sensitivity analysis: We can derive identified sets for LATT under approximate balance assumptions and then use randomization to calibrate the sensitivity parameters. Here we briefly sketch the analysis in the one-sided noncompliance case.

\begin{Bassumption}[One-sided noncompliance]\label{assump:IVoneSided}
$T_i \neq a$ for all $i=1,\ldots,N$.
\end{Bassumption}

Without always takers, the second term in equation \eqref{eq:mainIVequation} disappears, and the third term becomes $\overline{\beta}_1(c)$. Hence the only remaining term is a difference in non-treated potential outcomes, which we bound via the following assumption.

\begin{Bassumption}[$K$-approximate mean balance for $Y(0)$]\label{assump:IVapproxMeanBalance}
There is a known $K \geq 0$ such that $| \overline{U}_{z=1} - \overline{U}_{z=0} | \leq K$.
\end{Bassumption}

The next result bounds the realized LATT as a function of $K$. Here we let $\pi \coloneqq \overline{X}_{z=1} - \overline{X}_{z=0}$ denote the first stage difference in means.

\begin{theorem}\label{thm:IVoneSidedNoncomplianceSet}
Suppose \ref{assump:noDefiers} (no defiers), \ref{assump:relevance} (relevance), \ref{assump:IVoneSided} (one-sided noncompliance), and \ref{assump:IVapproxMeanBalance} ($K$-approximate mean balance for $Y(0)$) hold. Then the finite population identified set for $\overline{\beta}_1(c)$ is
\[
	\left[ \frac{\overline{Y}_{z=1} - \overline{Y}_{z=0}}{\overline{X}_{z=1} - \overline{X}_{z=0}} - \frac{K}{\pi}, \; \frac{\overline{Y}_{z=1} - \overline{Y}_{z=0}}{\overline{X}_{z=1} - \overline{X}_{z=0}} + \frac{K}{\pi} \right].
\]
\end{theorem}

The identified set in theorem \ref{thm:IVoneSidedNoncomplianceSet} is analogous to infinite population identified sets where instrument exogeneity is relaxed at the population level. For example, see \cite{BoundJaegerBaker1995} or \cite{ConleyHansenRossi2012}. The identified set in theorem \ref{thm:IVoneSidedNoncomplianceSet} can be further adjusted to impose the bounded outcome assumption \ref{assump:boundedSupport1}, like in theorem \ref{thm:KapproxiBalanceATEset}. Then, assuming the instrument is randomly assigned according to a known distribution, we can construct a function similar to $\underline{p}$ in equation \eqref{eq:defOfunderlinep} and use this to calibrate the value of $K$. We omit the details for brevity. The general two-sided noncompliance case is more complicated, since it involves more than just a single balance condition (i.e., relaxations of the three conditions in \ref{assump:IVexactBalance}). We conjecture that our analysis extends to this case but leave a full exploration to future work.

\section{Further Extensions}\label{sec:extensions}

\subsection{Sampling}\label{sec:sampling}

For most of this paper we assumed for simplicity that there was no sampling---all units in the finite population are observed, and the only uncertainty is about the unknown potential outcomes. Here we briefly discuss two extensions: An analysis of sampling by itself, and an analysis that combines both sampling and random assignment (e.g., as in \citealt{AbadieAtheyImbensWooldridge2020}). In both cases we reframe the classical question of inferring population quantities from sample data as an identification problem, which allows us to provide a single approach to quantify uncertainty from missing units and from missing potential outcomes. Keep in mind that this extension will generally make bounds wider, since it accounts for an additional source of uncertainty.

First consider the sampling setup from section \ref{sec:intro}. The population are the numbers $\Y = (Y_1,\ldots,Y_N)$. Now suppose we only observe $n < N$ of these units, and the goal is to learn about the population mean $\overline{Y} \coloneqq \frac{1}{N} \sum_{i=1}^N Y_i$ (rather than a treatment effect parameter). As before, $S_i \in \{0,1\}$ denotes whether unit $i$ is sampled or not. From an identification perspective, sampling is a missing data problem---we observe $Y_i$ when $S_i=1$ but have no data whatsoever on unit $i$ when $S_i=0$. Consequently, if all we know are that all outcomes $Y_i$ lie in known bounds (an assumption similar to \ref{assump:boundedSupport1}), then all we can say \emph{for sure} about the population mean is that it lies in domain-of-consensus bounds analogous to $\Theta_I(\infty)$; this is the motivation behind including sample indicators in definition \ref{def:identifiedSet}. However, we can shrink the identified set by making assumptions like
\[
	\left| \frac{1}{n} \sum_{i=1}^N S_i Y_i - \frac{1}{N-n} \sum_{i=1}^N (1-S_i) Y_i \right| \leq K,
\]
which is analogous to the $K$-approximate mean balance assumption \ref{assump:KapproxBalance}. Under this assumption, we can derive identified sets for $\overline{Y}$ as a function of the sensitivity parameter $K$. Finally, suppose we know the sample was obtained via simple random sampling (SRS), for example. This is the sampling analog of uniform randomization (\ref{assump:uniformRandomization})---all possible samples of size $n$ from $N$ have equal probability of being selected. Then, as in section \ref{sec:sensitivityAnalysis}, we can compute worst case design probabilities of imbalance, which we can use to calibrate the sensitivity parameter $K$. This allows us to perform a design-based sensitivity analysis for sampling.

Next consider the randomized experiment setting we focused on in this paper. Suppose we only observe a sample of $n < N$ units. In this case we need to address the identification problem that arises from missing potential outcomes as well as missing data on some units altogether (see \citealt{Manski1996} and \citealt{KlineTamer2018} for related identification analyses that combine experiments and sampling, but in the infinite population setting). This setting does not require any new conceptual ideas, and so we only discuss it briefly. Consider the average treated potential outcome. Let $n_1 < n$ denote the number of observed treated units. Assume both $n$ and $n_1$ are fixed a priori, and both sampling and randomization are performed independently according to simple random sampling and uniform randomization. We observe the average treated potential outcome among sampled units who are treated, $\frac{1}{n_1} \sum_{i : S_i=1, X_i=1} Y_i(1)$. We do not know the average $\frac{1}{N-n_1} \sum_{i : S_i=0 \text{ or } X_i=0} Y_i(1)$. However, we can make a $K$-approximate mean balance assumption that says this unobserved mean is not too far from the observed one. Then we can use our sampling and randomization assumptions to calibrate the value $K$. The same analysis can be done for the non-treated potential outcome, and they can be combined to do a design-based sensitivity analysis for the population ATE.

\subsubsection*{The Distinction Between Being Identified Versus Identifiable}

Our definition \ref{def:identifiedSet} treats sampling and randomization symmetrically---both lead to missing data which can be thought of as a finite population identification problem, and therefore analyzed using the tools of partial identification. This is perhaps unusual, since uncertainty due to sampling is traditionally separated from other types of uncertainty, as in Koopmans' (1949, page 132) original definition of identification. Nonetheless, as sketched above, definition \ref{def:identifiedSet} provides a foundation to quantify uncertainty from both missing units and missing potential outcomes. That said, here we briefly discuss a distinction between the two, which helps reconcile the difference between our definition \ref{def:identifiedSet} and traditional definitions of identification in infinite populations like Koopmans' \citeyearpar{Koopmans1949} which assume away sampling uncertainty.

Consider the set of all \emph{designs}, which are mappings from $(\Y(1),\Y(0),\W)$ into joint probabilities over sampling and treatment assignments. Say a parameter is \emph{point identifiable} with respect to a set of assumptions $\mathcal{P}$ if there exists a design such that, for any $(\Y(1),\Y(0),\W)$ satisfying the assumptions $\mathcal{P}$, and for any realization $(\S,\X)$ from that design, the subsequent identified set $\Theta_I$ is a singleton. Then the population mean $\overline{Y} \coloneqq \frac{1}{N} \sum_{i=1}^N Y_i(X_i)$ is point identifi\emph{able} with respect to \ref{assump:boundedSupport1}, since we can in principle sample all units with probability 1. But in any fixed dataset with $n < N$ it will only be partially identifi\emph{ed}. In contrast, ATE is not point identifiable with respect to \ref{assump:boundedSupport1} since we can never treat everyone and treat no-one. Thus, even though our definition of the identified set treats the ex post uncertainty due to missing units and missing potential outcomes symmetrically, our definition still allows these two kinds of uncertainty to be ex ante different, since one can in principle sample all units whereas we cannot even in principle treat all units.

\subsection{Variations on the Bounded Outcomes Assumption}\label{sec:boundedOutcomesVariations}

Obtaining nontrivial bounds on ATE usually requires some kind of assumption that restricts the range of possible potential outcomes. Throughout this paper we used the familiar uniform bound $[y_\text{min}, y_\text{max}]$ on potential outcomes, assumption \ref{assump:boundedSupport1}. It is important to recognize that this is a substantive identifying assumption, not a regularity condition---if these bounds are large then the researcher is explicitly allowing for a wide range of possible outcomes, which allows for large magnitudes of imbalance.

The specific form of the assumption we used can be replaced or augmented with a variety of similar assumptions, and all of our methods will continue to apply. Here we give just a few examples: (i) A simple extension is to allow unit specific bounds, $Y_i(x) \in [y_{\text{min},i}, y_{\text{max},i}]$. We use this version in one of our empirical applications, where the units are groups and the outcome depends on group size. (ii) One could impose a bounded unit-level treatment effect assumption: $| Y_i(1) - Y_i(0) | \leq M$ for all $i \in \mathcal{I}$, where $M$ is a known sensitivity parameter. This assumption implies unit-specific bounds on the unobserved potential outcomes: $Y_i(x) \in [Y_i - M, Y_i + M]$. (iii) Alternatively, one could assume the sum of unit-level treatment effect magnitudes is bounded: $\sum_{i=1}^N | Y_i(1) - Y_i(0) | \leq M$ for a known $M$. This would allow for some units to have very large treatment effects, so long as not too many do. This condition implies that ATE can be no larger than $M / N$. (iv) Or one could restrict the population variance of potential outcomes to be no larger than a known $M$. This assumption implies that $Y_i(x)$ is in the interval $\frac{1}{N_x} \sum_{i=1}^N Y_i \indicator(X_i = x) \pm 2 \sqrt{N_x M}$. Hence it does \emph{not} require users to specify a priori bounds $[y_\text{min},y_\text{max}]$; they only must choose $M$. (v) Finally, the covariate restrictions in section \ref{sec:covariates} can also be viewed as one way of restricting the range of possible potential outcomes.

\subsection{Distributional Balance and Parameters Beyond ATE}\label{sec:beyondATE}

Our analysis has focused on balance defined via differences in means. This corresponds to our focus on ATE, where balance in means is the ``fundamental identification condition'' (page 263, \citealt{HeckmanIchimuraTodd1998}). However, there are many other ways to quantify balance across the treatment and control groups; see chapter 14 of \cite{ImbensRubin2015}. For example, letting $\Prob^\text{true}(Y(x) \leq y \mid X=x) \coloneqq \frac{1}{N_x} \sum_{i : X_i=x} \indicator[Y_i(x) \leq y]$ (see appendix \ref{sec:popDistributions}), we could consider the assumption
\begin{equation}\label{eq:supNormBalance}
	\sup_{y \in \R} \big| \Prob^\text{true}(Y(x) \leq y \mid X=1) - \Prob^\text{true}(Y(x) \leq y \mid X=0) \big| \leq K
\end{equation}
which bounds the sup-norm distance between the population distributions of potential outcomes in the treatment and control groups. We could then derive identified sets for the parameter of interest under this assumption, for a fixed $K$. Given a randomization design, we could then compute the worst case design-probability that \eqref{eq:supNormBalance} holds, which would lead to a design-based sensitivity analysis. 

Different forms of balance might lead to different bounds than those based on mean balance. Thus the choice of balance metric in our analysis could be loosely thought of as analogous to the choice of the test statistic in classical randomization tests. Exactly how much the bounds vary with the choice of balance metric will likely depend on the parameter of interest. Alternative balance metrics may be more appropriate for studying parameters beyond ATE. For example, the sup-norm distance in equation \eqref{eq:supNormBalance} will likely work well for identification of quantile treatment effects (QTEs) since these are defined using inverses of the unconditional population cdfs $\Prob^\text{true}(Y(x) \leq y) \coloneqq \frac{1}{N} \sum_{i=1}^N \indicator[Y_i(x) \leq y]$. We leave a full exploration of these questions to future work.

\section{Empirical Applications}\label{sec:empirical}

In this section we illustrate our approach in two empirical applications with small, finite populations ($N=10$ and $N=17$). While our methods apply to any population size, and are feasible for larger population sizes (see appendix \ref{sec:LaLonde} for a third application with $N=722$), these two applications show that it is still possible to do meaningful inference in small datasets. In appendix \ref{sec:AlternativeEmpiricalCIs} we show three alternative frequentist confidence intervals for comparison.

\subsection{Getting Parents to Pick Their Kids Up on Time}\label{sec:Gneezy}

Our first application uses data from the well known paper \citet[\emph{Journal of Legal Studies}]{GneezyRustichini2000}, which has about 3600 Google Scholar citations as of March 2026. They study day care centers, where administrators were frustrated with parents showing up late to pick up their kids. They asked: Would a monetary fine incentivize parents to be on time? The population is 10 centers. The treatment is the introduction of a center-wide late fee. 6 centers were treated and 4 were not. The outcome variable is the number of late parents in a week. The logical lower bound is zero and the logical upper bound is five times the number of kids in that center (assuming every parent is late every day of the week). We assume that regardless of treatment, on average, each child has a late parent at most once per week. That is, we set $y_{\text{max},i}$ equal to the number of children in center $i$ (see (i) in section \ref{sec:boundedOutcomesVariations}). In the data there are between 28 and 37 kids per center. We let $y_{\text{min},i} = 0$ for all units.

The authors gathered baseline data on outcomes for 4 weeks. The fee was introduced at treated centers at the beginning of week 5. It was removed at the beginning of week 17. The authors gathered another 4 weeks of data after removal of the fee, for a total of 20 weeks of data. Their table 1 provides the full dataset. For simplicity we only use data from one post-treatment week, week 19. It would be interesting to study how to extend our results to use the panel dimension of this dataset, but we leave this to future work.

The ATE point estimate is $13$ late parents, suggesting that the addition of a fee \emph{increased} the number of late parents per week by 13. To quantify the uncertainty around this estimate, we conduct a design-based sensitivity analysis. Figure \ref{fig:Gneezy} in the introduction shows the results, which we discussed already. First consider the breakdown point, $\underline{p}(K^\text{bp}) = 65$\%. So there is at least a 65\% chance that the ATE is non-negative, according to our robust empirical Bayesian interpretation. Although this does not attain conventional levels of ``significance'', like 95\%, it is nonetheless a non-trivial inference given that this dataset only contains 10 units. This conclusion can also be seen in figure \ref{fig:GneezyDesignBased2}, which plots $\Theta_I(K(\alpha))$ as a function of $1-\alpha$. All sets with probabilities smaller than 64\% contain only positive values. Moreover, even for large probabilities, the sets $\Theta_I(K(\alpha))$ are still mostly in the positive region. For example, the 90\% set is about $[-6, 22]$. There is at least a 90\% chance that the true ATE is in this set. Overall, these findings suggest that there is reasonably strong evidence that the ATE is positive.

\subsection{The Long Run Adoption of Management Practices}\label{sec:Bloom}

\nocite{BloomData}

Our second application uses data from \citet[\emph{QJE}]{BloomEtAl2013} and \citet[\emph{AEJ: Applied}]{BloomEtAl2020}. These are influential papers with about 2650 total Google scholar citations as of March 2026. These papers asked whether large observed differences in productivity across firms are driven by variation in firms' management practices. To answer this, they ran a randomized experiment in a population of 17 woven cotton fabric firms. These were large and old firms, with an average of 270 employees per firm and an average age of 20 years old at baseline. Their control group received a one-month diagnostic about their management practices. The treatment group received the diagnostic plus four months of support for implementing the management changes. They consider many different outcomes of interest, but we will focus on the long run outcome from their 2020 paper. Specifically, the outcome variable is the proportion of 38 management practices adopted in 2017, which was about 8 to 9 years after they received treatment. Since this outcome is a proportion, we set $[y_\text{min}, y_\text{max}] = [0,1]$.

\begin{figure}[t]
  \centering

  \begin{subfigure}[c]{0.45\textwidth}
    \centering
\includegraphics[width=\linewidth]{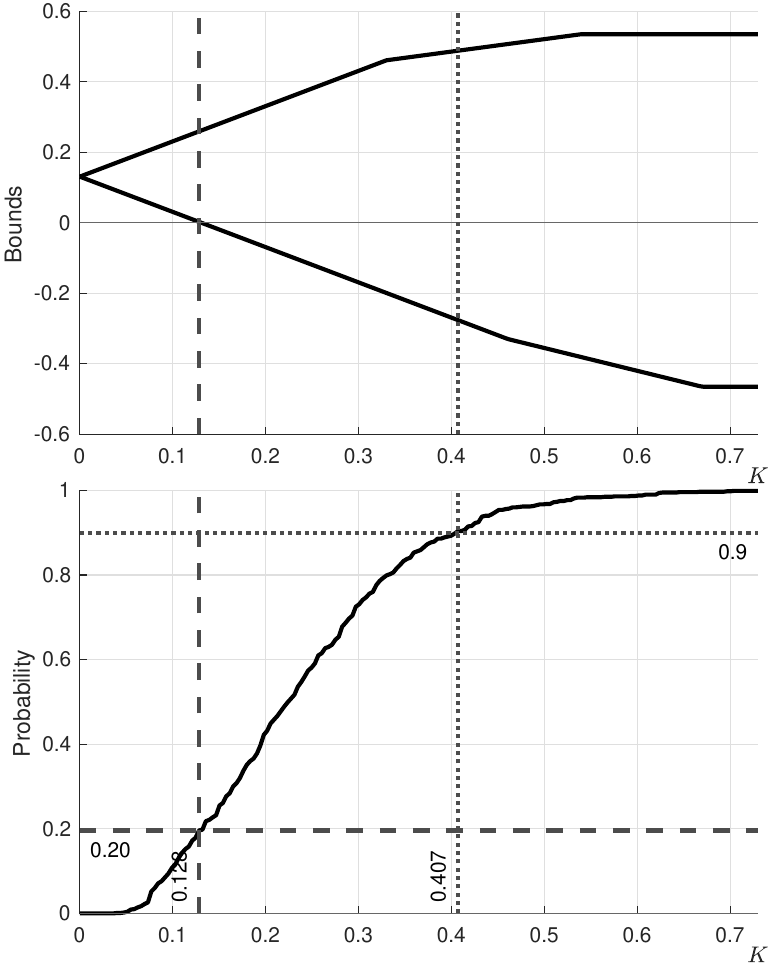}
\caption{Top: $\Theta_I(K)$, bounds on ATE as a function of the maximum magnitude of ex post imbalance, $K$. Bottom: $\underline{p}(K)$. \label{fig:BloomDesignBased}}
  \end{subfigure}
  \hfill
  \begin{subfigure}[c]{0.5\textwidth}
    \centering
\includegraphics[width=\linewidth]{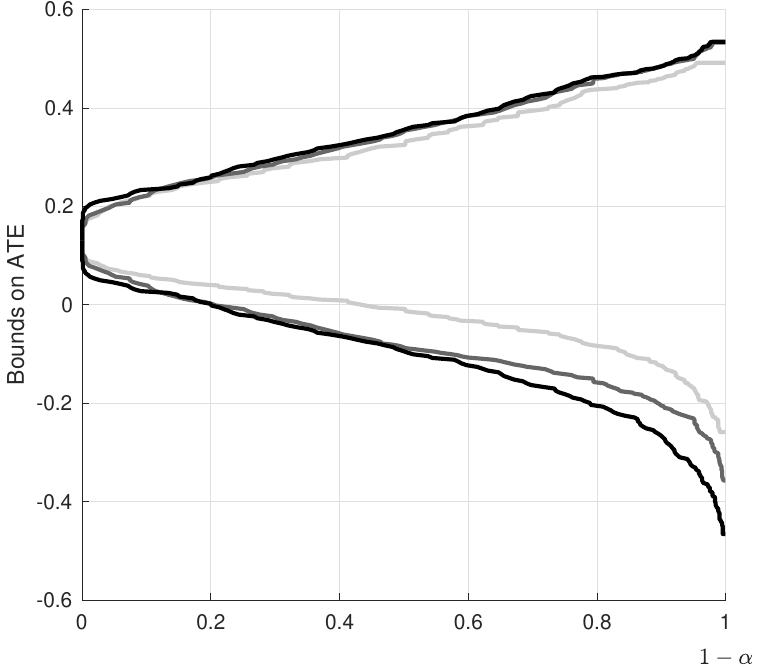}
\caption{$\Theta_I(K(\alpha))$, bounds on the ATE as a function of $1-\alpha$. The darkest line is $\lambda = 0$, the lightest line is $\lambda = 0.4$, and the line in the middle is $\lambda = 0.2$. \label{fig:BloomDesignBased2}}
  \end{subfigure}

\caption{Quantifying the uncertainty in the impact of management interventions on long run adoption of management practices.}
  \label{fig:Bloom}
\end{figure}

Some of the 17 firms in the population have multiple plants. Treatment was administered at and varies at the plant-level. This implies that there are non-treated plants at firms with a different treated plant. The authors use this data to study within-firm spillovers. For simplicity we ignore all data from non-treated plants at treated-firms. We use the authors' ``experimental'' dataset (as described in figure 1 of \citealt{BloomEtAl2020}), where each unit is a single plant. There are 11 treated plants and 6 control plants.\footnote{See pages 16--17 of \cite{BloomEtAl2013} for more details on the experimental design.}

The ATE point estimate is 0.13, suggesting that providing 4 months of support for changing management practices leads to a 13 percentage point increase in the proportion of practices that are still in place about one decade later. The authors emphasize that it is important to also measure the uncertainty associated with this point estimate, however:
\begin{quote}
``The major challenge of our experiment was its small cross-sectional sample size. We have data on only 28 plants across 17 firms. To address concerns over statistical inference in small samples, we implemented permutation tests whose properties are independent of sample size.'' (page 4, \citealt{BloomEtAl2013})
\end{quote}
That is, to deal with the small population size they performed exact tests of the sharp null of no unit-level treatment effects, assuming uniform randomization. To complement their results, we implement our design-based sensitivity analysis. Figure \ref{fig:Bloom} shows the main results. First consider the breakdown point, $\underline{p}(K^\text{bp}) = 20$\%. So there is at least a 20\% chance that the ATE is non-negative, according to our robust empirical Bayesian interpretation. So with this dataset it is unlikely that potential outcomes would be balanced enough to ensure that $K$ is small enough that we can rule out negative ATE values. This is also shown in the outer-most bounds of figure \ref{fig:BloomDesignBased2}, which plots $\Theta_I(K(\alpha))$ as a function of $1-\alpha$. Here we see that the sets all contain negative numbers for probabilities larger than 20\%. For example, the 90\% interval is $[-0.28, 0.49]$. So there is at least a 90\% chance that ATE is inside this interval.

We can obtain tighter bounds by adding assumptions about covariates, as in section \ref{sec:covariates}. Figure \ref{fig:BloomDesignBased2} also shows bounds on ATE that impose the predictive covariates restriction \ref{assump:approxCovariateLinearityALT} for $\lambda = 0$ (no covariate restrictions), $0.2$, and $0.4$, going from the darkest, outer-most bands to the lightest, inner-most bands. We stop at $\lambda = 0.4$ since larger values are generally falsified (and hence lead to empty identified sets). Following the authors' analysis (table 2 in \citealt{BloomEtAl2020}), we only use a single covariate, baseline management score in 2008, which is essentially a lagged outcome, and specify a simple linear model $q(W_i) = (1,W_i)'$. For this choice, the predictive covariates assumption tightens the bounds somewhat, bringing the breakdown point up to about 45\%, compared to its unconstrained value of 20\%. The bounds could potentially be further tightened by using additional covariates (e.g., those in their table 1), or by making additional assumptions about the unobserved potential outcomes, as in section \ref{sec:boundedOutcomesVariations}.

\section{Conclusion}\label{sec:conclusion}

In this paper we studied identification in finite populations. We showed that the prior conventional definition of identification implies that every potential outcome for every unit is point identified under an overlap condition alone, even in observational settings where treatment depends on potential outcomes. Hence we argued that it is not adequate for understanding what can be learned about treatment effects in real datasets.

As our first main contribution, we developed a definition of identification that is not subject to this problem. Moreover, unlike the traditional design-based approaches to inference in finite populations, our definition of identification does not pre-suppose a known distribution from which treatment, an instrument, or a ``shock,'' was drawn. This allows it to be used in a wide variety of settings, including both experimental and observational studies.

Building on our definition of identification, in our second main contribution we developed a new approach to quantifying uncertainty in finite populations, called a \emph{design-based sensitivity analysis}. This approach combines design distributions and our finite population identified sets to construct new uncertainty intervals that have (1) identification, (2) robust Bayesian, and (3) uniform frequentist interpretations.

\bibliographystyle{ecta-fullname}
\bibliography{finite_pop_ident}

\begin{thebibliography}{99}
\newcommand{\enquote}[1]{``#1''}
\expandafter\ifx\csname natexlab\endcsname\relax\def\natexlab#1{#1}\fi

\bibitem[\protect\citeauthoryear{Abadie, Athey, Imbens, and Wooldridge}{Abadie
  et~al.}{2020}]{AbadieAtheyImbensWooldridge2020}
\textsc{Abadie, Alberto, Susan Athey, Guido~W Imbens, and Jeffrey~M Wooldridge}
  (2020): \enquote{Sampling-based versus design-based uncertainty in regression
  analysis,} \emph{Econometrica}, 88 (1), 265--296.

\bibitem[\protect\citeauthoryear{Abadie, Athey, Imbens, and Wooldridge}{Abadie
  et~al.}{2023}]{AbadieAtheyImbensWooldridge2023}
---\hspace{-.1pt}---\hspace{-.1pt}--- (2023): \enquote{When should you adjust
  standard errors for clustering?} \emph{The Quarterly Journal of Economics},
  138 (1), 1--35.

\bibitem[\protect\citeauthoryear{Altman}{Altman}{1985}]{Altman1985}
\textsc{Altman, Douglas~G} (1985): \enquote{Comparability of randomised
  groups,} \emph{Journal of the Royal Statistical Society Series D: The
  Statistician}, 34 (1), 125--136.

\bibitem[\protect\citeauthoryear{Aronow, Robins, Saarinen, S{\"a}vje, and
  Sekhon}{Aronow et~al.}{2025}]{AronowEtAl2025}
\textsc{Aronow, Peter~M, James~M Robins, Theo Saarinen, Fredrik S{\"a}vje, and
  Jasjeet Sekhon} (2025): \enquote{Nonparametric identification is not enough,
  but randomized controlled trials are,} \emph{Observational Studies}, 11 (1),
  3--16.

\bibitem[\protect\citeauthoryear{Aronow and Samii}{Aronow and
  Samii}{2017}]{AronowSamii2017}
\textsc{Aronow, Peter~M. and Cyrus Samii} (2017): \enquote{{Estimating average
  causal effects under general interference, with application to a social
  network experiment},} \emph{The Annals of Applied Statistics}, 11 (4), 1912
  -- 1947.

\bibitem[\protect\citeauthoryear{Athey, Eckles, and Imbens}{Athey
  et~al.}{2018}]{AtheyEcklesImbens2018}
\textsc{Athey, Susan, Dean Eckles, and Guido~W Imbens} (2018): \enquote{Exact
  p-values for network interference,} \emph{Journal of the American Statistical
  Association}, 113 (521), 230--240.

\bibitem[\protect\citeauthoryear{Athey and Imbens}{Athey and
  Imbens}{2022}]{AtheyImbens2022}
\textsc{Athey, Susan and Guido~W Imbens} (2022): \enquote{Design-based analysis
  in difference-in-differences settings with staggered adoption,} \emph{Journal
  of Econometrics}, 226 (1), 62--79.

\bibitem[\protect\citeauthoryear{Bai, Huang, Romano, Shaikh, and
  Tabord-Meehan}{Bai et~al.}{2025{\natexlab{a}}}]{BaiEtAl2025}
\textsc{Bai, Yuehao, Xun Huang, Joseph~P. Romano, Azeem~M. Shaikh, and Max
  Tabord-Meehan} (2025{\natexlab{a}}): \enquote{A new design-based variance
  estimator for finely stratified experiments,} \emph{arXiv preprint
  arXiv:2503.10851}.

\bibitem[\protect\citeauthoryear{Bai, Shaikh, and Tabord-Meehan}{Bai
  et~al.}{2025{\natexlab{b}}}]{BaiShaikhTabordMeehan2025}
\textsc{Bai, Yuehao, Azeem~M Shaikh, and Max Tabord-Meehan}
  (2025{\natexlab{b}}): \enquote{A primer on the analysis of randomized
  experiments and a survey of some recent advances,} \emph{JPE: Microeconomics
  (forthcoming)}.

\bibitem[\protect\citeauthoryear{Baiocchi, Small, Lorch, and
  Rosenbaum}{Baiocchi et~al.}{2010}]{BaiocchiSmallLorchRosenbaum2010}
\textsc{Baiocchi, Mike, Dylan~S Small, Scott Lorch, and Paul~R Rosenbaum}
  (2010): \enquote{Building a stronger instrument in an observational study of
  perinatal care for premature infants,} \emph{Journal of the American
  Statistical Association}, 105 (492), 1285--1296.

\bibitem[\protect\citeauthoryear{Blandhol, Bonney, Mogstad, and
  Torgovitsky}{Blandhol et~al.}{2025}]{BlandholEtAl2025}
\textsc{Blandhol, Christine, John Bonney, Magne Mogstad, and Alexander
  Torgovitsky} (2025): \enquote{When is TSLS \emph{actually} LATE?} \emph{The
  Review of Economic Studies (forthcoming)}.

\bibitem[\protect\citeauthoryear{Bloom, Eifert, Mahajan, McKenzie, and
  Roberts}{Bloom et~al.}{2013}]{BloomEtAl2013}
\textsc{Bloom, Nicholas, Benn Eifert, Aprajit Mahajan, David McKenzie, and John
  Roberts} (2013): \enquote{Does management matter? Evidence from India,}
  \emph{The Quarterly Journal of Economics}, 128 (1), 1--51.

\bibitem[\protect\citeauthoryear{Bloom, Mahajan, McKenzie, and Roberts}{Bloom
  et~al.}{2020}]{BloomEtAl2020}
\textsc{Bloom, Nicholas, Aprajit Mahajan, David McKenzie, and John Roberts}
  (2020): \enquote{Do management interventions last? Evidence from India,}
  \emph{American Economic Journal: Applied Economics}, 12 (2), 198--219.

\bibitem[\protect\citeauthoryear{Bojinov, Rambachan, and Shephard}{Bojinov
  et~al.}{2021}]{BojinovRambachanShephard2021}
\textsc{Bojinov, Iavor, Ashesh Rambachan, and Neil Shephard} (2021):
  \enquote{Panel experiments and dynamic causal effects: A finite population
  perspective,} \emph{Quantitative Economics}, 12 (4), 1171--1196.

\bibitem[\protect\citeauthoryear{Bontemps and Magnac}{Bontemps and
  Magnac}{2017}]{BontempsMagnac2017}
\textsc{Bontemps, Christian and Thierry Magnac} (2017): \enquote{Set
  identification, moment restrictions, and inference,} \emph{Annual Review of
  Economics}, 9, 103--129.

\bibitem[\protect\citeauthoryear{Borusyak, Hull, and Jaravel}{Borusyak
  et~al.}{2025}]{BorusyakHullJaravel2024}
\textsc{Borusyak, Kirill, Peter Hull, and Xavier Jaravel} (2025):
  \enquote{Design-based identification with formula instruments: a review,}
  \emph{The Econometrics Journal}, 28 (1), 83--108.

\bibitem[\protect\citeauthoryear{Bound, Jaeger, and Baker}{Bound
  et~al.}{1995}]{BoundJaegerBaker1995}
\textsc{Bound, John, David~A Jaeger, and Regina~M Baker} (1995):
  \enquote{Problems with instrumental variables estimation when the correlation
  between the instruments and the endogenous explanatory variable is weak,}
  \emph{Journal of the American Statistical Association}, 90 (430), 443--450.

\bibitem[\protect\citeauthoryear{Caetano, Caetano, Callaway, and Dyal}{Caetano
  et~al.}{2026}]{CaetanoEtAl2026}
\textsc{Caetano, Carolina, Gregorio Caetano, Brantly Callaway, and Derek Dyal}
  (2026): \enquote{Causal inference for aggregated treatment,} \emph{arXiv
  preprint arXiv:2506.22885}.

\bibitem[\protect\citeauthoryear{Canay and Shaikh}{Canay and
  Shaikh}{2017}]{CanayShaikh2017}
\textsc{Canay, Ivan~A. and Azeem~M. Shaikh} (2017): \enquote{Practical and
  theoretical advances in inference for partially identified Models,} in
  \emph{Advances in {{Economics}} and {{Econometrics}}}, ed. by Bo~Honor{\'e},
  Ariel Pakes, Monika Piazzesi, and Larry Samuelson, Cambridge University
  Press, 271--306, 1 ed.

\bibitem[\protect\citeauthoryear{Card}{Card}{2022}]{Card2022}
\textsc{Card, David} (2022): \enquote{Design-based research in empirical
  microeconomics,} \emph{American Economic Review}, 112 (6), 1773--1781.

\bibitem[\protect\citeauthoryear{Cassel, S{\"a}rndal, and Wretman}{Cassel
  et~al.}{1977}]{CasselSarndalWretman1977}
\textsc{Cassel, Claes-Magnus, Carl-Erik S{\"a}rndal, and Jan~H{\aa}kan Wretman}
  (1977): \emph{Foundations of Inference in Survey Sampling}, John Wiley \&
  Sons.

\bibitem[\protect\citeauthoryear{Chen, Roth, and Spiess}{Chen
  et~al.}{2026}]{ChenRothSpiess2026}
\textsc{Chen, Jiafeng, Jonathan Roth, and Jann Spiess} (2026): \enquote{Testing
  monotonicity in a finite population,} \emph{arXiv preprint arXiv:2512.25032}.

\bibitem[\protect\citeauthoryear{Chesher and Rosen}{Chesher and
  Rosen}{2020}]{ChesherRosen2020}
\textsc{Chesher, Andrew and Adam~M Rosen} (2020): \enquote{Generalized
  instrumental variable models, methods, and applications,} \emph{Handbook of
  Econometrics}, 7, 1--110.

\bibitem[\protect\citeauthoryear{Christensen and Connault}{Christensen and
  Connault}{2023}]{ChristensenConnault2023}
\textsc{Christensen, Timothy and Benjamin Connault} (2023):
  \enquote{Counterfactual sensitivity and robustness,} \emph{Econometrica}, 91
  (1), 263--298.

\bibitem[\protect\citeauthoryear{Conley, Hansen, and Rossi}{Conley
  et~al.}{2012}]{ConleyHansenRossi2012}
\textsc{Conley, Timothy~G, Christian~B Hansen, and Peter~E Rossi} (2012):
  \enquote{Plausibly exogenous,} \emph{Review of Economics and Statistics}, 94
  (1), 260--272.

\bibitem[\protect\citeauthoryear{Deaner}{Deaner}{2025}]{Deaner2025}
\textsc{Deaner, Ben} (2025): \enquote{The trade-off between flexibility and
  robustness in instrumental variables analysis,} \emph{American Economic
  Review}, 115 (11), 3975--3998.

\bibitem[\protect\citeauthoryear{Ding}{Ding}{2017}]{Ding2017}
\textsc{Ding, Peng} (2017): \enquote{A paradox from randomization-based causal
  inference,} \emph{Statistical Science}, 331--345.

\bibitem[\protect\citeauthoryear{Ding}{Ding}{2024}]{Ding2024}
---\hspace{-.1pt}---\hspace{-.1pt}--- (2024): \emph{A First Course in Causal
  Inference}, CRC Press.

\bibitem[\protect\citeauthoryear{Ding}{Ding}{2025}]{Ding2025}
---\hspace{-.1pt}---\hspace{-.1pt}--- (2025): \enquote{What randomization can
  and cannot guarantee,} \emph{Observational Studies}, 11 (1), 27--40.

\bibitem[\protect\citeauthoryear{Ding, Li, and Miratrix}{Ding
  et~al.}{2017}]{LiDingMiratrix2017}
\textsc{Ding, Peng, Xinran Li, and Luke~W Miratrix} (2017): \enquote{Bridging
  finite and super population causal inference,} \emph{Journal of Causal
  Inference}, 5 (2), 20160027.

\bibitem[\protect\citeauthoryear{Eckles, Ignatiadis, Wager, and Wu}{Eckles
  et~al.}{2020}]{EcklesEtAl2020}
\textsc{Eckles, Dean, Nikolaos Ignatiadis, Stefan Wager, and Han Wu} (2020):
  \enquote{Noise-induced randomization in regression discontinuity designs,}
  \emph{arXiv preprint arXiv:2004.09458}.

\bibitem[\protect\citeauthoryear{Fan and Park}{Fan and
  Park}{2010}]{FanPark2010}
\textsc{Fan, Yanqin and Sang~Soo Park} (2010): \enquote{Sharp bounds on the
  distribution of treatment effects and their statistical inference,}
  \emph{Econometric Theory}, 26 (03), 931--951.

\bibitem[\protect\citeauthoryear{Florens and Simoni}{Florens and
  Simoni}{2021}]{FlorensSimoni2021}
\textsc{Florens, Jean-Pierre and Anna Simoni} (2021): \enquote{Revisiting
  identification concepts in Bayesian analysis,} \emph{Annals of Economics and
  Statistics},  (144), 1--38.

\bibitem[\protect\citeauthoryear{Giacomini and Kitagawa}{Giacomini and
  Kitagawa}{2021}]{giacomini2021robust}
\textsc{Giacomini, Raffaella and Toru Kitagawa} (2021): \enquote{Robust
  Bayesian inference for set-identified models,} \emph{Econometrica}, 89 (4),
  1519--1556.

\bibitem[\protect\citeauthoryear{Gneezy and Rustichini}{Gneezy and
  Rustichini}{2000}]{GneezyRustichini2000}
\textsc{Gneezy, Uri and Aldo Rustichini} (2000): \enquote{A fine is a price,}
  \emph{The Journal of Legal Studies}, 29 (1), 1--17.

\bibitem[\protect\citeauthoryear{Greenland}{Greenland}{1990}]{Greenland1990}
\textsc{Greenland, Sander} (1990): \enquote{Randomization, statistics, and
  causal inference,} \emph{Epidemiology}, 1 (6), 421--429.

\bibitem[\protect\citeauthoryear{Greenland and Robins}{Greenland and
  Robins}{1986}]{GreenlandRobins1986}
\textsc{Greenland, Sander and James~M Robins} (1986): \enquote{Identifiability,
  exchangeability, and epidemiological confounding,} \emph{International
  Journal of Epidemiology}, 15 (3), 413--419.

\bibitem[\protect\citeauthoryear{Greenland and Robins}{Greenland and
  Robins}{2009}]{GreenlandRobins2009}
---\hspace{-.1pt}---\hspace{-.1pt}--- (2009): \enquote{Identifiability,
  exchangeability and confounding revisited,} \emph{Epidemiologic Perspectives
  \& Innovations}, 6, 1--9.

\bibitem[\protect\citeauthoryear{Heckman, Ichimura, and Todd}{Heckman
  et~al.}{1998}]{HeckmanIchimuraTodd1998}
\textsc{Heckman, James~J, Hidehiko Ichimura, and Petra Todd} (1998):
  \enquote{Matching as an econometric evaluation estimator,} \emph{The Review
  of Economic Studies}, 65 (2), 261--294.

\bibitem[\protect\citeauthoryear{Heckman and Vytlacil}{Heckman and
  Vytlacil}{2007}]{HeckmanVytlacil2007}
\textsc{Heckman, James~J. and Edward~J. Vytlacil} (2007): \enquote{Econometric
  evaluation of social programs, part I: Causal models, structural models and
  econometric policy evaluation,} \emph{Handbook of Econometrics}, 6.

\bibitem[\protect\citeauthoryear{Holland}{Holland}{1986}]{Holland1986}
\textsc{Holland, Paul~W.} (1986): \enquote{Statistics and causal inference,}
  \emph{Journal of the American Statistical Association}, 81 (396), 945--960.

\bibitem[\protect\citeauthoryear{Hong, Leung, and Li}{Hong
  et~al.}{2020}]{HongLeungLi2020}
\textsc{Hong, Han, Michael~P Leung, and Jessie Li} (2020): \enquote{Inference
  on finite-population treatment effects under limited overlap,} \emph{The
  Econometrics Journal}, 23 (1), 32--47.

\bibitem[\protect\citeauthoryear{Hurwicz}{Hurwicz}{1950}]{Hurwicz1950}
\textsc{Hurwicz, Leonid} (1950): \enquote{Generalization of the concept of
  identification,} \emph{Statistical Inference in Dynamic Economic Models}, 10,
  245--57.

\bibitem[\protect\citeauthoryear{Imbens}{Imbens}{2018}]{Imbens2018}
\textsc{Imbens, Guido} (2018): \enquote{Understanding and misunderstanding
  randomized controlled trials: A commentary on Deaton and Cartwright,}
  \emph{Social Science \& Medicine}, 210, 50--52.

\bibitem[\protect\citeauthoryear{Imbens and Menzel}{Imbens and
  Menzel}{2021}]{ImbensMenzel2021}
\textsc{Imbens, Guido and Konrad Menzel} (2021): \enquote{A causal bootstrap,}
  \emph{The Annals of Statistics}, 49 (3), 1460--1488.

\bibitem[\protect\citeauthoryear{Imbens and Angrist}{Imbens and
  Angrist}{1994}]{ImbensAngrist1994}
\textsc{Imbens, Guido~W. and Joshua~D. Angrist} (1994): \enquote{Identification
  and estimation of local average treatment effects,} \emph{Econometrica}, 62
  (2), 467--475.

\bibitem[\protect\citeauthoryear{Imbens and Rosenbaum}{Imbens and
  Rosenbaum}{2005}]{ImbensRosenbaum2005}
\textsc{Imbens, Guido~W and Paul~R Rosenbaum} (2005): \enquote{Robust, accurate
  confidence intervals with a weak instrument: quarter of birth and education,}
  \emph{Journal of the Royal Statistical Society Series A: Statistics in
  Society}, 168 (1), 109--126.

\bibitem[\protect\citeauthoryear{Imbens and Rubin}{Imbens and
  Rubin}{2015}]{ImbensRubin2015}
\textsc{Imbens, Guido~W and Donald~B Rubin} (2015): \emph{Causal Inference for
  Statistics, Social, and Biomedical Sciences}, Cambridge University Press.

\bibitem[\protect\citeauthoryear{Kang, Peck, and Keele}{Kang
  et~al.}{2018}]{KangPeckKeele2018}
\textsc{Kang, Hyunseung, Laura Peck, and Luke Keele} (2018): \enquote{Inference
  for instrumental variables: A randomization inference approach,}
  \emph{Journal of the Royal Statistical Society Series A: Statistics in
  Society}, 181 (4), 1231--1254.

\bibitem[\protect\citeauthoryear{Keele, Small, and Grieve}{Keele
  et~al.}{2017}]{KeeleSmallGrieve2017}
\textsc{Keele, Luke, Dylan Small, and Richard Grieve} (2017):
  \enquote{Randomization-based instrumental variables methods for binary
  outcomes with an application to the `IMPROVE' trial,} \emph{Journal of the
  Royal Statistical Society Series A: Statistics in Society}, 180 (2),
  569--586.

\bibitem[\protect\citeauthoryear{Kline and Tamer}{Kline and
  Tamer}{2016}]{kline2016bayesian}
\textsc{Kline, Brendan and Elie Tamer} (2016): \enquote{Bayesian inference in a
  class of partially identified models,} \emph{Quantitative Economics}, 7 (2),
  329--366.

\bibitem[\protect\citeauthoryear{Kline and Tamer}{Kline and
  Tamer}{2018}]{KlineTamer2018}
---\hspace{-.1pt}---\hspace{-.1pt}--- (2018): \enquote{Identification of
  treatment effects with selective participation in a randomized trial,}
  \emph{The Econometrics Journal}, 21 (3), 332--353.

\bibitem[\protect\citeauthoryear{Kline and Tamer}{Kline and
  Tamer}{2023}]{KlineTamer2023}
---\hspace{-.1pt}---\hspace{-.1pt}--- (2023): \enquote{Recent developments in
  partial identification,} \emph{Annual Review of Economics}, 15, 125--150.

\bibitem[\protect\citeauthoryear{Kochenderfer and Wheeler}{Kochenderfer and
  Wheeler}{2019}]{KochenderferWheeler2019}
\textsc{Kochenderfer, Mykel~J and Tim~A Wheeler} (2019): \emph{Algorithms for
  Optimization}, MIT Press.

\bibitem[\protect\citeauthoryear{Koopmans}{Koopmans}{1949}]{Koopmans1949}
\textsc{Koopmans, Tjalling~C} (1949): \enquote{Identification problems in
  economic model construction,} \emph{Econometrica}, 125--144.

\bibitem[\protect\citeauthoryear{Koopmans and Reiers{\o}l}{Koopmans and
  Reiers{\o}l}{1950}]{KoopmansReiersol1950}
\textsc{Koopmans, Tjalling~C and Olav Reiers{\o}l} (1950): \enquote{The
  identification of structural characteristics,} \emph{The Annals of
  Mathematical Statistics}, 21 (2), 165--181.

\bibitem[\protect\citeauthoryear{Lewbel}{Lewbel}{2019}]{Lewbel2019}
\textsc{Lewbel, Arthur} (2019): \enquote{The identification zoo: Meanings of
  identification in econometrics,} \emph{Journal of Economic Literature}, 57
  (4), 835--903.

\bibitem[\protect\citeauthoryear{Li and Ding}{Li and Ding}{2017}]{LiDing2017}
\textsc{Li, Xinran and Peng Ding} (2017): \enquote{General forms of finite
  population central limit theorems with applications to causal inference,}
  \emph{Journal of the American Statistical Association}, 112 (520),
  1759--1769.

\bibitem[\protect\citeauthoryear{Lindley}{Lindley}{1980}]{Lindley1980}
\textsc{Lindley, D.~V.} (1980): \enquote{Randomization analysis of experimental
  data: The Fisher randomization test, comment,} \emph{Journal of the American
  Statistical Association}, 75 (371), 589--590.

\bibitem[\protect\citeauthoryear{Makarov}{Makarov}{1982}]{Makarov1982}
\textsc{Makarov, GD} (1982): \enquote{Estimates for the distribution function
  of a sum of two random variables when the marginal distributions are fixed,}
  \emph{Theory of Probability \& its Applications}, 26 (4), 803--806.

\bibitem[\protect\citeauthoryear{Manski}{Manski}{1990}]{Manski1990}
\textsc{Manski, Charles~F} (1990): \enquote{Nonparametric bounds on treatment
  effects,} \emph{The American Economic Review P\&P}, 80 (2), 319--323.

\bibitem[\protect\citeauthoryear{Manski}{Manski}{1996}]{Manski1996}
---\hspace{-.1pt}---\hspace{-.1pt}--- (1996): \enquote{Learning about treatment
  effects from experiments with random assignment of treatments,} \emph{Journal
  of Human Resources}, 31 (4), 709--733.

\bibitem[\protect\citeauthoryear{Manski}{Manski}{1997}]{Manski1997}
---\hspace{-.1pt}---\hspace{-.1pt}--- (1997): \enquote{Monotone treatment
  response,} \emph{Econometrica}, 65 (6), 1311--1334.

\bibitem[\protect\citeauthoryear{Manski}{Manski}{2003}]{Manski2003}
---\hspace{-.1pt}---\hspace{-.1pt}--- (2003): \emph{Partial Identification of
  Probability Distributions}, Springer.

\bibitem[\protect\citeauthoryear{Manski}{Manski}{2009}]{Manski2009}
---\hspace{-.1pt}---\hspace{-.1pt}--- (2009): \emph{Identification for
  Prediction and Decision}, Harvard University Press.

\bibitem[\protect\citeauthoryear{Manski and Pepper}{Manski and
  Pepper}{2018}]{ManskiPepper2018}
\textsc{Manski, Charles~F and John~V Pepper} (2018): \enquote{How do
  right-to-carry laws affect crime rates? Coping with ambiguity using
  bounded-variation assumptions,} \emph{Review of Economics and Statistics},
  100 (2), 232--244.

\bibitem[\protect\citeauthoryear{Masten and Poirier}{Masten and
  Poirier}{2021}]{MastenPoirier2021}
\textsc{Masten, Matthew~A and Alexandre Poirier} (2021): \enquote{Salvaging
  falsified instrumental variable models,} \emph{Econometrica}, 89 (3),
  1449--1469.

\bibitem[\protect\citeauthoryear{Matzkin}{Matzkin}{2007}]{Matzkin2007}
\textsc{Matzkin, Rosa~L} (2007): \enquote{Nonparametric identification,}
  \emph{Handbook of Econometrics}, 6, 5307--5368.

\bibitem[\protect\citeauthoryear{McKenzie, Bloom, Mahajan, and
  Roberts}{McKenzie et~al.}{2019}]{BloomData}
\textsc{McKenzie, David, Nick Bloom, Aprajit Mahajan, and John Roberts} (2019):
  \enquote{Replication Files for "Do Management Interventions Last? Evidence
  from India" 2017,} \emph{World Bank, Development Data Group}.

\bibitem[\protect\citeauthoryear{Molinari}{Molinari}{2020}]{Molinari2020}
\textsc{Molinari, Francesca} (2020): \enquote{Microeconometrics with partial
  identification,} \emph{Handbook of Econometrics}, 7, 355--486.

\bibitem[\protect\citeauthoryear{Moon and Schorfheide}{Moon and
  Schorfheide}{2012}]{moon2012bayesian}
\textsc{Moon, Hyungsik~Roger and Frank Schorfheide} (2012): \enquote{Bayesian
  and frequentist inference in partially identified models,}
  \emph{Econometrica}, 80 (2), 755--782.

\bibitem[\protect\citeauthoryear{Neyman}{Neyman}{1923, 1990}]{Neyman1923}
\textsc{Neyman, Jerzy} (1923, 1990): \enquote{On the application of probability
  theory to agricultural experiments. Essay on principles. Section 9.}
  \emph{Statistical Science}, 465--472.

\bibitem[\protect\citeauthoryear{Obradovi{\'c}}{Obradovi{\'c}}{2024}]{Obradovic2024}
\textsc{Obradovi{\'c}, Filip} (2024): \enquote{Identification of long-term
  treatment effects via temporal links, observational, and experimental data,}
  \emph{arXiv preprint arXiv:2411.04380}.

\bibitem[\protect\citeauthoryear{Poirier}{Poirier}{1998}]{poirier1998revising}
\textsc{Poirier, Dale~J} (1998): \enquote{Revising beliefs in nonidentified
  models,} \emph{Econometric Theory}, 14 (4), 483--509.

\bibitem[\protect\citeauthoryear{Pollmann}{Pollmann}{2023}]{Pollmann2023}
\textsc{Pollmann, Michael} (2023): \enquote{Causal inference for spatial
  treatments,} \emph{arXiv preprint arXiv:2011.00373}.

\bibitem[\protect\citeauthoryear{Rambachan and Roth}{Rambachan and
  Roth}{2025}]{RambachanRoth2025}
\textsc{Rambachan, Ashesh and Jonathan Roth} (2025): \enquote{Design-based
  uncertainty for quasi-experiments,} \emph{Journal of the American Statistical
  Association}.

\bibitem[\protect\citeauthoryear{Robins and Ritov}{Robins and
  Ritov}{1997}]{RobinsRitov1997}
\textsc{Robins, James~M and Ya'acov Ritov} (1997): \enquote{Toward a curse of
  dimensionality appropriate (CODA) asymptotic theory for semi-parametric
  models,} \emph{Statistics in Medicine}, 16 (3), 285--319.

\bibitem[\protect\citeauthoryear{Rosenbaum}{Rosenbaum}{1996}]{Rosenbaum1996}
\textsc{Rosenbaum, Paul~R.} (1996): \enquote{Identification of causal effects
  using instrumental variables: Comment,} \emph{Journal of the American
  Statistical Association}, 91 (434), 465--468.

\bibitem[\protect\citeauthoryear{Rosenbaum}{Rosenbaum}{2001}]{Rosenbaum2001}
\textsc{Rosenbaum, Paul~R} (2001): \enquote{Effects attributable to treatment:
  Inference in experiments and observational studies with a discrete pivot,}
  \emph{Biometrika}, 88 (1), 219--231.

\bibitem[\protect\citeauthoryear{Rosenberger and Lachin}{Rosenberger and
  Lachin}{2015}]{RosenbergerLachin2015}
\textsc{Rosenberger, William~F and John~M Lachin} (2015): \emph{Randomization
  in Clinical Trials: Theory and Practice}, John Wiley \& Sons.

\bibitem[\protect\citeauthoryear{Roth and Sant’Anna}{Roth and
  Sant’Anna}{2023}]{RothSantAnna2023}
\textsc{Roth, Jonathan and Pedro~HC Sant’Anna} (2023): \enquote{Efficient
  estimation for staggered rollout designs,} \emph{Journal of Political
  Economy: Microeconomics}, 1 (4), 669--709.

\bibitem[\protect\citeauthoryear{Rothenberg}{Rothenberg}{1971}]{Rothenberg1971}
\textsc{Rothenberg, Thomas~J} (1971): \enquote{Identification in parametric
  models,} \emph{Econometrica}, 577--591.

\bibitem[\protect\citeauthoryear{Royall and Pfeffermann}{Royall and
  Pfeffermann}{1982}]{RoyallPfeffermann1982}
\textsc{Royall, Richard~M and Dany Pfeffermann} (1982): \enquote{Balanced
  samples and robust Bayesian inference in finite population sampling,}
  \emph{Biometrika}, 69 (2), 401--409.

\bibitem[\protect\citeauthoryear{Rubin}{Rubin}{1978}]{Rubin1978}
\textsc{Rubin, Donald~B} (1978): \enquote{Bayesian inference for causal
  effects: The role of randomization,} \emph{The Annals of Statistics}, 34--58.

\bibitem[\protect\citeauthoryear{Sancibri{\'a}n}{Sancibri{\'a}n}{2024}]{Sancibrian2024}
\textsc{Sancibri{\'a}n, V{\'i}ctor} (2024): \enquote{Estimation uncertainty in
  repeated finite populations,} \emph{Working paper}.

\bibitem[\protect\citeauthoryear{S{\"a}vje}{S{\"a}vje}{2021}]{Savje2021}
\textsc{S{\"a}vje, Fredrik} (2021): \enquote{Randomization does not imply
  unconfoundedness,} \emph{arXiv preprint arXiv:2107.14197}.

\bibitem[\protect\citeauthoryear{S{\l}oczy{\'n}ski}{S{\l}oczy{\'n}ski}{2025}]{Sloczynski2025}
\textsc{S{\l}oczy{\'n}ski, Tymon} (2025): \enquote{When should we (not)
  interpret linear IV estimands as LATE?} \emph{The Review of Economic Studies
  (forthcoming)}.

\bibitem[\protect\citeauthoryear{Spini}{Spini}{2024}]{Spini2024}
\textsc{Spini, Pietro~Emilio} (2024): \enquote{Robustness, heterogeneous
  treatment effects and covariate shifts,} \emph{arXiv preprint
  arXiv:2112.09259}.

\bibitem[\protect\citeauthoryear{Startz and Steigerwald}{Startz and
  Steigerwald}{2023}]{StartzSteigerwald2023}
\textsc{Startz, Richard and Douglas~G Steigerwald} (2023): \enquote{Inference
  and extrapolation in finite populations with special attention to
  clustering,} \emph{Econometric Reviews}, 42 (4), 343--357.

\bibitem[\protect\citeauthoryear{Särndal, Thomsen, Hoem, Lindley,
  Barndorff-Nielsen, and Dalenius}{Särndal et~al.}{1978}]{SarndalEtAl1978}
\textsc{Särndal, Carl-Erik, Ib~Thomsen, Jan~M. Hoem, D.~V. Lindley,
  O.~Barndorff-Nielsen, and Tore Dalenius} (1978): \enquote{Design-based and
  Model-based inference in survey sampling [with discussion and reply],}
  \emph{Scandinavian Journal of Statistics}, 5 (1), 27--52.

\bibitem[\protect\citeauthoryear{Tamer}{Tamer}{2010}]{Tamer2010}
\textsc{Tamer, Elie} (2010): \enquote{Partial identification in econometrics,}
  \emph{Annual Review of Economics}, 2 (1), 167--195.

\bibitem[\protect\citeauthoryear{Thompson}{Thompson}{1997}]{Thompson1997}
\textsc{Thompson, Mary} (1997): \emph{Theory of Sample Surveys}, vol.~74, CRC
  Press.

\bibitem[\protect\citeauthoryear{Till{\'e}}{Till{\'e}}{2020}]{Tille2020}
\textsc{Till{\'e}, Yves} (2020): \emph{Sampling and Estimation From Finite
  Populations}, John Wiley \& Sons.

\bibitem[\protect\citeauthoryear{Wooldridge}{Wooldridge}{2023}]{Wooldridge2023}
\textsc{Wooldridge, Jeffrey~M} (2023): \enquote{What is a standard error? (And
  how should we compute it?),} \emph{Journal of Econometrics}, 237 (2), 105517.

\bibitem[\protect\citeauthoryear{Wu and Ding}{Wu and Ding}{2021}]{WuDing2021}
\textsc{Wu, Jason and Peng Ding} (2021): \enquote{Randomization tests for weak
  null hypotheses in randomized experiments,} \emph{Journal of the American
  Statistical Association}, 116 (536), 1898--1913.

\bibitem[\protect\citeauthoryear{Xu}{Xu}{2021}]{Xu2021}
\textsc{Xu, Ruonan} (2021): \enquote{Potential outcomes and finite-population
  inference for M-estimators,} \emph{The Econometrics Journal}, 24 (1),
  162--176.

\bibitem[\protect\citeauthoryear{Xu and Wooldridge}{Xu and
  Wooldridge}{2022}]{XuWooldridge2022}
\textsc{Xu, Ruonan and Jeffrey~M. Wooldridge} (2022): \enquote{A design-based
  approach to spatial correlation,} \emph{arxiv preprint arXiv:2211.14354}.

\bibitem[\protect\citeauthoryear{Young}{Young}{2019}]{Young2019}
\textsc{Young, Alwyn} (2019): \enquote{Channeling Fisher: Randomization tests
  and the statistical insignificance of seemingly significant experimental
  results,} \emph{The Quarterly Journal of Economics}, 134 (2), 557--598.

\bibitem[\protect\citeauthoryear{Zhao and Ding}{Zhao and
  Ding}{2021}]{ZhaoDing2021}
\textsc{Zhao, Anqi and Peng Ding} (2021): \enquote{Covariate-adjusted Fisher
  randomization tests for the average treatment effect,} \emph{Journal of
  Econometrics}, 225 (2), 278--294.

\end{thebibliography}


\begin{thebibliography}{33}
\newcommand{\enquote}[1]{``#1''}
\expandafter\ifx\csname natexlab\endcsname\relax\def\natexlab#1{#1}\fi

\bibitem[\protect\citeauthoryear{Aronow, Chang, and Lopatto}{Aronow
  et~al.}{2023}]{AronowChangLopatto2023}
\textsc{Aronow, Peter~M, Haoge Chang, and Patrick Lopatto} (2023):
  \enquote{Fast computation of exact confidence intervals for randomized
  experiments with binary outcomes,} \emph{arXiv preprint arXiv:2305.09906}.

\bibitem[\protect\citeauthoryear{Aronow, Green, and Lee}{Aronow
  et~al.}{2014}]{AronowGreenLee2014}
\textsc{Aronow, Peter~M, Donald~P Green, and Donald~KK Lee} (2014):
  \enquote{Sharp bounds on the variance in randomized experiments,} \emph{The
  Annals of Statistics}, 850--871.

\bibitem[\protect\citeauthoryear{Aronow, Robins, Saarinen, S{\"a}vje, and
  Sekhon}{Aronow et~al.}{2025}]{AronowEtAl2025}
\textsc{Aronow, Peter~M, James~M Robins, Theo Saarinen, Fredrik S{\"a}vje, and
  Jasjeet Sekhon} (2025): \enquote{Nonparametric identification is not enough,
  but randomized controlled trials are,} \emph{Observational Studies}, 11 (1),
  3--16.

\bibitem[\protect\citeauthoryear{Athey, Eckles, and Imbens}{Athey
  et~al.}{2018}]{AtheyEcklesImbens2018}
\textsc{Athey, Susan, Dean Eckles, and Guido~W Imbens} (2018): \enquote{Exact
  p-values for network interference,} \emph{Journal of the American Statistical
  Association}, 113 (521), 230--240.

\bibitem[\protect\citeauthoryear{Athey and Imbens}{Athey and
  Imbens}{2017}]{AtheyImbens2017}
\textsc{Athey, Susan and Guido~W Imbens} (2017): \enquote{The econometrics of
  randomized experiments,} \emph{Handbook of Economic Field Experiments}, 1,
  73--140.

\bibitem[\protect\citeauthoryear{Basse, Feller, and Toulis}{Basse
  et~al.}{2019}]{BasseFellerToulis2019}
\textsc{Basse, Guillaume~W, Avi Feller, and Panos Toulis} (2019):
  \enquote{Randomization tests of causal effects under interference,}
  \emph{Biometrika}, 106 (2), 487--494.

\bibitem[\protect\citeauthoryear{Berger}{Berger}{1985}]{Berger1985}
\textsc{Berger, James~O} (1985): \emph{Statistical Decision Theory and Bayesian
  Analysis, second edition}, Springer-Verlag.

\bibitem[\protect\citeauthoryear{Berger, Bernardo, and Sun}{Berger
  et~al.}{2024}]{BergerBernardoSun2024}
\textsc{Berger, James~O, Jose-Miguel Bernardo, and Dongchu Sun} (2024):
  \emph{Objective Bayesian Inference}, World Scientific.

\bibitem[\protect\citeauthoryear{Bowley}{Bowley}{1926}]{Bowley1926}
\textsc{Bowley, Arthur~Lyon} (1926): \enquote{Measurement of the precision
  attained in sampling,} \emph{Bulletin de l'Institut International de
  Statistique}, 22, 6--62.

\bibitem[\protect\citeauthoryear{Ding}{Ding}{2017}]{Ding2017comment}
\textsc{Ding, Peng} (2017): \enquote{Rejoinder: A paradox from
  randomization-based causal inference,} \emph{Statistical Science}, 32 (3),
  362--366.

\bibitem[\protect\citeauthoryear{Ding}{Ding}{2024}]{Ding2024}
---\hspace{-.1pt}---\hspace{-.1pt}--- (2024): \emph{A First Course in Causal
  Inference}, CRC Press.

\bibitem[\protect\citeauthoryear{Ding}{Ding}{2025}]{Ding2025}
---\hspace{-.1pt}---\hspace{-.1pt}--- (2025): \enquote{What randomization can
  and cannot guarantee,} \emph{Observational Studies}, 11 (1), 27--40.

\bibitem[\protect\citeauthoryear{Fisher}{Fisher}{1930}]{Fisher1930}
\textsc{Fisher, Ronald~A} (1930): \enquote{Inverse probability,}
  \emph{Mathematical Proceedings of the Cambridge Philosophical Society}, 26
  (4), 528--535.

\bibitem[\protect\citeauthoryear{Haavelmo}{Haavelmo}{1944}]{Haavelmo1944}
\textsc{Haavelmo, Trygve} (1944): \enquote{The probability approach in
  econometrics,} \emph{Econometrica}, iii--115.

\bibitem[\protect\citeauthoryear{Horowitz}{Horowitz}{1990}]{Horowitz1990}
\textsc{Horowitz, Joseph} (1990): \enquote{A uniform law of large numbers and
  empirical central limit theorem for limits of finite populations,}
  \emph{Statistics \& Probability Letters}, 10 (2), 159--166.

\bibitem[\protect\citeauthoryear{Imbens and Menzel}{Imbens and
  Menzel}{2021}]{ImbensMenzel2021}
\textsc{Imbens, Guido and Konrad Menzel} (2021): \enquote{A causal bootstrap,}
  \emph{The Annals of Statistics}, 49 (3), 1460--1488.

\bibitem[\protect\citeauthoryear{Imbens and Xu}{Imbens and
  Xu}{2024}]{ImbensXu2024}
\textsc{Imbens, Guido and Yiqing Xu} (2024): \enquote{Lalonde (1986) after
  nearly four decades: Lessons learned,} \emph{arXiv preprint
  arXiv:2406.00827}.

\bibitem[\protect\citeauthoryear{Imbens and Rubin}{Imbens and
  Rubin}{2015}]{ImbensRubin2015}
\textsc{Imbens, Guido~W and Donald~B Rubin} (2015): \emph{Causal Inference for
  Statistics, Social, and Biomedical Sciences}, Cambridge University Press.

\bibitem[\protect\citeauthoryear{LaLonde}{LaLonde}{1986}]{LaLonde1986}
\textsc{LaLonde, Robert~J} (1986): \enquote{Evaluating the econometric
  evaluations of training programs with experimental data,} \emph{The American
  Economic Review}, 604--620.

\bibitem[\protect\citeauthoryear{Li and Ding}{Li and Ding}{2016}]{LiDing2016}
\textsc{Li, Xinran and Peng Ding} (2016): \enquote{Exact confidence intervals
  for the average causal effect on a binary outcome,} \emph{Statistics in
  Medicine}, 35 (6), 957--960.

\bibitem[\protect\citeauthoryear{Little}{Little}{2004}]{Little2004}
\textsc{Little, Roderick~J} (2004): \enquote{To model or not to model?
  Competing modes of inference for finite population sampling,} \emph{Journal
  of the American Statistical Association}, 99 (466), 546--556.

\bibitem[\protect\citeauthoryear{Loh, Richardson, and Robins}{Loh
  et~al.}{2017}]{LohRichardsonRobins2017}
\textsc{Loh, Wen~Wei, Thomas~S Richardson, and James~M Robins} (2017):
  \enquote{An apparent paradox explained,} \emph{Statistical Science}, 32 (3),
  356--361.

\bibitem[\protect\citeauthoryear{Neyman}{Neyman}{1923, 1990}]{Neyman1923}
\textsc{Neyman, Jerzy} (1923, 1990): \enquote{On the application of probability
  theory to agricultural experiments. Essay on principles. Section 9.}
  \emph{Statistical Science}, 465--472.

\bibitem[\protect\citeauthoryear{Rigdon and Hudgens}{Rigdon and
  Hudgens}{2015}]{RigdonHudgens2015}
\textsc{Rigdon, Joseph and Michael~G Hudgens} (2015): \enquote{Randomization
  inference for treatment effects on a binary outcome,} \emph{Statistics in
  Medicine}, 34 (6), 924--935.

\bibitem[\protect\citeauthoryear{Ritzwoller, Romano, and Shaikh}{Ritzwoller
  et~al.}{2024}]{RitzwollerRomanoShaikh2024}
\textsc{Ritzwoller, David~M, Joseph~P Romano, and Azeem~M Shaikh} (2024):
  \enquote{Randomization inference: Theory and applications,} \emph{arXiv
  preprint arXiv:2406.09521}.

\bibitem[\protect\citeauthoryear{Robins}{Robins}{1988}]{Robins1988}
\textsc{Robins, James~M} (1988): \enquote{Confidence intervals for causal
  parameters,} \emph{Statistics in Medicine}, 7 (7), 773--785.

\bibitem[\protect\citeauthoryear{Robins and Ritov}{Robins and
  Ritov}{1997}]{RobinsRitov1997}
\textsc{Robins, James~M and Ya'acov Ritov} (1997): \enquote{Toward a curse of
  dimensionality appropriate (CODA) asymptotic theory for semi-parametric
  models,} \emph{Statistics in Medicine}, 16 (3), 285--319.

\bibitem[\protect\citeauthoryear{Scott and Wu}{Scott and
  Wu}{1981}]{ScottWu1981}
\textsc{Scott, Alastair and Chien-Fu Wu} (1981): \enquote{On the asymptotic
  distribution of ratio and regression estimators,} \emph{Journal of the
  American Statistical Association}, 76 (373), 98--102.

\bibitem[\protect\citeauthoryear{Student}{Student}{1908}]{Student1908}
\textsc{Student} (1908): \enquote{Probable error of a correlation coefficient,}
  \emph{Biometrika}, 302--310.

\bibitem[\protect\citeauthoryear{Wu and Thompson}{Wu and
  Thompson}{2020}]{WuThompson2020}
\textsc{Wu, Changbao and Mary~E Thompson} (2020): \emph{Sampling Theory and
  Practice}, Springer.

\bibitem[\protect\citeauthoryear{Wu and Ding}{Wu and Ding}{2021}]{WuDing2021}
\textsc{Wu, Jason and Peng Ding} (2021): \enquote{Randomization tests for weak
  null hypotheses in randomized experiments,} \emph{Journal of the American
  Statistical Association}, 116 (536), 1898--1913.

\bibitem[\protect\citeauthoryear{Young}{Young}{2019}]{Young2019}
\textsc{Young, Alwyn} (2019): \enquote{Channeling Fisher: Randomization tests
  and the statistical insignificance of seemingly significant experimental
  results,} \emph{The Quarterly Journal of Economics}, 134 (2), 557--598.

\bibitem[\protect\citeauthoryear{Zhao and Ding}{Zhao and
  Ding}{2021}]{ZhaoDing2021}
\textsc{Zhao, Anqi and Peng Ding} (2021): \enquote{Covariate-adjusted Fisher
  randomization tests for the average treatment effect,} \emph{Journal of
  Econometrics}, 225 (2), 278--294.

\end{thebibliography}

\appendix

\allowdisplaybreaks

\section{Proofs of Main Results}\label{sec:appendix}

\begin{proof}[Proof of theorem \ref{thm:KapproxiBalanceATEset}]
Here we use the notation of appendix \ref{sec:popDistributions}. By iterated expectations, $\Exp[Y(x)] = \Exp(Y \mid X=x) \Prob^\text{data}(X=x) + \Exp[Y(x) \mid X=1-x] \Prob^\text{data}(X= 1-x)$. By \ref{assump:boundedSupport1} and \ref{assump:KapproxBalance}, $\Exp[Y(x) \mid X=1-x] \in [\max\{y_\text{min}, \Exp(Y \mid X=x) - K \}, \min \{ y_\text{max}, \Exp(Y \mid X=x) + K \} ]$. Substitute to get the bounds in the theorem. Sharpness follows as in \cite{Manski1990}, with the minor addition that knowledge of unit identifiers in $\P^\text{data}$ has no identifying power. Specifically, any values of $\Exp[Y(1) \mid X=0] \in [y_\text{min},y_\text{max}]$ and $\Exp[Y(0) \mid X=1] \in [y_\text{min},y_\text{max}]$ are consistent with $\P^\text{data}$ because these averages depend solely on values of potential outcomes that are not present in the data. 
\end{proof}

\begin{proof}[Proof of theorem \ref{thm:randomizationUseless}]
From the proof of theorem \ref{thm:KapproxiBalanceATEset} we know that for any element $\theta \in \Theta_I(\infty)$, there exists a population matrix $\P$ that is consistent with \ref{assump:boundedSupport1} and $\P^\text{data}$ and has $\theta = \theta(\P)$. We now also know, however, that $\X$ is a realization from the distribution $\Prob_\text{design}$. However, this knowledge does not constrain the unobserved values of potential outcomes in any way, since $\Prob_\text{design}$ does not depend on potential outcomes. Hence it is still consistent with $\P$ being the true value of the population matrix.
\end{proof}

\begin{proof}[Proof of proposition \ref{prop:designMomentConditions}]
Let $\Prob_\mathcal{D}$ denote the probability measure of $\mathcal{D}$. For any given $\Prob_\mathcal{D}$, there is a unique solution to $\Exp_\text{design}[m(\mathcal{D},\theta)] = 0$, by assumption. Let $f(\Prob_\mathcal{D})$ denote this unique solution. This unique solution is assumed to equal $\theta^\text{true}$; hence $f(\Prob_\mathcal{D}) = \theta^\text{true}$. Since $m$ is known, $f$ is known. Thus, by definition \ref{def:conventionalIdentification}, $\theta^\text{true}$ is point identified.
\end{proof}

For any $N$-vector $\A = (A_1,\ldots,A_N)$, let
\[
	\text{DIM}(\A,\x^\text{new}) \coloneqq \frac{1}{N_1} \sum_{i=1}^N A_i \indicator(x_i^\text{new}=1) - \frac{1}{N_0} \sum_{i=1}^N A_i \indicator(x_i^\text{new} = 0)
\]
denote the difference-in-means of the $A_i$ variables across the treatment and control groups, when $\x^\text{new}$ is the assigned treatment vector. Then equation \eqref{eq:Ktrue} can be written as $K^\text{true}(x) = | \text{DIM}(\Y(x), \X) |$. Let $K^\text{true,new}(x) \coloneqq | \text{DIM}(\Y(x), \X^\text{new}) |$.

\begin{proof}[Proof of proposition \ref{prop:underlinePisAcdf}]
$p(\cdot, \Y(1),\Y(0))$ is the cdf for the random variable $\max \{ \allowbreak K^\text{true,new}(1), \allowbreak K^\text{true,new}(0) \}$, and hence has all the properties of a cdf. We will show that these properties are preserved once we take the infimum over $\Y(1)$ and $\Y(0)$.

\textbf{1. Monotonicity:} Since $p(\cdot,\Y(1),\Y(0))$ is a cdf, it is monotonic: $p(K_1, \Y(1),\Y(0) ) \leq p(K_2, \Y(1),\Y(0))$ for any $K_1 \leq K_2$, for all $\Y(1), \Y(0)$. Taking the infimum preserves the inequality, to get $\underline{p}(K_1) \leq \underline{p}(K_2)$ for any realization $\X$ of $\X^\text{new}$.

\textbf{2. $\underline{p}$ is right continuous:} $p(\cdot,\Y(1),\Y(0))$ is monotonic and right continuous, which implies that it is upper semicontinuous everywhere. Moreover, the pointwise infimum of an upper semicontinuous function is still upper semicontinuous. Thus $\underline{p}(\cdot)$ is upper semicontinuous. And from part 1, $\underline{p}(\cdot)$ is monotonically increasing. Now apply lemma \ref{lemma:upperSemi}.

\textbf{3. Limits:} Convergence to 1 as $K \rightarrow \infty$: Let $K \geq y_\text{max} - y_\text{min}$. Then \ref{assump:boundedSupport1}
implies that for all matrices of $\Y(1),\Y(0)$, for all realizations $\X$ of $\X^\text{new}$, $| \overline{Y}_1(x) - \overline{Y}_0(x) | \leq K$, where $\overline{Y}_g(x) \coloneqq \frac{1}{N_g} \sum_{i=1}^N Y_i(x) \indicator(X_i = g)$ for $g \in \{0,1\}$. Thus $p(K,\Y(1),\Y(0)) = 1$ for all such $K$. Since this holds for all $\Y(1),\Y(0)$, taking the infimum does not change the result.

Convergence to 0 as $K \rightarrow 0$: We want to show that, for each realization $\X$ of $\X^\text{new}$, $\lim_{K \rightarrow 0} \underline{p}(K) = 0$. We'll show something stronger: There is an $\varepsilon > 0$ such that $\underline{p}(K) = 0$ for $K \in [0,\varepsilon)$. Recall that $p(K, \Y(1),\Y(0)) = \Prob_\text{design}(K^\text{true,new}(1) \leq K, K^\text{true,new}(0) \leq K)$. And $\underline{p}$ is the infimum of this term over all $\Y(1)$ and $\Y(0)$ satisfying \ref{assump:boundedSupport1}. Thus it suffices to find (i) a single choice of $(\Y(1),\Y(0))$ that is consistent with the data and (ii) a small enough value of $K$ such that, for that choice, $K^\text{true,new}(x) \leq K$ holds with probability zero for at least one $x \in \{0,1\}$. 

Without loss of generality, consider only $\Y(1)$; if we can find an $\varepsilon > 0$ and a $\Y(1)$ that is consistent with the data and such that $K$-approximate mean balance in $\Y(1)$ is impossible for $K \in [0,\varepsilon)$, then it does not matter how the missing values of $\Y(0)$ are filled in, we will still have $\underline{p}(K) = 0$ for $K \in [0,\varepsilon)$. This follows from $\Prob_\text{design}(K^\text{true,new}(1) \leq K, K^\text{true,new}(0) \leq K) \leq \Prob_\text{design}(K^\text{true,new}(1) \leq K)$.

Order the observations so the first $N_1$ units are treated. We observe $\Y(1)_{1:N_1} = \Y_{1:N_1}$ in the data. Suppose there exist values $\Y(1)_{(N_1+1):N}^* \in [y_\text{min},y_\text{max}]^{N_0}$ such that $\text{DIM}(\Y(1)^*,\x^\text{new}) \neq 0$ for all $\x^\text{new} \in \supp(\X^\text{new})$, where $\Y(1)^* \coloneqq (\Y_{1:N_1},\Y(1)_{(N_1+1):N}^*)$. Let
\[
	\varepsilon \coloneqq \min_{\x^\text{new} \in \supp(\X^\text{new})} \; | \text{DIM}(\Y(1)^*,\x^\text{new}) |.
\]
Since $\supp(\X^\text{new})$ is finite, $\varepsilon > 0$. Then for any $K \in [0,\varepsilon)$, the event $\{ \x^\text{new} \in \supp(\X^\text{new}) : | \text{DIM}(\Y(1)^*,\x^\text{new}) | \leq K \}$ is the empty set. Hence $\Prob_\text{design}( | \text{DIM}(\Y(1)^*,\X^\text{new}) | \leq K) = 0$ for all $K \in [0,\varepsilon)$. Consequently, the infimum over all possible completions $\Y(1)_{(N_1+1):N} \in [y_\text{min},y_\text{max}]^{N_0}$ will also equal zero.

Now suppose by way of contradiction that such a completion $\Y(1)^*$ does not exist. Then for \emph{all} completions $\Y(1)_{(N_1+1):N} \in [y_\text{min},y_\text{max}]^{N_0}$ there is some $\x^\text{new} \in \supp(\X^\text{new})$ such that $\text{DIM}(\Y(1),\x^\text{new}) = 0$. That is,
{\footnotesize
\begin{equation}\label{eq:contradiction}
	[y_\text{min},y_\text{max}]^{N_0} \subseteq \bigcup_{\x^\text{new} \in \supp(\X^\text{new})} \left\{ \Y(1)_{(N_1+1):N} \in [y_\text{min},y_\text{max}]^{N_0} : \text{DIM} \big( (\Y_{1:N_1},\Y(1)_{(N_1+1):N}) , \x^\text{new} \big) = 0 \right\}.
\end{equation}
}
The right hand side of \eqref{eq:contradiction} is a finite union of hyperplanes, which therefore has empty interior, while the left hand side has non-empty interior since $y_\text{min} < y_\text{max}$. This is a contradiction.
\end{proof}

\begin{proof}[Proof of theorem \ref{thm:ValidCoverage}]
For brevity, let $\mathcal{C} = \mathcal{C}(\Y(1) \times \X^\text{new} + \Y(0) \times (\bold{1}-\X^\text{new}),\X^\text{new})$ denote the random confidence set. Recall its realizations are sets $\Theta_I(K(\alpha))$. First note that
\[
	\Prob_\text{design}( K^\text{true,new}(x) \leq K(\alpha) \text{ for $x \in \{0,1\}$}) \leq \Prob_\text{design} \big(\mathcal{C} \ni \theta(\Y(1),\Y(0)) \big).
\]
This follows because for any realization $\X$ of $\X^\text{new}$, $K^\text{true}(x) \leq K(\alpha)$ for each $x \in \{0,1\}$ implies that \ref{assump:KapproxBalance} holds, and hence theorem \ref{thm:KapproxiBalanceATEset} gives $\text{ATE} \in \Theta_I(K(\alpha))$. So the inequality follows by monotonicity of probability measures. Next, define $K^*(\alpha) \coloneqq \inf \{ K \geq 0 : p(K,\Y(1),\Y(0)) \geq 1-\alpha \}$. This is a non-stochastic, infeasible, ``oracle'' choice of $K(\alpha)$. We have $K^*(\alpha) \leq K(\alpha)$ for all realizations $\X$ (recall that $K(\alpha)$ is random here). That follows because $\underline{p}(K) \leq p(K,\Y(1),\Y(0))$ for all $K$, for all realizations of $\X$ ($\underline{p}$ is also random here) and by definition of $K(\alpha)$. This implies that
\[
	\Prob_\text{design}( K^\text{true,new}(x) \leq K^*(\alpha) \text{ for $x \in \{0,1\}$}) \leq \Prob_\text{design}( K^\text{true,new}(x) \leq K(\alpha) \text{ for $x \in \{0,1\}$}).
\]
Finally, by definition, $p(K,\Y(1),\Y(0)) \coloneqq \Prob_\text{design}( K^\text{true,new}(x) \leq K \text{ for $x \in \{0,1\}$})$. Thus
\[
	\Prob_\text{design}( K^\text{true,new}(x) \leq K^*(\alpha) \text{ for $x \in \{0,1\}$})
	= p(K^*(\alpha), \Y(1),\Y(0)) \geq 1-\alpha.
\]
The last inequality follows by the definition of $K^*(\alpha)$, and since $p(\cdot,\Y(1),\Y(0))$ is right continuous. Putting everything together gives $\Prob_\text{design} \big(\mathcal{C} \ni \theta(\Y(1),\Y(0)) \big) \geq 1-\alpha$. This holds for all $\Y(1)$ and $\Y(0)$ and therefore the inequality holds for the infimum as well.
\end{proof}

\begin{proof}[Proof of theorem \ref{thm:randomizationGood}]
\textbf{Part 1.} We want to show that $\underline{p}(K^\text{bp}) \xrightarrow{p} 1$. There are two steps. First we use lemma \ref{lemma:convergenceOfIdentifiedSet} to show that the finite population breakdown point converges to a limiting breakdown point. Then we combine this step with lemma \ref{lemma:underlinepGoesTo1} to get the final result. Suppose $\mu(1) - \mu(0) > 0$; the less than zero case can be handled symmetrically. 

\begin{enumerate}
\item Define the limit breakdown point $K_\infty^\text{bp} \coloneqq \sup \{ K \geq 0 : 0 \notin \Theta_{I,\infty}(K) \}$. Since $\Theta_{I,\infty}(K)$ is an interval and $\mu(1) - \mu(0) > 0$, $K_\infty^\text{bp}$ can alternatively be written as the unique smallest solution to $\text{LB}_K^\infty(1) - \text{UB}_K^\infty(0) = 0$ (defined in lemma \ref{lemma:convergenceOfIdentifiedSet}). Similarly, $K^\text{bp} \coloneqq \sup \{ K \geq 0 : 0 \notin \Theta_I(K) \}$ can be written as the unique smallest solution to $\text{LB}_K(1) - \text{UB}_K(0) = 0$. Both the finite $N$ and limiting bound functions are continuous in $K$. And lemma \ref{lemma:convergenceOfIdentifiedSet} showed that the finite $N$ bound functions converge in probability to the limiting bound functions uniformly in $K$. Consequently, $K^\text{bp} \xrightarrow{p} K_\infty^\text{bp}$.

\item Since $\mu(1) - \mu(0) \neq 0$, $K_\infty^\text{bp} > 0$. This combined with $K^\text{bp} \xrightarrow{p} K_\infty^\text{bp}$ implies that there is an $\varepsilon > 0$ such that $\indicator(K^\text{bp} \geq \varepsilon) \xrightarrow{p} 1$. So $\underline{p}(K^\text{bp}) = \underline{p}(K^\text{bp}) \indicator(K^\text{bp} < \varepsilon) + \underline{p}(K^\text{bp}) \indicator(K^\text{bp} \geq \varepsilon) = O_p(1) o_p(1) + \underline{p}(K^\text{bp}) \indicator(K^\text{bp} \geq \varepsilon)$. Finally, $1 \geq \underline{p}(K^\text{bp}) \indicator(K^\text{bp} \geq \varepsilon) \geq \underline{p}(\varepsilon) \indicator(K^\text{bp} \geq \varepsilon)$ by monotonicity of $\underline{p}$, and since, as a cdf, it is bounded above by 1. The last line converges in probability to 1 since $\varepsilon > 0$ and by lemma \ref{lemma:underlinepGoesTo1}. Thus $\underline{p}(K^\text{bp}) \indicator(K^\text{bp} \geq \varepsilon) \xrightarrow{p} 1$. 
\end{enumerate}

\noindent \textbf{Part 2.} Fix $\alpha \in (0,1)$. Since $\Theta_I(K)$ is an interval, $\sup_{\theta \in \Theta_I(K(\alpha))} | \theta - \text{ATE}_N |$ equals
\[
	\max \left\{  \left| \big( \text{LB}_{K(\alpha)}(1) - \text{UB}_{K(\alpha)}(0) \big) - \text{ATE}_N \right|, \;
		\left| \big( \text{UB}_{K(\alpha)}(1) - \text{LB}_{K(\alpha)}(0) \big) - \text{ATE}_N \right| \right\}.
\]
We will consider the first term only; the proof for the second term is analogous. We have $| \big( \text{LB}_{K(\alpha)}(1) - \text{UB}_{K(\alpha)}(0) \big) - \text{ATE}_N | \leq | \big( \text{LB}_{K(\alpha)}(1) - \text{UB}_{K(\alpha)}(0) \big) - \big( \mu(1) - \mu(0) \big) | + | \text{ATE}_N - \big( \mu(1) - \mu(0) \big) |$. The second term goes to zero by assumption. So consider the first term: $| \big( \text{LB}_{K(\alpha)}(1) - \text{UB}_{K(\alpha)}(0) \big) - \big( \mu(1) - \mu(0) \big) | \leq | \big( \text{LB}_{K(\alpha)}(1) - \text{UB}_{K(\alpha)}(0) \big) \allowbreak - \big( \text{LB}_{K(\alpha)}^\infty(1) - \text{UB}_{K(\alpha)}^\infty(0) \big) | + | \big( \text{LB}_{K(\alpha)}^\infty(1) - \text{UB}_{K(\alpha)}^\infty(0) \big) - \big( \mu(1) - \mu(0) \big) |$. The first term here goes to zero in probability by lemma \ref{lemma:convergenceOfIdentifiedSet}. The second term equals $ | \mu(1) \rho + \max \{ y_\text{min}, \mu(1) - K(\alpha) \} (1-\rho) -  \mu(0) (1-\rho) - \min \{ y_\text{max}, \mu(0) + K(\alpha) \} \rho - \mu(1) + \mu(0) |$. Recall that $K(\alpha) \coloneqq \inf \{ K \geq 0 : \underline{p}(K) \geq 1-\alpha \}$. So lemma \ref{lemma:underlinepGoesTo1} implies that $K(\alpha) \rightarrow 0$ as $N \rightarrow \infty$, since $\alpha$ is strictly between 0 and 1. This implies that this second term goes to zero as $N \rightarrow \infty$. Thus we have shown that $\left| \big( \text{LB}_{K(\alpha)}(1) - \text{UB}_{K(\alpha)}(0) \big) - \text{ATE}_N \right| \xrightarrow{p} 0$ as $N \rightarrow \infty$. 
\end{proof}

\begin{proof}[Proof of lemma \ref{lemma:firstIVeq}]
By algebra, $\overline{Y}_{z=1} - \overline{Y}_{z=0} = \big( \overline{U}_{z=1} - \overline{U}_{z=0} \big) + \sum_{i=1}^N \beta_i \big( X_i(1) \allowbreak \indicator(Z_i=1) /N_1 - X_i(0) \indicator(Z_i = 0) / N_0 \big)$ and $\overline{X}_{z=1} - \overline{X}_{z=0} = \big( \overline{T}_1(a) - \overline{T}_0(a) \big) + \overline{T}_1(c)$. Similarly, we can write $\sum_{i=1}^N \beta_i \big( X_i(1) \indicator(Z_i=1) /N_1 - X_i(0) \indicator(Z_i = 0) / N_0 \big) = \Big( \overline{T}_1(a) \overline{\beta}_1(a) - \overline{T}_0(a) \overline{\beta}_0(a) \Big) + \overline{T}_1(c) \overline{\beta}_1(c)$. Putting these derivations together gives $\overline{Y}_{z=1} - \overline{Y}_{z=0} = \big( \overline{U}_{z=1} - \overline{U}_{z=0} \big) + \Big( \overline{T}_1(a) \overline{\beta}_1(a) - \overline{T}_0(a) \overline{\beta}_0(a) \Big) + \overline{T}_1(c) \overline{\beta}_1(c)$. Dividing by the first stage gives equation \eqref{eq:mainIVequation}.
\end{proof}

\begin{proof}[Proof of proposition \ref{prop:IVpointIdent}]
Follows immediately from lemma \ref{lemma:firstIVeq}.
\end{proof}

\begin{proof}[Proof of theorem \ref{thm:IVoneSidedNoncomplianceSet}]
When there are no always takers, equation \eqref{eq:mainIVequation} simplifies to $\text{WaldEstimand} \coloneqq \frac{\overline{Y}_{z=1} - \overline{Y}_{z=0}}{\overline{X}_{z=1} - \overline{X}_{z=0}} = \frac{\overline{U}_{z=1} - \overline{U}_{z=0}}{\overline{X}_{z=1} - \overline{X}_{z=0} } + \overline{\beta}_1(c)$. Thus we no longer have to worry about balance in always takers, simply because they don't exist. We only have to worry about balance in $Y_i(0)$ across the treatment and control groups. Solving this equation for LATT gives $\overline{\beta}_1(c) = \text{WaldEstimand} - (\overline{U}_{z=1} - \overline{U}_{z=0})/ \pi$ where recall that $\pi \coloneqq \overline{X}_{z=1} - \overline{X}_{z=0}$ denotes the first stage, and relevance ensures that we are not dividing by zero. This equation and \ref{assump:IVapproxMeanBalance} immediately show that the bounds in the theorem statement are valid. Sharpness obtains because the unknown $Y_i(0)$ values are completely unconstrained, so long as they satisfy \ref{assump:IVapproxMeanBalance}. Hence any value in the interval is attainable.
\end{proof}

\makeatletter\@input{KM2aux.tex}\makeatother
\end{document}


\maketitle
\appendix
\allowdisplaybreaks

\setcounter{lemma}{1}
\setcounter{section}{1}
\setcounter{equation}{7}

This document contains appendices to accompany the main text. Appendix \ref{sec:history} briefly discusses the history of finite and infinite population analysis, and how a typical design-based inference analysis proceeds. Appendix \ref{sec:popDistributions} provides an alternative but equivalent representation of the finite populations matrix $\P^\text{true}$ in terms of random variables. Appendix \ref{sec:sampleSurveyGuide} gives a parallel exposition to section \ref{sec:introPractitionerGuideRCTs}, focusing on survey sampling rather than causal inference. Appendix \ref{sec:conventionalAppendix} analyzes the implications of the conventional definition of finite population identification in instrumental variable models. Appendix \ref{sec:moreBayesian} gives additional discussion of the Bayesian interpretation of section \ref{sec:interpretation}. Appendix \ref{sec:MILPappendix} develops the mixed integer linear programming (MILP) characterization of our main optimization problem in equation \eqref{eq:defOfunderlinep}. Appendix \ref{sec:numericalAppendix} gives additional details on the numerical illustration of section \ref{sec:numericalIllustration}. Appendix \ref{sec:AlternativeEmpiricalCIs} shows three alternative frequentist confidence intervals for our main empirical applications. Appendix \ref{sec:LaLonde} provides a third empirical application, with $N=722$. Appendix \ref{sec:simulations} studies the frequentist properties of our uncertainty intervals. Appendix \ref{sec:lemmas} describes various lemmas used in our main proofs.

\section{A Brief History of Finite versus Infinite Population Analysis}\label{sec:history}

\nocite{Neyman1923}

\noindent The distinction between two types of populations goes back to the beginning of statistics:
\begin{enumerate}
\item They can be \emph{finite}, as in Neyman (1923), or \cite{Bowley1926} who considered ``a sample containing $n$ persons or things [that] is selected from a universe containing $N$....'' (page 3).

\item Or they can be \emph{infinite}, as in \cite{Fisher1930}, or \cite{Student1908} who assumed ``a random sample has been obtained from an indefinitely large* population.... *Note that the indefinitely large population need not actually exist'' (page 302).
\end{enumerate}
Inference in finite populations came to be called ``design-based inference'' or ``randomization inference'' while inference in infinite populations came to be called ``model-based inference'' or ``super-population inference'' (e.g., \citealt{WuThompson2020}). Econometric theory has traditionally focused on super-population inference, going back to \cite{Haavelmo1944} who explained that ``when we describe $s$ as a random variable with a certain probability distribution for each fixed set of values of the variables $x$, we are thinking of a class of hypothetical, infinite populations...'' (page 51). Likewise, much of statistics has also focused on the super-population approach, with survey sampling theory as the main exception (see \citealt{Little2004}).

\subsubsection*{The Traditional Template of Design-Based Inference}

The recent literature on design-based inference in explicitly finite populations (see section \ref{sec:litReview}) mostly follows a template largely laid out by Neyman (1923). This template has three steps:
\begin{enumerate}
\item Describe the finite \emph{population}.

\item Describe the \emph{design}, which explains how the data is obtained.

\item Derive properties of the \emph{induced design distribution} of statistics that depend on the population and the data.
\end{enumerate}
For example, in the context of survey sampling, the population is a set of numbers $\Y = (Y_1,\ldots,Y_N)$ (step 1). A classical design is simple random sampling: Sample exactly $n$ units from $\{ 1,\ldots,N \}$ uniformly at random (step 2). Let $\S = (S_1,\ldots,S_N)$ be binary random variables indicating which units were sampled. A classical theorem is that the sample mean (a function of $\S$ and $\Y$) is design-unbiased for the finite population mean when $\S$ is chosen uniformly at random: $\Exp_\text{design} \left[ \frac{1}{n} \sum_{i=1}^N S_i Y_i \right] = \frac{1}{N} \sum_{i=1}^N Y_i$ (step 3). As we discussed in section \ref{sec:intro}, there is a noticeable omission in this template: It does not involve any identification analysis. In super-population based econometrics, identification analysis is often a key step, which has led to a vast and influential literature. By developing a formal definition of identification in finite populations, we hope to bring similar ideas from super-population identification theory to bear on the problem of quantifying uncertainty in finite populations.

\section{A Random Variable Representation of a Finite Population}\label{sec:popDistributions}

The matrix $\P^\text{true}$ fully describes the population, but it is often helpful to use an equivalent representation of the population in terms of random variables on $\mathcal{I}$. Specifically, for $x \in \{0,1\}$ let $Y_{(\cdot)}(x) : \mathcal{I} \rightarrow \mathcal{Y}$ denote the function which tells us the $x$-potential outcome for unit $i$ when evaluated at $i \in \mathcal{I}$. Here $\mathcal{Y} \subseteq \R$ denotes the set of logically possible values of potential outcomes. Likewise, define $X_{(\cdot)} : \mathcal{I} \rightarrow \{ 0,1 \}$ and $W_{(\cdot)} : \mathcal{I} \rightarrow \mathcal{W}$. Here $\mathcal{W}$ denotes the set of logically possible values of covariates. Finally, define $I_{(\cdot)}: \mathcal{I} \rightarrow \mathcal{I}$ as a unit identifier variable, $I_i = i$. Let $\mathbb{U}$ be the uniform probability measure on $(\mathcal{I}, \text{PowerSet}(\mathcal{I}))$, so $\mathbb{U}(\{ i \}) = 1/N$. Through all of these functions on $\mathcal{I}$, $\mathbb{U}$ induces a discrete probability distribution on $\mathcal{I} \times \mathcal{Y} \times \mathcal{Y} \times \{0,1\} \times \mathcal{W}$. Denote this induced probability distribution by $\mathbb{P}^\text{true}$ and call it the \emph{population distribution} of the variables. Use $\Prob$ to denote alternative possible values of this population distribution. As usual, we will drop the explicit argument of the functions $(I_{(\cdot)}, Y_{(\cdot)}(1), Y_{(\cdot)}(0), X_{(\cdot)}, W_{(\cdot)})$ and simply write them as the random variables $(I, Y(1),Y(0),X,W)$. We can similarly define $Y_{(\cdot)} : \mathcal{I} \rightarrow \mathcal{Y}$ by $Y_i \coloneqq Y_i(1) X_i + Y_i(0) (1-X_i)$ for all $i \in \mathcal{I}$. We then let $\Prob^\text{data}$ denote the induced distribution of $(I_{(\cdot)}, Y_{(\cdot)},X_{(\cdot)},W_{(\cdot)})$. As before we also drop the arguments of these functions and write them as the random variables $(I,Y,X,W)$.

This notation allows us to use probability theory concepts to describe features of the population $\P^\text{true}$ and the data $\P^\text{data}$, such as using expected values to denote average potential outcomes. For example, $\text{ATE} = \Exp[Y(1) - Y(0)]$, where $\Exp$ refers to a population level expectation; that is, with respect to the discrete distribution $\Prob^\text{true}$. Likewise this lets us use $\Exp(Y \mid X=x)$ to denote the average observed outcome in group that received treatment value $x$. Finally, note that the two representations $\P^\text{true}$ and $\Prob^\text{true}$ are equivalent because we can always recover $\P^\text{true}$ from $\Prob^\text{true}$ by conditioning on $I=i$ for all $i \in \mathcal{I}$.

\section{Our Approach to Quantifying Uncertainty in Survey Samples}\label{sec:sampleSurveyGuide}

We sketched our approach in section \ref{sec:introPractitionerGuideRCTs}, in the context of randomized experiments. Here we present an even simpler exposition of our approach, in the context of quantifying uncertainty due to non-sampled units. This is conceptually the same as simply focusing on a single potential outcome in a randomized experiment, but we include for readers who would like to see it explicitly. Also see section \ref{sec:sampling} for further discussion of sampling.

Consider the population of six units in table \ref{table:examplePopulationSurvey} (left). Here $W_i$ denote the value of some variable for unit $i$. Suppose our goal is to learn the finite population mean, $\mu \coloneqq \frac{1}{N} \sum_{i=1}^N W_i$. In this population, $\mu = 0.417$. 

\begin{table}[h]
  \centering
  \begin{subtable}[t]{0.4\textwidth}
  \vspace{0pt}
    \centering
\begin{tikzpicture}[remember picture,
                    baseline=(current bounding box.north)]

\node (mytable) {
  \begin{tabular}{c|c}
  $i$ & $W_i$  \\[0.5em]
  \hline
  1 & \cellnode{r1c2}{0.1}  \\
  2 & \cellnode{r2c2}{0}   \\
  3 & \cellnode{r3c2}{0.2}  \\
  4 & \cellnode{r4c2}{0.8}  \\
  5 & \cellnode{r5c2}{0.9} \\
  6 & \cellnode{r6c2}{0.5} \\
  \end{tabular}
};/

\end{tikzpicture}
  \end{subtable}
  \begin{subtable}[t]{0.4\textwidth}
  \vspace{-10pt}
    \centering
\begin{tikzpicture}[remember picture,
                    baseline=(current bounding box.north)]

\node (mytable) {
  \begin{tabular}{c|cc}
  $i$ & $W_i$ & $S_i$ \\[0.5em]
  \hline
  1 & \cellnode{r1c2}{0.1} & 1 \\
  2 & \cellnode{r2c2}{0}  & 1 \\
  3 & \cellnode{r3c2}{0.2} & 1 \\[0.5em]
  4 & \cellnode{r4c2}{?} &  0 \\
  5 & \cellnode{r5c2}{?} &  0 \\
  6 & \cellnode{r6c2}{?} &  0 \\
  \end{tabular}
};

\node[fit=(r1c2)(r2c2)(r3c2),
      draw,
      inner sep=2.5pt] {};

\node[fit=(r4c2)(r5c2)(r6c2),
      draw,
      dashed,
      inner sep=4pt] {};

\end{tikzpicture}
  \end{subtable}
  \caption{Left: Example population, $N=6$. Right: The observed data from that population.}
  \label{table:examplePopulationSurvey}
\end{table}

Now, suppose that we sampled $n=3$ of these $N=6$ units. Let $S_i=1$ if unit $i$ is sampled, and $S_i=0$ otherwise. The data, shown in the right table, only reveals the value of $W_i$ for half of the population. What can we say about $\mu$? Without \emph{any} assumptions, we cannot say anything about $\mu$ whatsoever, because the values of $W_i$ for the non-sampled units can be anything. Suppose, though, that we are willing to assume that all values of this variable must be between 0 and 1. Then, without any additional assumptions, all we can say \emph{for sure} about $\mu$ is that it is in the set
\begin{align*}
	\left[ \frac{(0.1+0+0.2) + (0+0+0)}{6}, \; \frac{(0.1+0+0.2) + (1+1+1)}{6} \right] = [0.05, 0.55].
	\eqqcolon \Theta_I(\infty)
\end{align*}
This is the finite population identified set for $\mu$, obtained by filling in the unobserved values in table \ref{table:examplePopulationSurvey} (right) with values between $[0,1]$ to either maximize or minimize the corresponding $\mu$ value.

That result allows the sampled and non-sampled groups to be arbitrarily different (subject to the constraint $W_i \in [0,1]$ for all $i$). What if we thought they were similar in some way? Specifically, suppose there is a known $K \geq 0$ such that
\[
	\left| \frac{1}{n} \sum_{i=1}^N W_i \indicator(S_i = 1) - \frac{1}{N-n} \sum_{i=1}^N W_i \indicator(S_i = 0) \right| \leq K.
\]
Like our main ``approximate mean balance'' assumption for causal inference, this assumption states that the average of the unobserved values in the dashed box is not farther than $K$ from the observed sample mean $(0.1+0+0.2)/3 = 0.1$ in the solid box. The bounds on $\mu$ above can be modified to incorporate this restriction; denote them by $\Theta_I(K)$.

This set $\Theta_I(K)$ allows us to examine the sensitivity of conclusions about $\mu$ to assumptions about the magnitude of the \emph{realized} imbalance between the sampled and non-sampled groups. But what value of $K$ is plausible? We can use a random sampling assumption to calibrate our beliefs about the plausible values of $K$. Specifically, suppose the following assumption holds.

\setcounter{Aassumption}{3}
\begin{Aassumption}[Random sampling]\label{assump:randomSampling}
The sample size $n$ is fixed a priori. $\S = (S_1,\ldots,S_N)$ is a single realization from the distribution that puts equal weight on all possible ${N \choose n}$ samples.
\end{Aassumption}

Suppose we knew all values of $W_i$. Then under assumption \ref{assump:randomSampling}, we can compute the \emph{ex ante} probability that the difference between the average value of $W_i$ in the sampled and non-sampled groups is at most $K$. This can be done for any $K$. Since we do not actually know all values of the variable, we can find the smallest possible value of this probability, across all logically possible completions of table \ref{table:examplePopulationSurvey} (right). Denote this smallest probability by $\underline{p}(K)$. Let $K(\alpha)$ denote the smallest magnitude of $K$ such that $\underline{p}(K) \geq 1-\alpha$. We show that the sets $\Theta_I(K(\alpha))$ can be interpreted as identified sets, robust Bayesian credible sets, or uniform frequentist design-based confidence sets. We recommend that researchers report plots of $\Theta_I(K(\alpha))$ versus $1-\alpha$ (akin to figure \ref{fig:GneezyDesignBased2}), which can then be interpreted any of these ways, depending on personal preference. 

Overall, in this section we showed how uncertainty due to non-sampled units can be quantified using the same tools that we developed to quantify uncertainty due to missing potential outcomes. 
 
\section{Conventional Identification of Instrumental Variable Models}\label{sec:conventionalAppendix}

In this section we provide additional details for the implications of proposition \ref{prop:weirdWeird} for the instrumental variable setting. Suppose all randomness comes from assignment of an instrument $Z_i^\text{new}$ with distribution $\Prob_\text{design}$. Let $X_i(z)$ denote potential treatments. Then proposition \ref{prop:weirdWeird} immediately implies that $X_i(z)$ is point identified for any $i$ and $z$ with $\Prob_\text{design}(Z_i^\text{new} = z) > 0$. Consequently, $Y_i(x)$ will be point identified for any $i$ and $x$ with $\Prob_\text{design}(X_i(Z_i^\text{new}) = x) > 0$. In particular, the assignment of $Z_i^\text{new}$ can be endogenous in the sense that $\Prob_\text{design}(X_i(Z_i^\text{new}) = x)$ can be a function $f(Y_i(1),Y_i(0),X_i(1),X_i(0))$, as long as it is not degenerate on 0.

For example, consider the 2 by 2 setting where treatment and the instrument are binary. Then the type of each unit $i$ is point identified (and this does not require assuming no-defiers). That follows since $p_i$ is point identified for all $i \in \mathcal{I}$, so units with $p_i = 0$ are never takers, units with $p_i = 1$ are always takers, and units with $p_i \in (0,1)$ are either compliers or defiers. For compliers or defiers, both $X_i(z)$ potential outcomes are point identified, and hence the type for each $i$ is point identified. Moreover, for compliers or defiers, both $Y_i(x)$ potential outcomes are point identified, because these units have ex ante positive probability of treatment and non-treatment. 

$Y_i(1)$ is point identified for always taker types but $Y_i(0)$ is not point identified for always takers because these units have $\Prob_\text{design}(X_i^\text{new}=0) = 0$, and hence there is not even a single realization of instrument assignment under which we can ever observe $Y_i(0)$. Similarly, $Y_i(0)$ is point identified for never-taker types, but $Y_i(1)$ is not.

\section{Additional Discussion of the Bayesian Interpretation}\label{sec:moreBayesian}

In this section we provide additional discussion of the Bayesian interpretation of $\underline{p}$ and $\Theta_I(K(\alpha))$ discussed in section \ref{sec:interpretation}. Specifically, we discuss (1) how $\underline{p}$ can be interpreted as ``robust empirical'' Bayes beliefs about imbalance and (2) the differences between our approach and the classical Bayesian approach. Recall that the classical, ``fully'' Bayesian approach first places a prior on the entire matrix $(\Y(1),\Y(0))$, and then updates this prior once the data are observed.

Recall from section \ref{sec:designBasedMainSection} that each specification of the matrix $(\Y(1),\Y(0))$ leads to a different randomization-induced distribution of the magnitude of ex post imbalance, $K^\text{true,max}$, summarized by the cdf $p(\cdot,\Y(1),\Y(0))$. A prior over $(\Y(1),\Y(0))$, combined with the randomization assumption \ref{assump:uniformRandomization}, implies a single prior on $K^\text{true,max}$. In our approach, we do not specify a distribution over $(\Y(1),\Y(0))$. 

Instead, we consider ``robust'' Bayesian inference, across the class of all possible priors on $(\Y(1),\Y(0))$ that satisfy the bounded outcome assumption \ref{assump:boundedSupport1}. Each of these priors leads to a distribution over $K^\text{true,max}$. By construction (see equation \eqref{eq:defOfunderlinep}), $\underline{p}$ first order stochastically dominates each of these other distributions over $K^\text{true,max}$ that arise from a prior that is consistent with \ref{assump:boundedSupport1} and \ref{assump:uniformRandomization}. Consequently, $\Theta_I(K(\alpha))$ is a credible set for ATE with respect to \emph{any} prior distribution on $(\Y(1),\Y(0))$ consistent with the maintained assumptions. This is the sense in which $\underline{p}$ is a ``robust'' distribution over imbalance, using traditional jargon from the Bayesian literature (e.g., section 4.7 of \citealt{Berger1985}).

We conclude with two remarks on jargon: $\underline{p}$ is ``empirical'' because it depends on the observed data $(\Y,\X)$ through the restriction that $Y_i = Y_i(1) X_i + Y_i(0) (1-X_i)$ for all $i$, as imposed in equation \eqref{eq:defOfunderlinep}. This is somewhat analogous to how the data is used to determine priors in the empirical Bayes' literature, although the mathematical settings and details are different. Finally, $\underline{p}$ is an ``objective'' distribution in the sense that it is constructed based on the ``objective'' implications of the observed data and the randomly assigned treatment assumption. See \cite{BergerBernardoSun2024} for a survey of objective Bayesian methods.

\section{MILP Computation of $\underline{p}$}\label{sec:MILPappendix}

In this section we develop the MILP representation of the optimization problem in equation \eqref{eq:defOfunderlinep}. Order the units so that the first $N_1$ indices correspond to treated units while the remaining $N_0$ units correspond to the untreated units. For brevity, we will use $a_i$ to denote the unknown value of $Y_i(1)$, for $i > N_1$, and $b_i$ to denote the unknown value of $Y_i(0)$ for $i \leq N_1$. So
\begin{align}
	&\underline{p}(K) \notag \\
	&\coloneqq \inf_{
	\substack{
	\Y(1)_{N_1+1:N} \in [y_\text{min},y_\text{max}]^{N_0} \\
	\Y(0)_{1:N_1} \in [y_\text{min},y_\text{max}]^{N_1} 
	}
	} \; p\big( K,(\Y_{1:N_1}, \Y(1)_{N_1+1:N}), (\Y(0)_{1:N_1}, \Y_{N_1+1:N}) \big) \notag \\
	&= \inf_{
	\substack{
	\a \in [y_\text{min},y_\text{max}]^{N_0} \\
	\b \in [y_\text{min},y_\text{max}]^{N_1} 
	}
	} \; p\big( K,(\Y_{1:N_1}, \a), (\b, \Y_{N_1+1:N}) \big) \notag \\
	&=
	\inf_{
	\substack{	
	a_i \in [y_\text{min}, y_\text{max}] : X_i = 0 \\
	b_i \in [y_\text{min}, y_\text{max}] : X_i = 1
	}
	}
	\; 
	\frac{1}{| \mathcal{X}_{N_1} |} \sum_{\x^\text{new} \in \mathcal{X}_{N_1}}
\resizebox{0.6\linewidth}{!}{$	
	\indicator \Big( \big| \text{Diff}_1(\Y,\a,\x^\text{new}) \big| \leq K \Big)
	\cdot \indicator \Big( \big| \text{Diff}_0(\Y,\b,\x^\text{new}) \big| \leq K \Big)
$}
	\label{eq:mainOptProb}
\end{align}
where the last line follows by \ref{assump:uniformRandomization} and
\begin{multline*}
	\text{Diff}_1(\Y,\a,\x^\text{new}) \coloneqq \\
	\frac{1}{N_1} \sum_{i=1}^N x_i^\text{new} \big( Y_i \indicator(X_i = 1) + a_i \indicator(X_i = 0) \big) - \frac{1}{N_0} \sum_{i=1}^N (1-x_i^\text{new}) \big( Y_i \indicator(X_i = 1) + a_i \indicator(X_i = 0) \big)
\end{multline*}
is the difference in average values of the $Y_i(1)$ potential outcome between the treatment and control group, for the new realization of treatment assignment, and
\begin{multline*}
	\text{Diff}_0(\Y,\b,\x^\text{new}) \coloneqq \\
	\frac{1}{N_1} \sum_{i=1}^N x_i^\text{new} \big( Y_i \indicator(X_i = 0) + b_i \indicator(X_i = 1) \big) - \frac{1}{N_0} \sum_{i=1}^N (1-x_i^\text{new}) \big( Y_i \indicator(X_i = 0) + b_i \indicator(X_i = 1) \big)
\end{multline*}
is the difference in average values of the $Y_i(0)$ potential outcome between the treatment and control group, for the new realization of treatment assignment. Here we used the representation
\[
	Y_i(x) \indicator(X_i^\text{new} = g) =
	\big( Y_i \indicator(X_i = x) + Y_i(x) \indicator(X_i \neq x) \big) \indicator(X_i^\text{new} = g).
\]
For simplicity, we'll ignore the second indicator in equation \eqref{eq:mainOptProb}, the $\text{Diff}_0$ term. All of the derivations below generalize to accommodate this additional indicator but require extra notation. So, dropping that indicator, equation \eqref{eq:mainOptProb} becomes
\[
	\inf_{
	\substack{	
	a_i \in [y_\text{min}, y_\text{max}] : X_i = 0
	}
	}
	\; \frac{1}{| \mathcal{X}_{N_1} |} \sum_{\x^\text{new} \in \mathcal{X}_{N_1}}
	\indicator \Big( \big| \text{Diff}_1(\Y,\a,\x^\text{new}) \big| \leq K \Big).
\]
Index elements of $\mathcal{X}_{N_1}$ by $j$. Define
\begin{align*}
	z_j^{1+} &\coloneqq \indicator \Big( \text{Diff}_1(\Y,\a,\x^\text{new}) \leq K \Big) \\
	z_j^{1-} &\coloneqq \indicator \Big( - \text{Diff}_1(\Y,\a,\x^\text{new}) \leq K \Big)
\end{align*}
where we leave the arguments of these functions implicit. Then the optimization problem can be written as
\[
	\inf_{
	\substack{	
	a_i \in [y_\text{min}, y_\text{max}] : X_i = 0
	}
	}
	\; \frac{1}{| \mathcal{X}_{N_1} |} \sum_{\x^\text{new} \in \mathcal{X}_{N_1}}
	z_j^{1+} z_j^{1-}
\]
We use a ``big $M$'' approach: Since outcomes are bounded (\ref{assump:boundedSupport1}), there are $M_1^{1+}$ and $M_2^{1+}$ large enough such that, uniformly over $\Y,\a,\x^\text{new}$,
\begin{align*}
	M_1^{1+} z_j^{1+} \geq K - \text{Diff}_1(\Y,\a,\x^\text{new})  \\
	M_2^{1+} (1 - z_j^{1+}) \geq \text{Diff}_1(\Y,\a,\x^\text{new}) - K
\end{align*}
for all $j$. Similarly for $z_j^{1-}$. So now we have
\[
	\inf_{
	\substack{	
	a_i \in [y_\text{min}, y_\text{max}] : X_i = 0 \\
	z_j^{1+}, z_j^{1-} \in \{ 0,1 \} \text{ all $j$}
	}
	}
	\; \frac{1}{| \mathcal{X}_{N_1} |} \sum_{\x^\text{new} \in \mathcal{X}_{N_1}}
	z_j^{1+} z_j^{1-}
\]
subject to
\begin{align*}
	M_1^{1+} z_j^{1+} \geq K - \text{Diff}_1(\Y,\a,\x^\text{new})  \\[0.5em]
	M_2^{1+} (1 - z_j^{1+}) \geq \text{Diff}_1(\Y,\a,\x^\text{new}) - K
\end{align*}
and two similar inequalities for $z_j^{1-}$. In the final step, let $z_j^3$ satisfy
\[
	z_j^3 \leq z_j^{1+}
	\qquad
	z_j^3 \leq z_j^{1-}
	\qquad
	z_j^{1+} + z_j^{1-} \leq 1 + z_j^3.
\]
Thus our optimization problem becomes
\[
	\inf_{
	\substack{	
	a_i \in [y_\text{min}, y_\text{max}] : X_i = 0 \\
	z_j^{1+}, z_j^{1-}, z_j^3 \in \{ 0,1 \} \text{ all $j$}
	}
	}
	\; \frac{1}{| \mathcal{X}_{N_1} |} \sum_{\x^\text{new} \in \mathcal{X}_{N_1}}
	z_j^3
\]
subject to
\begin{align*}
	M_1^{1+} z_j^{1+} \geq K - \text{Diff}_1(\Y,\a,\x^\text{new})  \\[0.5em]
	M_2^{1+} (1 - z_j^{1+}) \geq \text{Diff}_1(\Y,\a,\x^\text{new}) - K \\[0.5em]
	z_j^3 \leq z_j^{1+}
	\qquad
	z_j^3 \leq z_j^{1-}
	\qquad
	z_j^{1+} + z_j^{1-} \leq 1 + z_j^3
\end{align*}
and two similar `$M$' inequalities for $z_j^{1-}$. This is a mixed integer linear program.

\section{Additional Numerical Illustration Details}\label{sec:numericalAppendix}

This appendix describes additional results to accompany our numerical illustration in section \ref{sec:numericalIllustration}.

\subsection*{The data generating process}

Let $N_\text{max}$ denote the largest population size we will consider. In our illustration below we set $N_\text{max} = 400$. The data generating process is defined in two steps: (1) Define potential outcomes, and (2) Assign treatment. In the first step, we define the population, the table of $Y_i(x)$ values for all $x \in \{0,1\}$ and $i \in \{1,\ldots,N_\text{max} \}$ (for example, see table \ref{table:examplePopulation}). Since we require bounded outcomes, define potential outcomes as $Y_i(x) = T(\beta x + U_i)$ where $T : \mathbb{R} \rightarrow [y_\text{min},y_\text{max} ]$ is a transformation function, and $\beta \in \mathbb{R}$. We set $T(\cdot) = \Phi(\cdot)$, the standard normal cdf, so $[y_\text{min}, y_\text{max} ] = [0,1]$. This implies that treatment effects are in $[-1,1]$. Here the maximum logical value of $K$ is 1.

Next, we generate $U_i$ by taking $N_\text{max}$ iid draws from a standard normal distribution. If this were an infinite population analysis, this choice would imply that non-treated potential outcomes were uniformly distributed on $[0,1]$ and treated potential outcomes had the cdf $\Phi[ \Phi^{-1}(y) - \beta]$. In this infinite population, the average treatment effect is
\begin{align*}
	\text{ATE}
	&= \mathbb{E}[Y(1)] - 0.5 \\
	&= \int_{-\infty}^\infty \Phi(\beta + u) \phi(u) \; du - 0.5.
\end{align*}
We pick $\beta$ so that this infinite population $\text{ATE} = 0.25$, a moderately sized value of ATE; this gives $\beta = 0.9648$. Note that the finite population ATE will generally be different from this infinite population value of ATE. Given a vector of $U_i$ draws, we now have the full science table---the values of $Y_i(x)$ for all $i \in \{1,\ldots,N_\text{max} \}$ and $x \in \{0,1\}$.

In step two we must assign treatment to each unit. We consider two different scenarios: (1) A single assignment of treatment and (2) Repeated, re-assignments of treatment. The second case is the standard frequentist thought experiment in the finite population design-based inference literature; we discuss it in appendix \ref{sec:simulations}. Here we will focus on the first case, a single assignment of treatment, which will produce a single dataset for each population size, analogous to the single dataset encountered in empirical settings.

For simplicity we focus on the case where $N_1 = N_0$, so there are an equal number of units in the treatment and control groups. We assign treatment uniformly at random (as in \ref{assump:uniformRandomization}). Since the $i$ draws themselves were iid, meaning that the order of the indices $i$ is uninformative about $Y_i(x)$ values, it is sufficient to simply set $X_i=1$ for $i \leq N_\text{max}/2$ and $X_i = 0$ for $i > N_\text{max}/2$ (recalling that we chose $N_\text{max}$ to be even). For choices $N < N_\text{max}$ we construct the dgp by deleting units appropriately. Specifically, let $N$ be even. Then among all units with $X_i=1$, keep the first $N/2$ units and delete the rest. Likewise for the $X_i=0$ group. This approach ensures that our sequence of finite populations is nested---they differ only because the larger populations have more units. We consider five population sizes, $N \in \{ 10, 20, 40, 100, 400 \}$.

\subsection*{Robustness to the batch size}

\begin{figure}[t]
    \centering
    \begin{minipage}{0.45\textwidth}
        \includegraphics[width=0.85\linewidth]{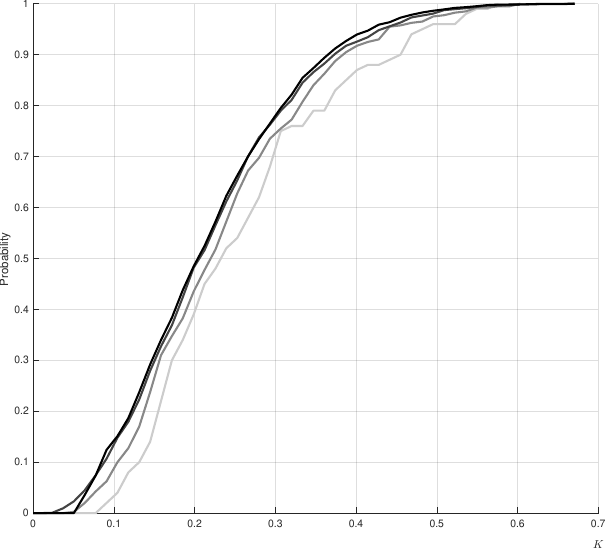}
    \end{minipage}
    \begin{minipage}{0.45\textwidth}
        \includegraphics[width=\linewidth]{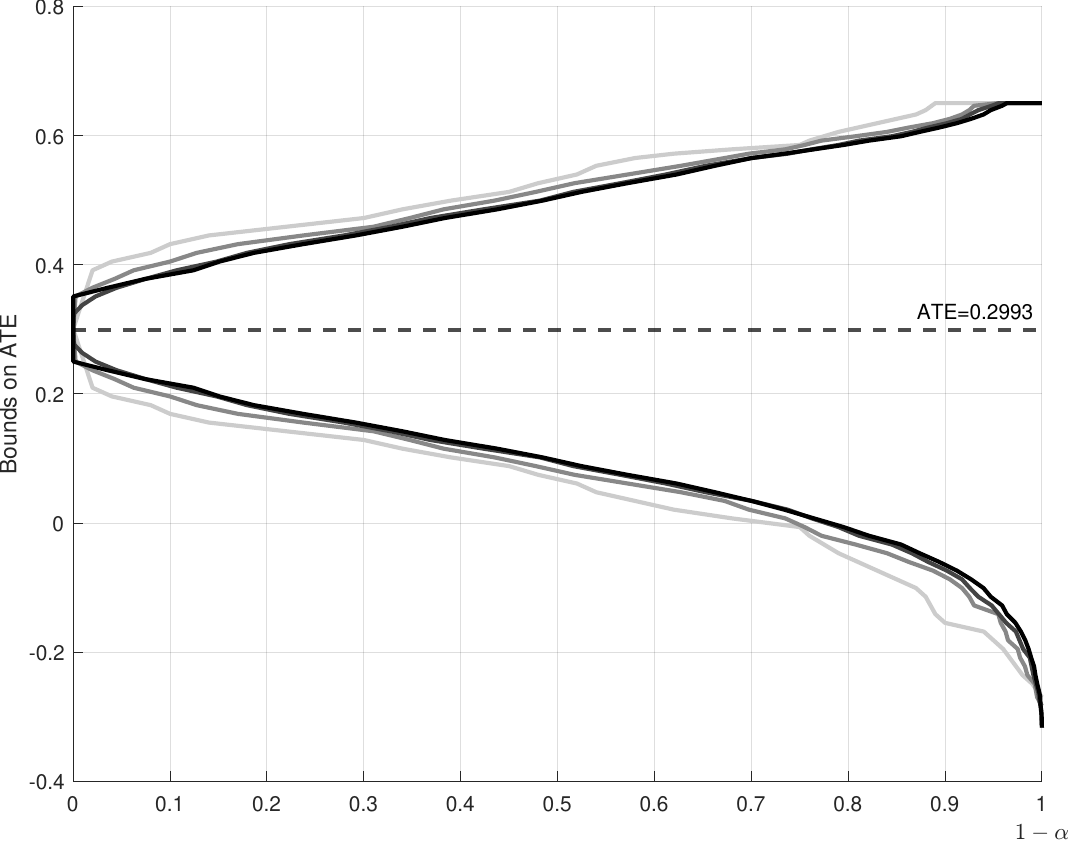}
    \end{minipage}
    \caption{Robustness to batch size: $\underline{p}$ vs $K$ (left), $\Theta_I(K(\alpha))$ vs $1-\alpha$ (right). $N=20$. The lightest gray line is $B=100$ while the darkest line is $B=6400$.}
    \label{fig:robustness_batch_size}
\end{figure}

As discussed in section \ref{sec:computation}, we approximate the objective function by sampling elements from $\mathcal{X}_{N_1}$. Call the number of sampled elements the \emph{batch size} and denote it by $B$. We consider $B \in \{ 100, 400, 1600, 6400 \}$. We do this for three values of the population size, $N \in\{ 20, 40, 100 \}$. Figure \ref{fig:robustness_batch_size} shows the results for $N=20$; similar results obtain for larger values of $N$. Overall we see that the results do not change much between the two largest batch sizes considered; even the relatively small value $B=400$ yields values of $\underline{p}$ and $\Theta_I(K(\alpha))$ that are quite similar to those for $B=6400$. 

\section{Alternative CI's for our Main Empirical Applications}\label{sec:AlternativeEmpiricalCIs}

\begin{figure}[t]
\centering
    \includegraphics[width=0.8\linewidth]{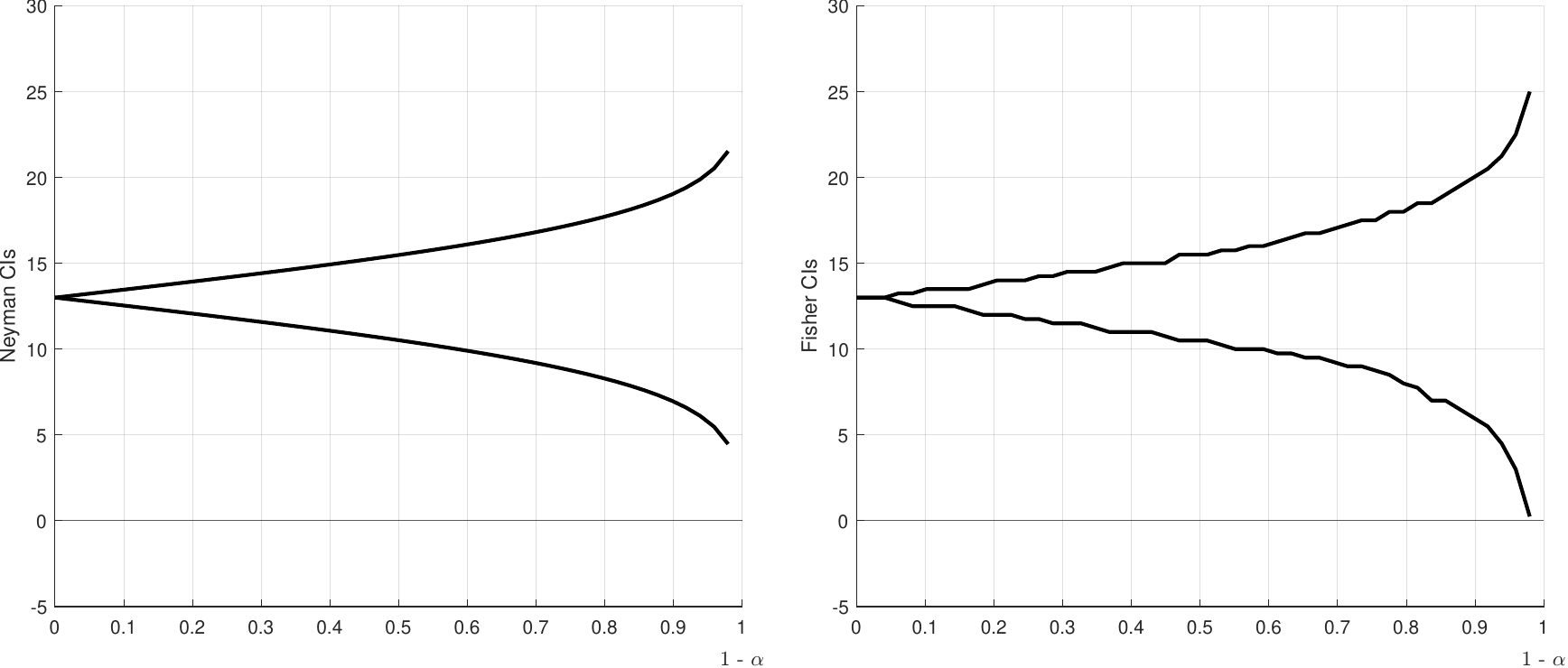}
\caption{Confidence intervals for the impact of a late fee on the number of parents who show up late. Left: The Neyman CI. Right: The Fisher CI. \label{fig:GneezyDesignBased3}}
\end{figure}

\begin{figure}[t]
\centering
\includegraphics[width=0.9\linewidth]{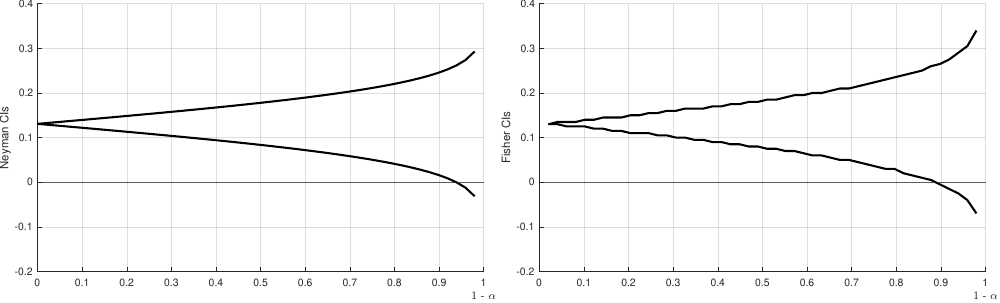}
\caption{Confidence intervals for the impact of management interventions on long run adoption of management practices. Left: The Neyman CI. Right: The Fisher CI. \label{fig:BloomDesignBased3}}
\end{figure}

Figure \ref{fig:GneezyDesignBased3} shows the Neyman and Fisher CI's, as described in appendix \ref{sec:simulations}, for our first empirical application in section \ref{sec:empirical}. Both approaches give much tighter intervals than $\Theta_I(K(\alpha))$. However, keep in mind that (1) they are only interpreted as confidence intervals and not identified sets or credible sets, unlike $\Theta_I(K(\alpha))$, which have all three interpretations, and (2) they are not valid for fixed $N$ with heterogeneous treatment effects. Likewise, figure \ref{fig:BloomDesignBased3} shows the Neyman and Fisher CI's for our second empirical application.

Figure \ref{fig:HoeffdingCIempirical} shows the Hoeffding CI's, discussed in appendix \ref{sec:simulations}, for both empirical applications. Unlike the Neyman and Fisher CI's, these CI's allow for heterogeneous treatment effects and are valid for fixed $N$. These intervals are always at least 3.5 times wider than ours, however, and can be up to 20 times wider, depending on the nominal coverage probability. Moreover, unlike our intervals, the Hoeffding CI's contain zero for \emph{all} values of $\alpha$ in both applications. Finally, like the Neyman and Fisher CI's, the Hoeffding CI does not have an identification or Bayesian interpretation, to our best knowledge.

\begin{figure}[t]
\centering
    \includegraphics[width=0.49\linewidth]{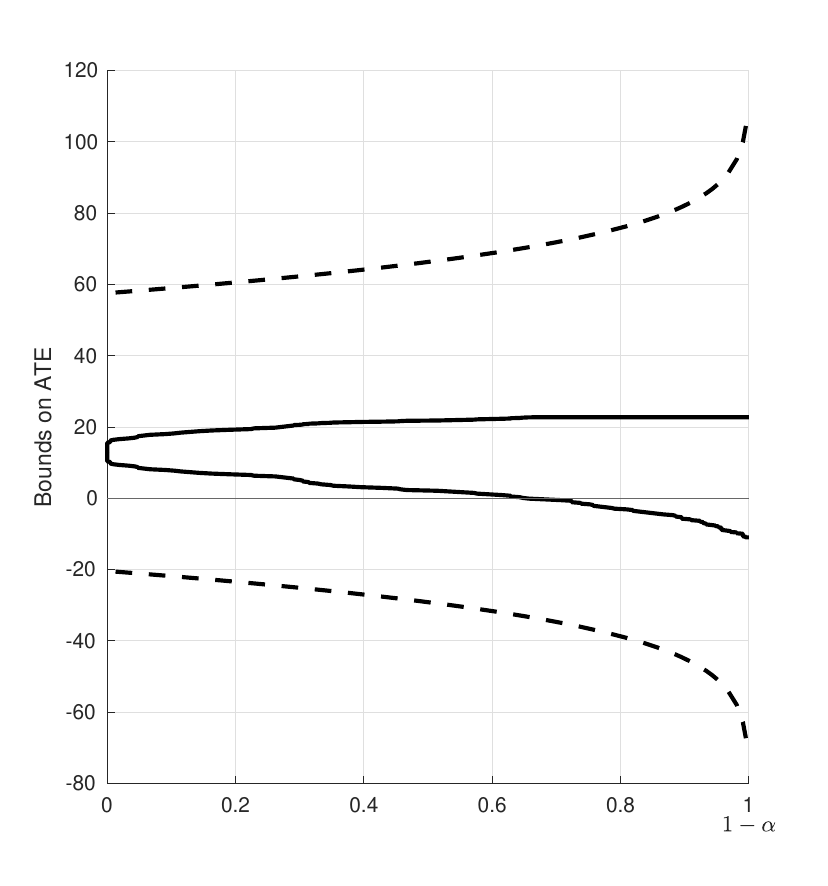}
    \includegraphics[width=0.49\linewidth]{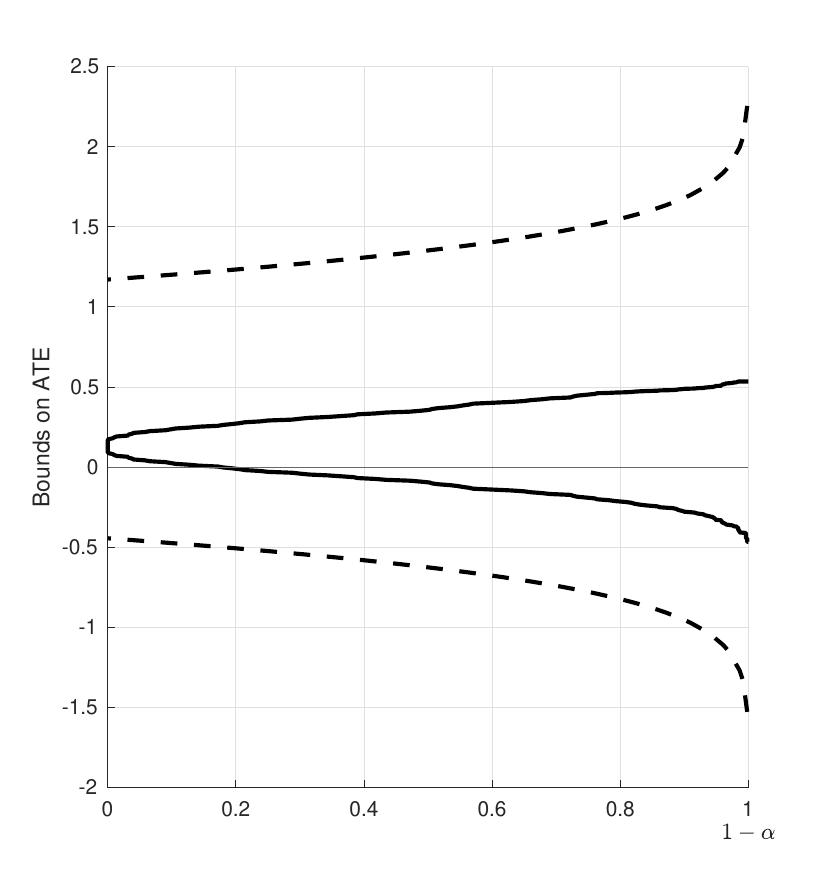}
\caption{Hoeffding confidence intervals (dashed). Our $\Theta_I(K(\alpha))$ is shown as solid lines for comparison. Left: The impact of a late fee on the number of parents who show up late. Right: The impact of management interventions on long run adoption of management practices. \label{fig:HoeffdingCIempirical}}
\end{figure}

\section{The National Supported Work Demonstration Project}\label{sec:LaLonde}

In this section we illustrate our approach with a third empirical application, to data from the National Supported Work demonstration project, as studied in \cite{LaLonde1986}. This is a very well known application, so we do not describe it here. See \cite{ImbensXu2024} for a detailed discussion. We consider this application primarily to illustrate the feasibility of our method with larger (but still finite) population sizes. Here $N = 722$ with $N_1 = 297$ and $N_0 = 425$. We set $[y_\text{min}, y_\text{max}]$ to be the range of observed earnings, which is $[0, \$60308]$. We consider this to be a relatively weak assumption, because it allows all unobserved potential outcomes to be anywhere in this range. We could obtain tighter bounds by further restricting the possible values of potential outcomes, as in section \ref{sec:boundedOutcomesVariations}. We omit this for brevity.

\begin{figure}[t]
  \centering

  \begin{subfigure}[c]{0.45\textwidth}
    \centering
\includegraphics[width=\linewidth]{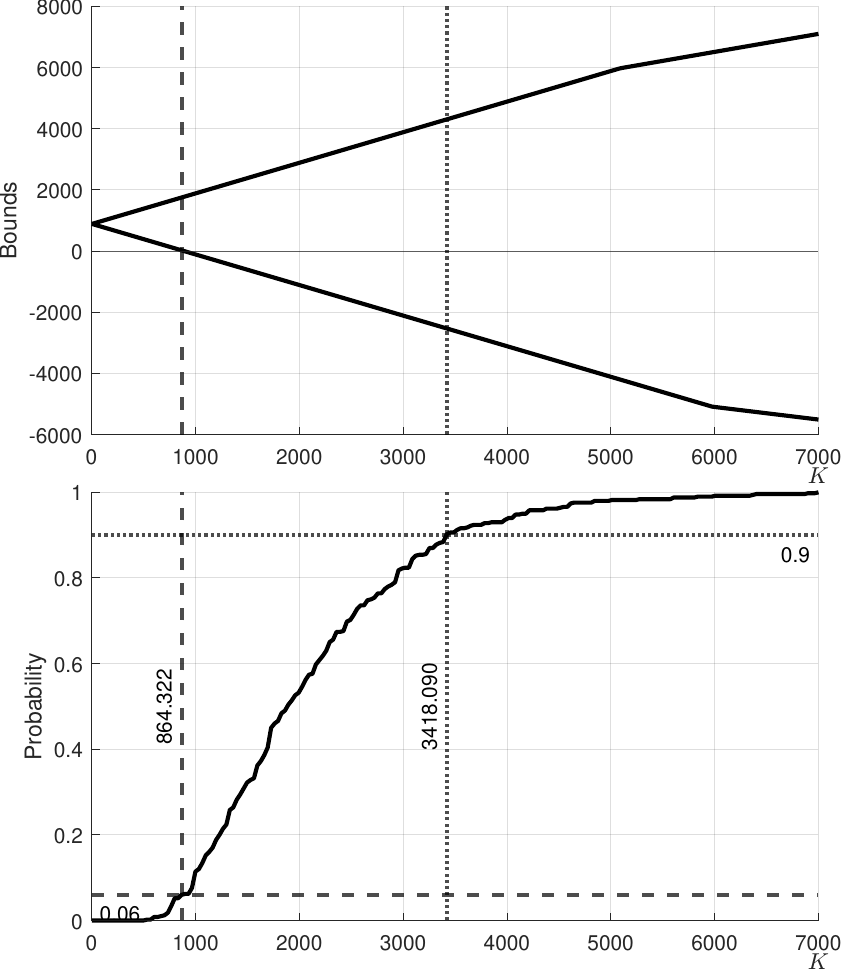}
\caption{Top: $\Theta_I(K)$, bounds on ATE as a function of the maximum magnitude of ex post imbalance, $K$. Bottom: $\underline{p}(K)$. \label{fig:LaLonde1}}
  \end{subfigure}
  \hfill
  \begin{subfigure}[c]{0.45\textwidth}
    \centering
\includegraphics[width=\linewidth]{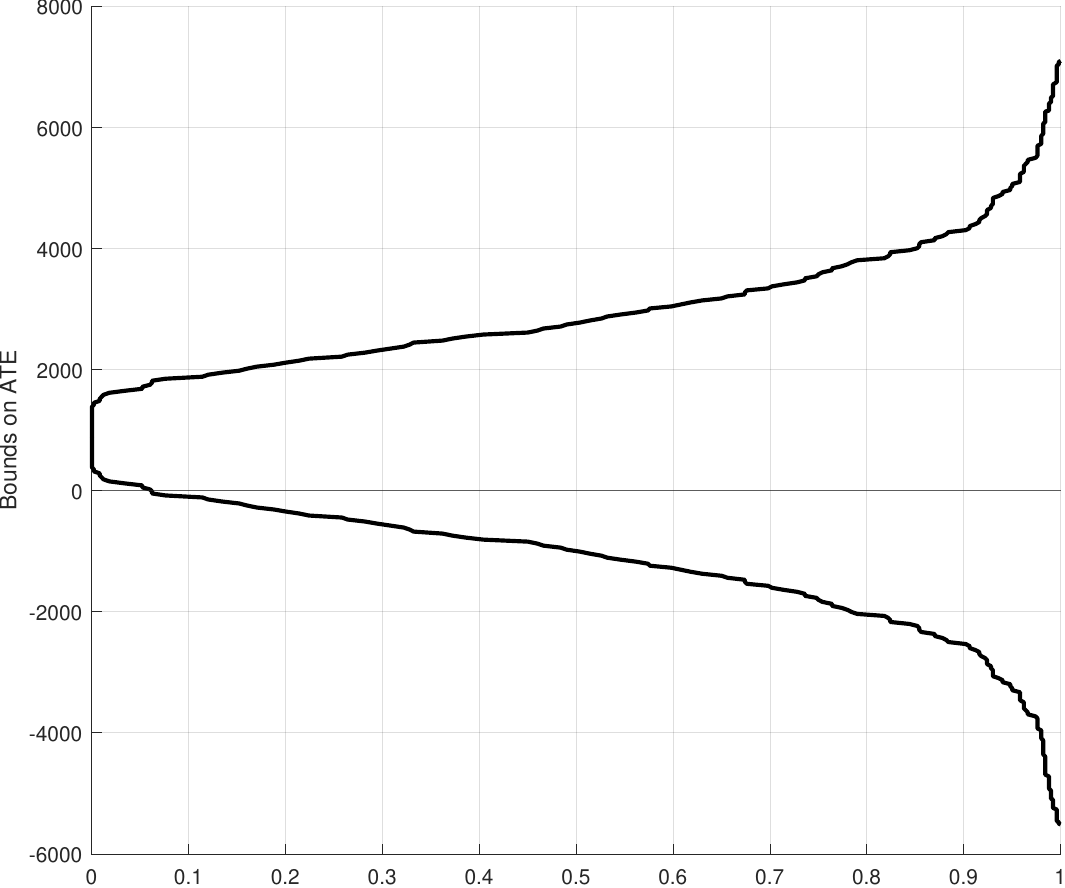}
\caption{$\Theta_I(K(\alpha))$, bounds on the ATE as a function of $1-\alpha$. \label{fig:LaLonde2}}
  \end{subfigure}

  \caption{Design-based sensitivity analysis for the impact of job training on earnings.}
  \label{fig:LaLonde}
\end{figure}

\begin{figure}[tbh]
\centering
    \includegraphics[width=0.75\linewidth]{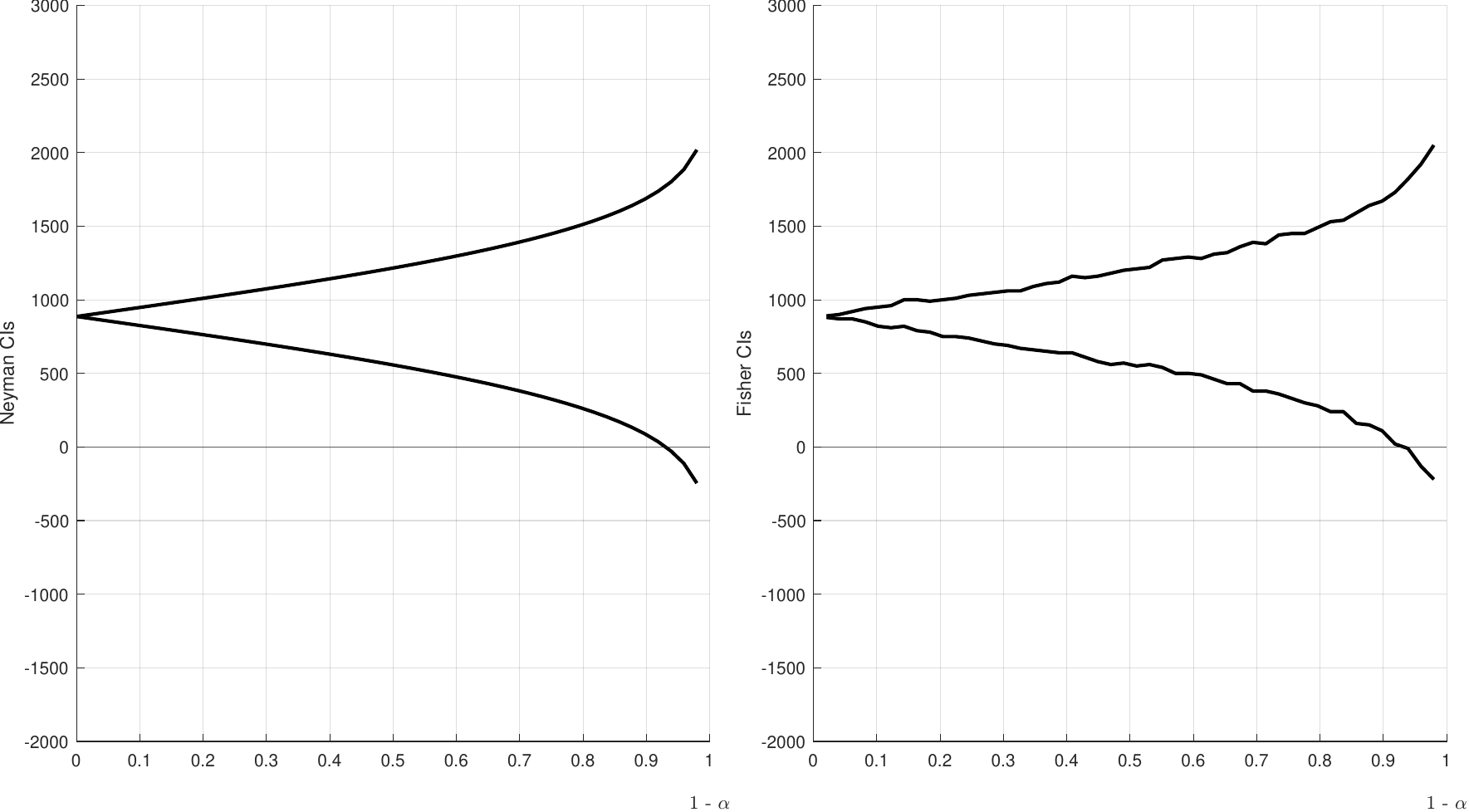}
\caption{Confidence intervals for the impact job training on earnings. Left: The Neyman CI. Right: The Fisher CI.}
    \label{fig:LaLonde3}
\end{figure}

Figure \ref{fig:LaLonde} shows the results. Here $\underline{p}(K^\text{bp}) = 0.11$, so there is at least an 11\% chance that ATE is positive, according to our robust empirical Bayesian approach. Since this probability is very low, there is a lot of uncertainty about the impact of job training on earnings. Figure \ref{fig:LaLonde2} shows the ATE bounds obtained by combining the two plots in figure \ref{fig:LaLonde1}. For reference, figure \ref{fig:LaLonde3} shows the Neyman and Fisher CI's as a function of their nominal coverage probability. 

\section{Frequentist Simulations}\label{sec:simulations}

Theorem \ref{thm:ValidCoverage} in section \ref{sec:interpretation} showed that $\Theta_I(K(\alpha))$ is a uniformly valid $100 (1-\alpha)$\% design-based confidence set. In this section we provide simulation evidence of this property. We also compare $\Theta_I(K(\alpha))$ with several alternative approaches to constructing confidence sets. Keep in mind, though, that these other approaches are not directly comparable: First, they do not share the identification or Bayesian interpretations of our construction. Second, some rely on stronger assumptions, such as validity of a large-$N$ asymptotic approximation or treatment effect homogeneity.

As noted earlier, our confidence sets are uniformly valid over all possible vectors of potential outcomes $(\Y(1),\Y(0))$ consistent with the data. This is well known to be an important property in the infinite population partial identification literature. A consequence of this uniformity, though, is that for some dgps (that is, vectors $\Y(1)$ and $\Y(0)$) our confidence sets will have actual coverage probabilities that are strictly larger than the nominal coverage probability; we will see this in the simulations below.

\subsection*{Alternative approaches to constructing design-based confidence intervals}

Here we describe four approaches for computing design-based confidence intervals from the prior literature. The first is called the \emph{Neyman CI}. To define it, let
\[
	\widehat{V}_x \coloneqq \frac{1}{N_x - 1} \sum_{i=1}^N (X_i)^x (1-X_i)^{1-x} (Y_i - \overline{Y}_x)^2
\]
be the variance in observed outcomes in treatment group $x \in \{0,1\}$. Then let
\[
	\widehat{V}_\text{Neyman} \coloneqq \frac{\widehat{V}_1}{N_1} + \frac{\widehat{V}_0}{N_0}.
\]
Let $\widehat{\text{ATE}} \coloneqq \overline{Y}_1 - \overline{Y}_0$ denote the difference-in-means point estimate. Then the Neyman CI is
\begin{multline*}
	[L_\text{Neyman}(\alpha), U_\text{Neyman}(\alpha)] \coloneqq \\\left[ \widehat{\text{ATE}} - \Phi^{-1}(1-\alpha/2) \sqrt{\widehat{V}_\text{Neyman}}, \;\; \widehat{\text{ATE}} + \Phi^{-1}(1-\alpha/2) \sqrt{\widehat{V}_\text{Neyman}} \; \right].
\end{multline*}
Here $\Phi$ is the standard normal cdf, so $\Phi^{-1}(0.975) = 1.96$. See section 6.6.1 of \cite{ImbensRubin2015} for more discussion of the Neyman CI.

The second standard approach is called the \emph{Fisher CI}. This approach is based on inverting an exact test of the sharp null hypothesis
\begin{equation}\label{eq:usualSharpNull}
	H_0 : Y_i(1) - Y_i(0) = c
	\qquad \text{for all $i=1,\ldots,N$}
\end{equation}
where $c \in \mathbb{R}$. This confidence interval collects all values of $c$ such that the exact test does not reject at level $\alpha$. This approach requires choosing a test statistic. Following section 5.7 of \cite{ImbensRubin2015}, we use $| \overline{Y}_1 - \overline{Y}_0 - c |$ in our simulations. Chapter 5 of \cite{ImbensRubin2015} provides further discussion of Fisher's exact test.

Both of these approaches are well known in the literature, but have known limitations: When viewed as a confidence interval for ATE, the Fisher CI does not rely on asymptotics, but is only valid in finite populations if treatment effects are homogeneous, a feature which is imposed by the sharp null hypothesis in equation \ref{eq:usualSharpNull}. The Neyman CI allows for arbitrary heterogeneous treatment effects, but is not valid for fixed finite populations; its justification is based on large $N$ asymptotics. A third approach has occasionally been discussed (\citealt{LohRichardsonRobins2017}, \citealt{Ding2017comment}), which we call the \emph{generalized Fisher CI}. Let 
\[
	\mathcal{Y}_\text{avg}(c) \coloneqq \left\{ (\Y(1),\Y(0)) \in [y_\text{min}, y_\text{max}]^{2N} : \frac{1}{N} \sum_{i=1}^N \big( Y_i(1) - Y_i(0) \big) = c \right\}.
\]
Then the null hypothesis that $\text{ATE} = c$ can be written as $H_0 : (\Y(1),\Y(0)) \in \mathcal{Y}_\text{avg}(c)$. For a given test statistic $T$ and observed data $\Y$ and $\X$, define
\begin{equation}\label{eq:generalizedPval}
\resizebox{0.98\linewidth}{!}{$	
	\text{pval}(\Y(1),\Y(0), \Y, \X)
	\coloneqq \Prob_\text{design}\Big( T \big( \{ Y_i(1) X_i^\text{new} + Y_i(0) (1-X_i^\text{new}) : i \in \mathcal{I} \} , \X^\text{new} \big) > T(\Y,\X) \Big).
$}
\end{equation}
Define the \emph{generalized Fisherian $p$-value} as
\[
	\overline{\text{pval}}(c,\Y,\X) \coloneqq 
	\sup_{
	\substack{
	(\Y(1),\Y(0)) \in \mathcal{Y}_\text{avg}(c) \\
	\Y = \Y(1) \times \X + \Y(0) \times (\bold{1}-\X)
	}
	}
	\; \text{pval}(\Y(1),\Y(0),\Y,\X).
\]
Then the set 
\[
	\mathcal{C}(\alpha,\Y,\X) \coloneqq \{ c \in [y_\text{min} - y_\text{max},y_\text{max} - y_\text{min}] : \overline{\text{pval}}(c,\Y,\X) \leq \alpha \}
\]
is a valid $100(1-\alpha)$\% design-based confidence set for ATE. Unlike the Fisher CI, it allows for heterogeneous treatment effects. And unlike the Neyman CI, it does not rely on asymptotics; it is valid for any fixed $N$. However, we are unaware of any papers which actually attempt to compute $\overline{\text{pval}}$ in equation \eqref{eq:generalizedPval}, except in the case where outcomes are binary (see \citealt{RigdonHudgens2015}, \citealt{LiDing2016}, and \citealt{AronowChangLopatto2023}). Instead, the literature either ignores this approach (e.g., it is not mentioned in the surveys of \citealt{Young2019} or \citealt{RitzwollerRomanoShaikh2024}), or says ``computing [$\overline{\text{pval}}$]...is almost intractable'' (\citealt{Ding2017comment}, page 362) or ``is often computationally infeasible'' (\citealt{Ding2024}, page 38), or ``for a continuous outcome, there will typically be too many populations in [the parameter space] to compute exact $p$-values for them all, necessitating the use of asymptotics'' (\citealt{LohRichardsonRobins2017}, page 360), or that ``to construct confidence intervals [for ATE], we do need to make large sample approximations'' (\citealt{AtheyImbens2017}, page 90). Our experience with using mixed integer linear programming and genetic algorithms to compute $\underline{p}(K)$ in equation \eqref{eq:defOfunderlinep}, a conceptually different but computationally similar function, suggests that computing $\overline{\text{pval}}$ may be more feasible than previously thought, especially in smaller populations. In particular, the prior literature on computation of exact Fisherian $p$-values argues that explicit ``modeling'' assumptions on potential outcomes are required to make computation feasible (e.g., examples 1--3 on page 363 of \citealt{Ding2017comment}). Our approach instead suggests that the bounded outcomes assumption \ref{assump:boundedSupport1} alone may be enough to sufficiently constrain the space of $\P^\text{true}$ matrices we need to search over. That said, we leave a full implementation of computing $\overline{\text{pval}}$ (perhaps via the genetic algorithm, similar to our other computational strategies) to a future draft.

A fourth approach uses concentration inequalities to construct design-based confidence sets that are valid for any fixed $N$ and allow for heterogeneous treatment effects. See \citet[pages 311--312]{RobinsRitov1997}, \citet[proposition 3]{AronowEtAl2025}, and \citet[section 2.2]{Ding2025}. The first two papers consider binary outcomes. The latter considers bounded outcomes and proposes the interval
\[
	\widehat{\text{ATE}} \pm 4 (y_\text{max} - y_\text{min}) \sqrt{ \frac{1}{2N} \log \frac{4}{\alpha} },
\]
(written here in our notation, and assuming for simplicity that $\Prob_\text{design}(X_i=1) = 0.5$ for all $i$, as occurs under \ref{assump:uniformRandomization} with $N_1 = N_0$) where
\[
	\widehat{\text{ATE}} \coloneqq \frac{1}{N} \sum_{i=1}^N 2 (2 X_i - 1 ) Y_i
\]
and then uses Hoeffding's inequality to verify finite $N$ validity. Hence we call this the \emph{Hoeffding CI}. Unlike our uncertainty intervals, or any of the three alternative CI's discussed above, the Hoeffding CI only uses data to select the mid-point of the interval; the width of the CI is deterministic and known a priori. Consequently, it appears to be substantially conservative. For example, in our empirical applications the Hoeffding CI is always at least 3.5 times wider than our uncertainty intervals, and can be up to 20 times wider, depending on the nominal coverage probability---see figure \ref{fig:HoeffdingCIempirical}. This is not surprising since, if $[y_\text{min}, y_\text{max}] = [0,1]$, for example, the population size must be at least as big as $N_\text{min}(\alpha) \coloneqq 8 \log ( 4 /\alpha)$ for the Hoeffding CI's width to be less than 2, the logical range of ATE. This function is decreasing in $\alpha$, attaining its smallest value of $N_\text{min}(1) = 11.09$ at $\alpha = 1$, which implies that the Hoeffding CI always has width larger than 2 with $N=10$, for example. That said, a major benefit of the Hoeffding CI is that it is trivially easy to compute, regardless of $N$, whereas our intervals are computationally intensive. Overall, because of its conservativeness, we do not consider the Hoeffding CI in our simulations below.
 
Finally, a large literature has constructed variations on both the Neyman and Fisher CIs. For example, see \cite{Robins1988}, \cite{AronowGreenLee2014}, \cite{WuDing2021}, \cite{ZhaoDing2021}, and \cite{ImbensMenzel2021}. When treatment effects are heterogeneous, these variations are also justified based on asymptotics. Moreover, they often lead to CIs that are narrower than the traditional Neyman CI, for example, by replacing $\widehat{V}_\text{Neyman}$ with a less conservative alternative. That will generally lead to CIs which are shorter than the Neyman CI. This will tend to exacerbate the finite $N$ under-coverage of the Neyman CI that we demonstrate below. Hence we do not consider these alternatives here. Several papers use conditional inference to construct CIs that are valid in fixed finite populations without imposing homogeneous treatment effects (e.g., \citealt{AtheyEcklesImbens2018} or \citealt{BasseFellerToulis2019}), but these methods are typically specific to a certain setting, like networks, and so we also do not consider this approach.

\subsection*{Simulation design}

\begin{table}[t]
\centering
\begin{minipage}{0.3\linewidth}
\centering
\begin{tabular}{c|cc}
$i$ & \(Y_i(0)\) & \(Y_i(1)\) \\
\hline
1  & 0.706 & 0.931 \\
2  & 0.062 & 0.275 \\
3  & 0.420 & 0.770 \\
4  & 0.309 & 0.671 \\
5  & 0.649 & 0.907 \\
6  & 0.660 & 0.912 \\
7  & 0.657 & 0.911 \\
8  & 0.358 & 0.719 \\
9  & 0.274 & 0.634 \\
10 & 0.278 & 0.638
\end{tabular}
\end{minipage}
\begin{minipage}{0.3\linewidth}
\centering
\begin{tabular}{c|cc}
$i$ & \(Y_i(0)\) & \(Y_i(1)\) \\
\hline
1  & 0 & 0 \\
2  & 0 & 0 \\
3  & 0 & 0 \\
4  & 0 & 0 \\
5  & 0 & 0 \\
6  & 0 & 0 \\
7  & 0 & 0 \\
8  & 0 & 0 \\
9  & 0 & 0 \\
10 & 0 & 1
\end{tabular}
\end{minipage}
\begin{minipage}{0.3\linewidth}
\centering
\begin{tabular}{c|cc}
$i$ & \(Y_i(0)\) & \(Y_i(1)\) \\
\hline
1 & 0.0499 & 0.0499 \\
2 & 0.0646 & 0.0646 \\
3 & 0.2540 & 0.2540 \\
4 & 0.0811 & 0.0811 \\
5 & 0.0847 & 0.0847 \\
6 & 0.1080 & 0.1080 \\
7 & 0.0269 & 0.0269 \\
8 & 0.0450 & 0.0450 \\
9 & 0.0547 & 0.0547 \\
10 & 0.1040 & 1
\end{tabular}
\end{minipage}
\vspace{2.5mm}
\caption{The values $\Y(1)$ and $\Y(0)$ for the three populations we use in the simulations. The left-most table is the same as in section \ref{sec:numericalIllustration}.}
\label{table:adversarialScienceTab1}
\end{table}

We focus on populations with $N=10$, but qualitatively similar results obtain for larger population sizes. Assume treatment is assigned uniformly at random, with an equal number of units in the treatment and control groups ($N_1 = N_0$). There are therefore ${N \choose N_1} = 252$ different assignments of treatment. This allows us to compute the exact design-distribution of all three CIs under consideration, without simulation error. Specifically, we compute
\[
	\Prob_\text{design} \big( \text{CI} \ni \overline{\Y(1)} - \overline{\Y(0)} \big) \coloneqq \frac{1}{S} \sum_{s=1}^S \indicator[ \text{CI}_s \ni \overline{\Y(1)} - \overline{\Y(0)} ],
\]
where recall that $\overline{\Y(x)} = \frac{1}{N} \sum_{i=1}^N Y_i(x)$, where $S=252$ and $s$ indexes elements of the set $\mathcal{X}_{N_1}$, and $\text{CI}_s$ is the confidence set for the $s$th assignment of treatment, for one of the four methods under consideration. Here we use an equal weight because we assume treatment is assigned uniformly at random.

We consider three different populations. Table \ref{table:adversarialScienceTab1} shows all three corresponding science tables. The first is the same finite population described in section \ref{sec:numericalIllustration}. The second and third populations both have the following feature: The unit level treatment effect is zero for all but one unit. In the second population potential outcomes are binary, and if we exclude the 10th unit who has a treatment effect, there is no variation in outcomes. The third population is similar, because all unit level treatment effects are zero except the tenth unit. However now we have introduced some variation in outcomes. We did this by generating $Y_i(0)$ from $\text{Beta}(2.5, 30)$. These two populations can be thought of as perturbations from populations where the sharp null of no treatment effect holds. In all of these populations, we set $[y_\text{min}, y_\text{max}] = [0,1]$. In particular, we do not assume potential outcomes are binary for the second population.

\subsection*{Results}

\begin{figure}[t]
    \centering
        \includegraphics[width=0.99\linewidth]{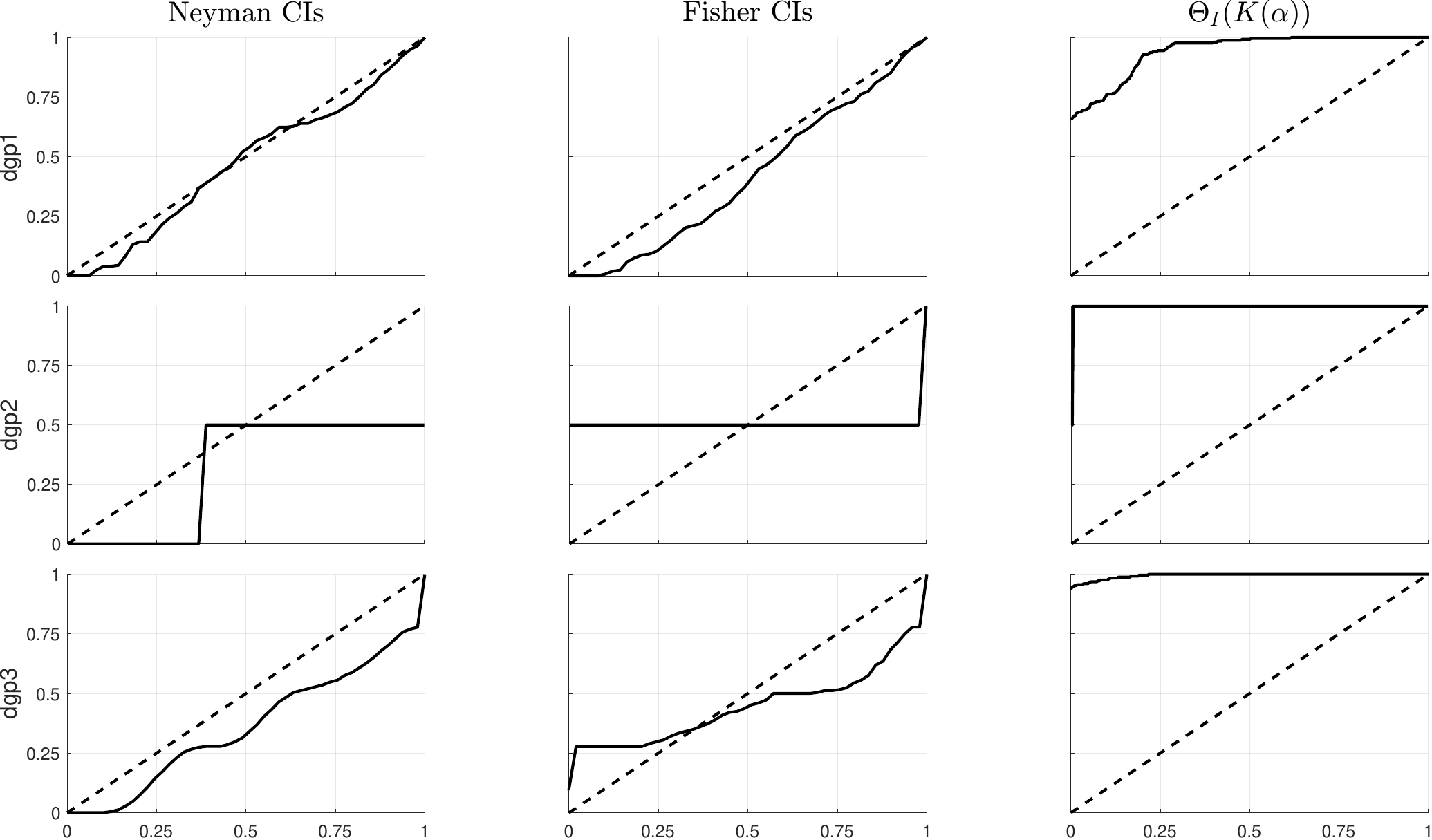}  
\caption{Confidence interval coverage probabilities for a population with $N=10$. Each row is a different dgp (our first, second, and third dgps in table \ref{table:adversarialScienceTab1}). First column is the Neyman CI. Second column is the Fisher CI. Third column is for $\Theta_I(K(\alpha))$.}
    \label{fig:simulations1}
\end{figure}

Figure \ref{fig:simulations1} shows the results. Each row is a different dgp while each column is a different method for constructing confidence intervals. Each plot shows actual coverage probabilities as a function of the nominal coverage probability, $1-\alpha$. Consider the Neyman and Fisher CI's first. In all three dgps these approaches are invalid, in the sense that their actual coverage probability is often lower than the nominal coverage. While this under-coverage is not too severe in the first dgp, it is worse in the third dgp and substantially worse in the second dgp.

\begin{figure}[t]
\centering
        \includegraphics[width=0.7\linewidth]{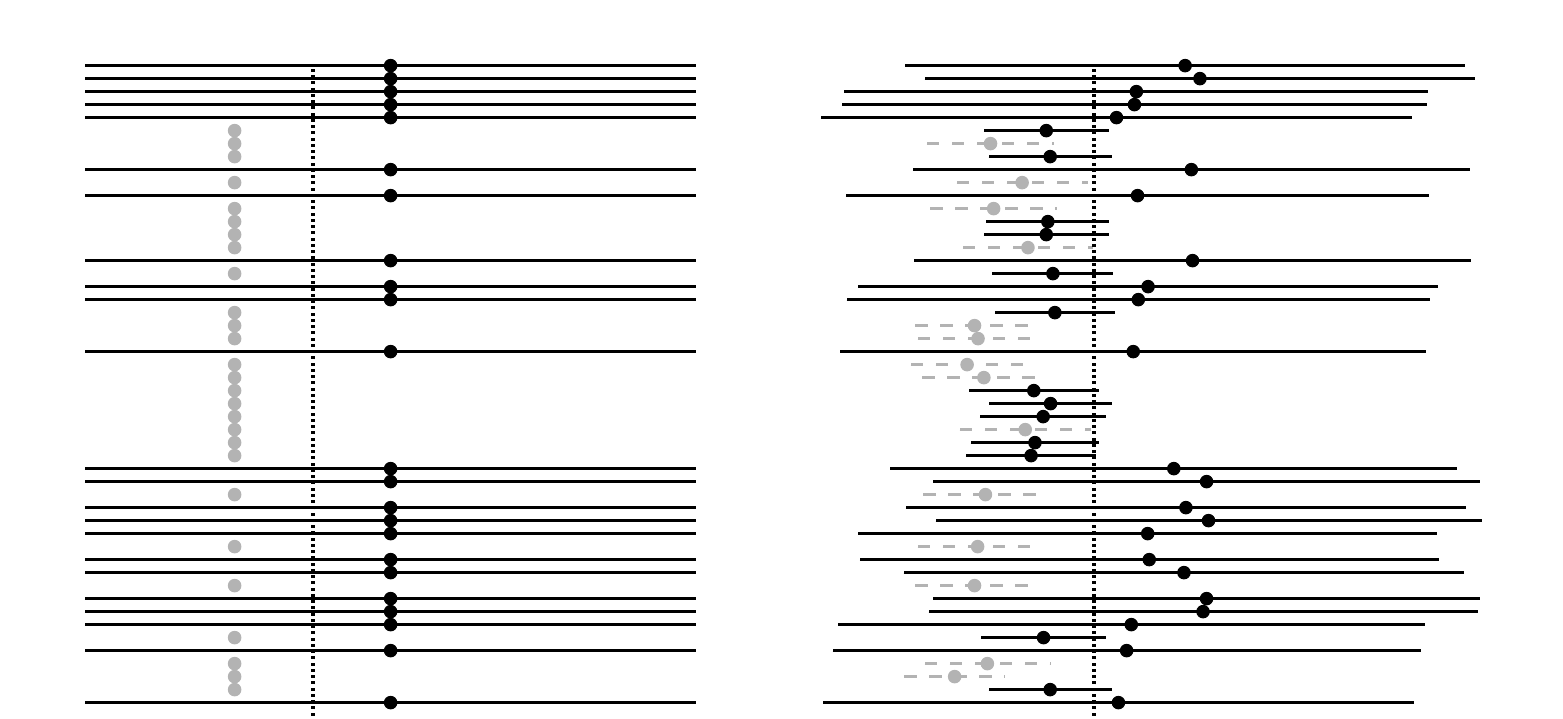}
\caption{Neyman CI's for 50 simulation draws. The circle is the center of the CI, the difference-in-means point estimand. Left plot shows the second dgp in table \ref{table:adversarialScienceTab1} while the right plot shows the third dgp. The true value of ATE is shown as a dashed vertical line. CIs that do not cover this true value are shown in bold.}
    \label{fig:simulations2}
\end{figure}

To understand why the Neyman and Fisher CI's perform poorly for these dgps, consider the third dgp. There is an ex ante 50\% probability that the 10th unit is \emph{not} treated. In this case, the variance estimate $\widehat{V}_\text{Neyman}$ is far too small, because $\widehat{V}_1$ is too small. The second dgp is a particularly extreme case because $\widehat{V}_\text{Neyman} = 0$ for these realizations of treatment assignment. This is why the coverage probability never exceeds 0.5 in that dgp. Figure \ref{fig:simulations2} illustrates this under-coverage by showing the Neyman CI's for 50 simulation draws in both dgps. Although the Fisher CI is not explicitly based on a variance estimator like $\widehat{V}_1$, it suffers from a similar problem: If the 10th unit is not treated, the randomization distribution under any sharp null which imposes homogeneous effects will be too far from the true value of ATE.

\begin{figure}[t]
    \centering
    \includegraphics[width=0.5\linewidth]{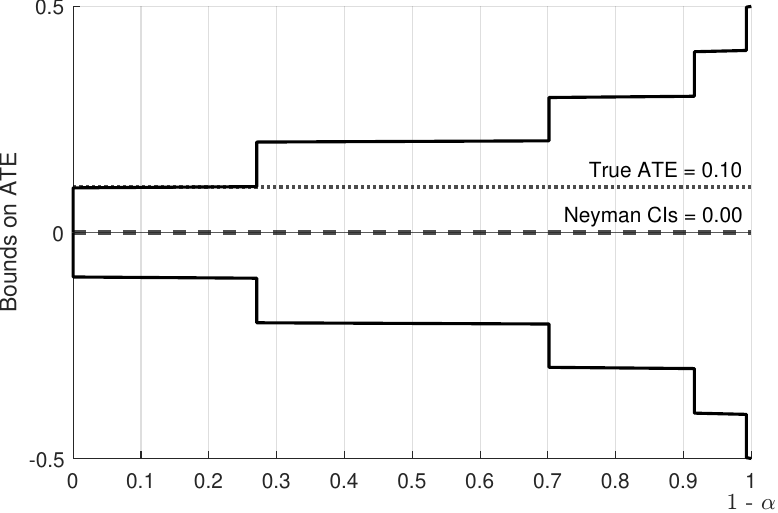}
    \caption{$\Theta_I(K(\alpha))$ for a single realization of the data from our second dgp, when unit 10 is not treated, along with the Neyman CI. In this dataset there is no observed variation in outcomes.}
    \label{fig:simulations4}
\end{figure}

\begin{figure}[tbh]
\centering
        \includegraphics[width=0.32\textwidth]{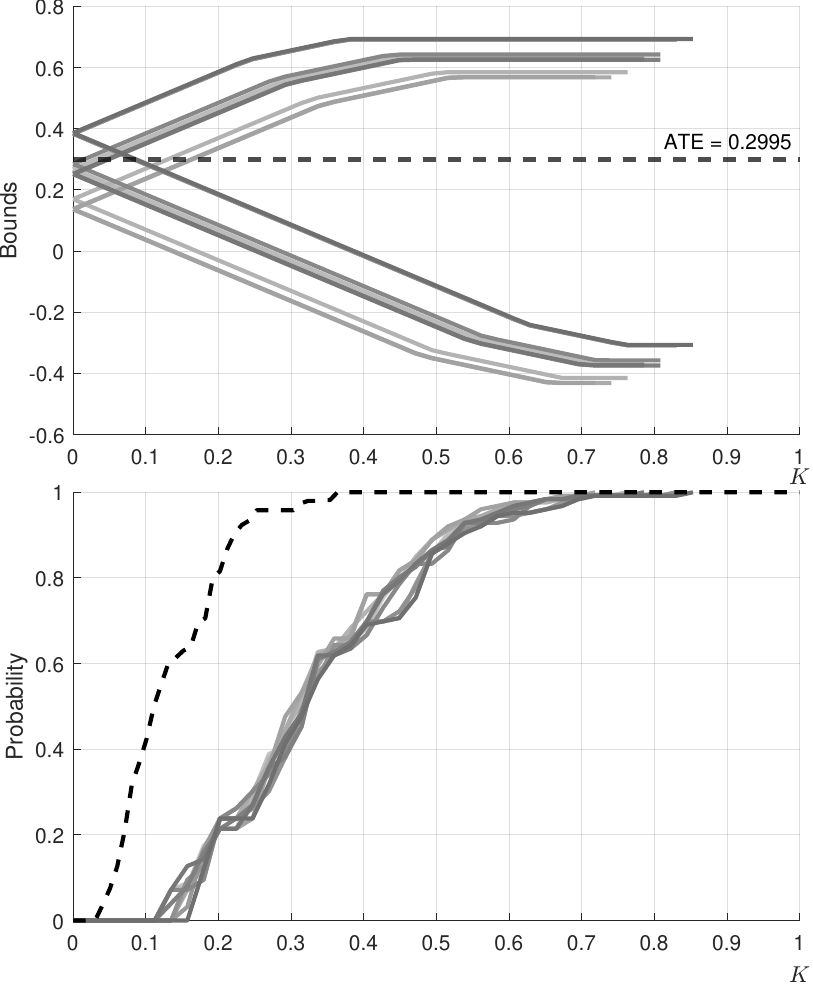}
        \includegraphics[width=0.32\textwidth]{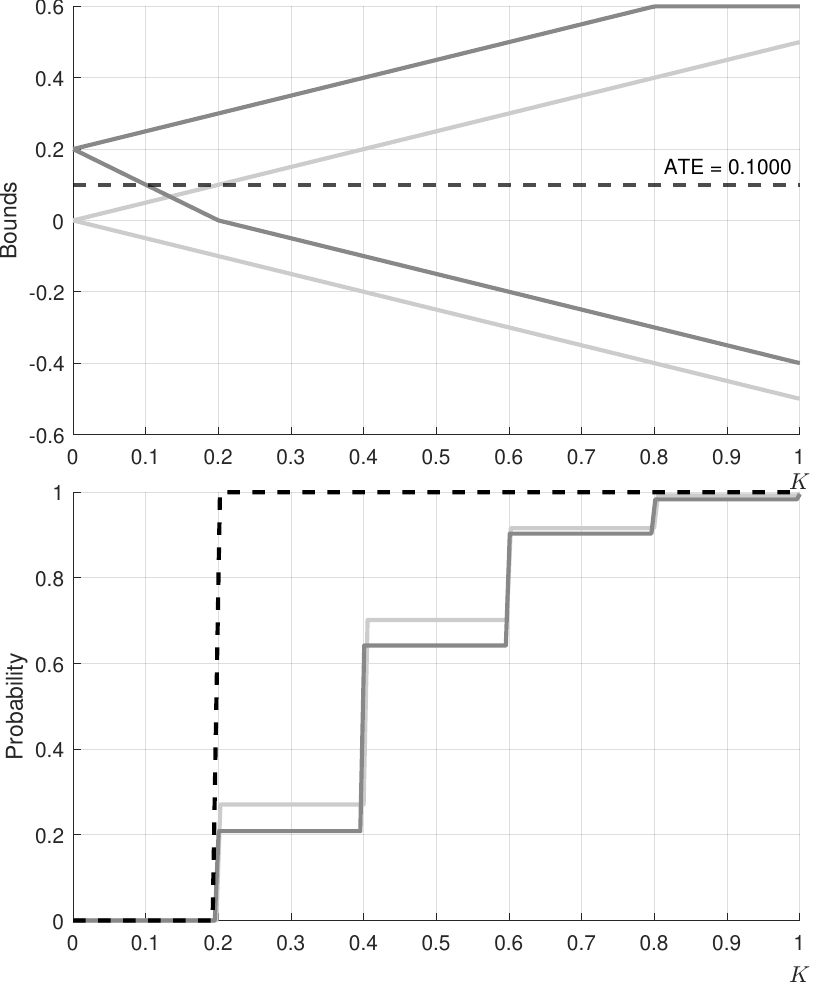} 
        \includegraphics[width=0.32\textwidth]{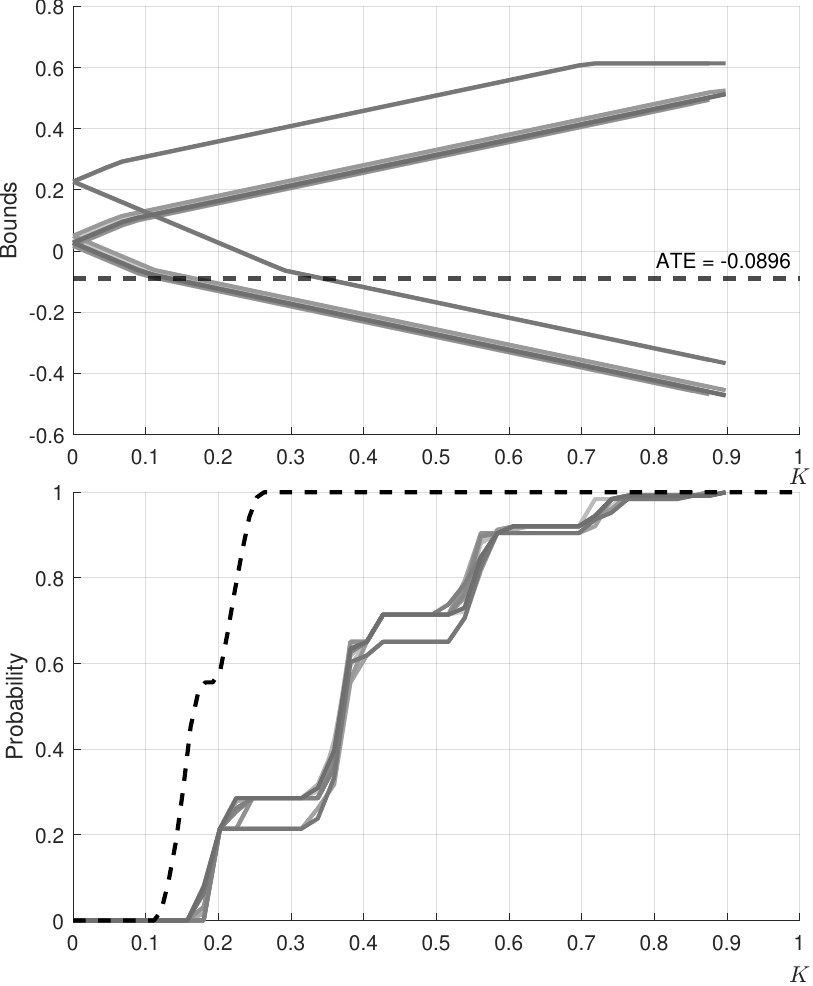}
\caption{Example draws from each simulation. Left: The first dgp. Middle: The second dgp. Right: The third dgp. The dashed line in the bottom plots show the function $p(K,\Y(1),\Y(0))$, the true probability of balance in each dgp. The different realizations then show $\underline{p}(K)$, the worst case probabilities of balance, based on the observed data and the maintained assumptions. We show 10 draws for the first and third dgps. The second dgp only has two possible realizations of the data and hence we show the two draws from this dgp.}
    \label{fig:fig_22}
\end{figure}

Next consider our approach, in the last column. There we see that for all three dgps our intervals are valid---their actual coverage probability is at least as large as their nominal coverage probability, as guaranteed by theorem \ref{thm:ValidCoverage}. Intuitively, even in the extreme case of a dataset where all outcomes are zero, as in the second dgp when the 10th unit is not treated, our procedure still accounts for the possibility that the unobserved potential outcomes can take any value in $[0,1]$; that is, the unobserved potential outcomes might have additional variation above and beyond that observed in the data. Consequently, our set $\Theta_I(K(\alpha))$ is nondegenerate for all values of $\alpha \in (0,1)$, unlike the Neyman CI. Figure \ref{fig:simulations4} illustrates this case. The price of this uniform validity is that there will be many dgps in which the procedure will be conservative, because the procedure is also hedging against the possibility that the dgp is something different from the true but unknown dgp. This can be seen in figure \ref{fig:fig_22}, which plots example draws of $\Theta_I(K)$ and $\underline{p}$ for each of the three dgps. It also plots the true function $p(K,\Y(1),\Y(0))$, showing that there is a substantial gap between this true but infeasible function and the worst case but feasible version $\underline{p}$, which is the source of the over-coverage in these particular dgps. That said, keep in mind (1) that our bounds are informative, as in our empirical applications in section \ref{sec:empirical}, and (2) that our procedure is consistent (theorem \ref{thm:randomizationGood} part 2).

Finally, although we focused on populations with $N=10$ here, similar results can be shown for any large but finite $N$. In particular, large population sizes $N$ do not guarantee that the Neyman CI or Fisher CI will have appropriate coverage. This follows because for any $N$ there exist dgps (vectors $\Y(1)$ and $\Y(0)$) that will lead these CIs to under-cover. This is a kind of non-uniformity property (see, for example, section 4.5.2 of \citealt{Ding2024}); in contrast, recall that our confidence sets are uniformly valid.

\section{Lemmas}\label{sec:lemmas}

This section contains various lemmas used throughout the proofs. The following lemma follows from standard analysis derivations; we omit its proof.

\begin{lemma}\label{lemma:upperSemi}
Let $f : \R \rightarrow \R$ be monotonically increasing and upper semicontinuous at $x_0$. Then $f$ is right continuous at $x_0$.
\end{lemma}

The next lemma is Theorem B(ii) on page 99 of \cite{ScottWu1981}.

\begin{lemma}[Finite Population LLN]\label{lemma:ScottWu1981}
Consider a sequence of non-random vectors $\X^N = (X_1^N,\ldots,X_N^N)$ and random vectors $\S = (S_1,\ldots,S_N)$ (superscript omitted here for simplicity). Suppose each $\S$ satisfies $\Prob_\text{design}(\S = \s) = 1 / \binom{N}{N_1}$ if $\sum_{i=1}^N s_i = N_1$ and zero otherwise. Here $N_1 \leq N$ is a non-random constant that can change along the sequence. Then
\begin{enumerate}
\item ${\var}_S(\frac{1}{N_1} \sum_{i=1}^N S_i X_i^N) = \frac{N-N_1}{N} \frac{1}{N_1} \frac{1}{N-1} \sum_{i=1}^N \left(X_i^N - \frac{1}{N} \sum_{i=1}^N X_i^N\right)^2$.

\item If ${\var}_S(\frac{1}{N_1} \sum_{i=1}^N S_i X_i^N) \rightarrow 0$ as $N_1,N \to \infty$, then $\frac{1}{N_1} \sum_{i=1}^N S_i X_i^N - \frac{1}{N} \sum_{i=1}^N X_i^N \xrightarrow{p} 0$ as $N_1,N \rightarrow\infty$. 
\end{enumerate}
\end{lemma}

\noindent In lemma \ref{lemma:ScottWu1981}, $\xrightarrow{p}$ refers to \emph{design} probabilities; all randomness arises from $\Prob_\text{design}$. In part 1, the term $V_N \coloneqq \frac{1}{N-1} \sum_{i=1}^N \left(X_i^N - \frac{1}{N} \sum_{i=1}^N X_i^N\right)^2$ is the population variance of the vector $\X^N$. So we can write ${\var}_S \left( \frac{1}{N_1} \sum_{i=1}^N S_i X_i^N \right) = \left( 1 - \frac{N_1}{N} \right) \frac{1}{N_1} V_N$. The term in parentheses here is usually called the \emph{finite population correction}, and is always between 0 and 1. Consequently, a sufficient condition for this term to go to zero is $N_1, N \rightarrow \infty$ and $\sup_{N: N \geq 1} \sup_{i=1,\ldots,N} | X_i^N | \leq C$ for some finite $C > 0$. This second condition implies that $V_N$ is uniformly bounded by $C$. This is useful because the bounded outcomes assumption \ref{assump:boundedSupport1} provides exactly this second condition, when we apply this LLN to potential outcomes below.

For $g, x \in \{0,1\}$, let $\overline{Y}_g^\text{new}(x) = \frac{1}{N_g} \sum_{i=1}^N Y_i(x) \indicator(X_i^\text{new} = g)$ be the random variable analog to $\overline{Y}_g(x)$ (the value corresponding to the realized $\X$ in the data). For any vector $\A = (A_1,\ldots,A_N)$ of length $N$, let $\overline{\A} = \frac{1}{N} \sum_{i=1}^N A_i$ denote its finite mean.

We also use the following result, which is a minor variation on lemma 2.1(a) in \cite{Horowitz1990}.

\begin{lemma}[Finite Population Hoeffding Inequality]\label{lemma:Horowitz1990}
Suppose \ref{assump:boundedSupport1} (bounded outcomes) and \ref{assump:uniformRandomization} (uniform randomization) hold. Then for any $K > 0$ and $g \in \{0,1\}$,
\[
	\Prob_\text{design}( | \overline{Y}_g^\text{new}(x) - \overline{\Y(x)} | > K) \leq 2 \exp \left[ \frac{- 2 K^2}{N_g (y_\text{max} - y_\text{min})^2} \right].
\]
\end{lemma}

\begin{proof}[Proof of lemma \ref{lemma:Horowitz1990}]
Horowitz's result was about simple random sampling rather than random assignment. However, suppose we think of the treatment group as a simple random sample of $N_1$ units from the population of $N$ units. Then for each of the ``sampled'' (treated) units we can take the average of their $Y_i(1)$ variables. This is precisely a realization $\overline{Y}_1(1)$ of $\overline{Y}_1^\text{new}(1)$. And the population mean is $\overline{\Y(1)}$. We can now immediately apply Horowitz's result. The same observation applies to $\overline{Y}_0(1)$, $\overline{Y}_1(0)$, and $\overline{Y}_0(0)$.
\end{proof}

The following lemma shows that the finite population identified set converges to a limiting identified set.

\begin{lemma}\label{lemma:convergenceOfIdentifiedSet}
Suppose the assumptions of theorem \ref{thm:randomizationGood} hold. Let
\begin{align*}
	\text{LB}_K^\infty(1) &\coloneqq \mu(1) \rho + \max \{ y_\text{min}, \mu(1) - K \} (1-\rho) \\
	\text{LB}_K^\infty(0) &\coloneqq \mu(0) (1-\rho) + \max \{ y_\text{min}, \mu(0) - K \} \rho  \\
	\text{UB}_K^\infty(1) &\coloneqq \mu(1) \rho + \min \{ y_\text{max}, \mu(1) + K \} (1-\rho) \\
	\text{UB}_K^\infty(0) &\coloneqq \mu(0) (1-\rho) + \min \{ y_\text{max}, \mu(0) + K \} \rho.
\end{align*}
Then $\sup_{K \geq 0} | \text{UB}_K(x) - \text{UB}_K^\infty(x) | \xrightarrow{p} 0$ for each $x \in \{0,1\}$, and likewise for the lower bound. Consequently, if we define $\Theta_{I,\infty}(K) \coloneqq [\text{LB}_K^\infty(1) - \text{UB}_K^\infty(0), \; \text{UB}_K^\infty(1) - \text{LB}_K^\infty(0)]$ then $\Theta_I(K) \xrightarrow{p} \Theta_{I,\infty}(K)$ uniformly in $K$, where the notion of set-convergence here means that the difference in the endpoints converge in probability uniformly in $K$, since both sets are intervals.
\end{lemma}

\begin{proof}[Proof of lemma \ref{lemma:convergenceOfIdentifiedSet}]
$N_1 / N \rightarrow \rho$ implies that $\frac{1}{N} \sum_{i=1}^N X_i = \frac{1}{N} N_1 \rightarrow \rho$ deterministically, because $\sum_{i=1}^N X_i$ is nonrandom by the uniform randomization assumption. The finite population LLN gives $\overline{Y}_1 \coloneqq \frac{1}{N_1} \sum_{i=1}^N Y_i X_i = \frac{1}{N_1} \sum_{i=1}^N Y_i(1) X_i \xrightarrow{p} \mu(1)$ and $\overline{Y}_0 \coloneqq \frac{1}{N_0} \sum_{i=1}^N Y_i (1- X_i) = \frac{1}{N_0} \sum_{i=1}^N Y_i(0) (1 - X_i) \xrightarrow{p} \mu(0)$. Recall that
\begin{align*}
	\text{LB}_K(x) 
	&\coloneqq \overline{Y}_x \frac{N_x}{N} + \max \left\{ y_\text{min}, \overline{Y}_x - K \right\} \frac{N_{1-x}}{N} \\
	\text{UB}_K(x)
	&\coloneqq \overline{Y}_x \frac{N_x}{N} + \min \{ y_\text{max}, \overline{Y}_x + K \} \frac{N_{1-x}}{N}.
\end{align*}
Hence, by the continuous mapping theorem,
\begin{align*}
	\text{LB}_K(1) &\xrightarrow{p} \mu(1) \rho + \max \{ y_\text{min}, \mu(1) - K \} (1-\rho) \\
	\text{LB}_K(0) &\xrightarrow{p} \mu(0) (1-\rho) + \max \{ y_\text{min}, \mu(0) - K \} \rho  \\
	\text{UB}_K(1) &\xrightarrow{p} \mu(1) \rho + \min \{ y_\text{max}, \mu(1) + K \} (1-\rho) \\
	\text{UB}_K(0) &\xrightarrow{p} \mu(0) (1-\rho) + \min \{ y_\text{max}, \mu(0) + K \} \rho.
\end{align*}
This gives pointwise-in-$K$ convergence. To obtain uniform convergence, consider
\begin{align*}
	&| \text{UB}_K(0) - \text{UB}_K^\infty(0) | \\
	&= \left| \left( \overline{Y}_0 \frac{N_0}{N} + \min \{ y_\text{max}, \overline{Y}_0 + K \} \frac{N_1}{N} \right) - \big( \mu(0) (1-\rho) + \min \{ y_\text{max}, \mu(0) + K \} \rho \big) \right| \\
	&= \left| \left( \overline{Y}_0 \frac{N_0}{N} - \mu(0) (1-\rho) \right) 
	+ \frac{N_1}{N} \left( \min \{ y_\text{max}, \overline{Y}_0 + K \} - \min \{ y_\text{max}, \mu(0) + K \} \rho \frac{N}{N_1} \right) \right| .
\end{align*}
The first term does not depend on $K$ and converges to zero in probability as above. In the second term, the coefficient $N_1/N \rightarrow \rho$. So consider the term in parentheses:
\begin{align*}
	&\left| \min \{ y_\text{max}, \overline{Y}_0 + K \} - \min \{ y_\text{max}, \mu(0) + K \} \rho \frac{N}{N_1} \right| \\
	&\leq \left| \min \{ y_\text{max}, \overline{Y}_0 + K \} - \min \{ y_\text{max}, \mu(0) + K \} \right| + \left| \min \{ y_\text{max}, \mu(0) + K \} \left( 1 - \rho \frac{N}{N_1} \right) \right|.
\end{align*}
For any $y$, $a$, and $b$, $| \min \{ y, a \} - \min \{y, b \} | \leq | a- b|$. Hence the first term is bounded above by $| \overline{Y}_0 - \mu(0) |$ and therefore converges to zero in probability uniformly over $K$. Consider the second term: $| \min \{ y_\text{max}, \mu(0) + K \} | \in [0, y_\text{max} ]$ and therefore is uniformly bounded in $K$. And $N/N_1 \rightarrow 1/\rho$. Hence the second term converges to zero in probability uniformly over $K$. A similar proof applies to the other three bound functions.
\end{proof}

The following lemma shows that the worst case design distribution is degenerate at zero, asymptotically. This is a consequence of randomization---we have exact balance asymptotically.

\begin{lemma}\label{lemma:underlinepGoesTo1}
Suppose the assumptions of theorem \ref{thm:randomizationGood} hold. Fix $K > 0$. For any sequence $\X^N = (X_1^N,\ldots,X_N^N)$ of realizations from $\X^\text{new}$, $\underline{p}(K) \rightarrow 1$ as $N \rightarrow \infty$.
\end{lemma}

\begin{proof}[Proof of lemma \ref{lemma:underlinepGoesTo1}]
Recall that
\[
	\underline{p}(K) \coloneqq \inf_{
	\substack{
	\Y(1) \in [y_\text{min},y_\text{max}]^{N_1} \\
	\Y(0) \in [y_\text{min},y_\text{max}]^{N_0} \\
	\Y = \X \Y(1) + (\bold{1}-\X) \Y(0)
	}
	} \; p (K, \Y(1), \Y(0) ).
\]
So, omitting the constraints for brevity,
\begin{align*}
	\underline{p}(K)
	&= \inf_{\Y(1),\Y(0)} \Prob_\text{design} \left( | \overline{Y}_1^\text{new}(x) - \overline{Y}_0^\text{new}(x) | \leq K \text{ for $x=0$ and $x=1$} \right) \\
	&= \inf_{\Y(1),\Y(0)} \Big( 1 - \Prob_\text{design} \left( | \overline{Y}_1^\text{new}(x) - \overline{Y}_0^\text{new}(x) | > K \text{ for $x=0$ or $x=1$} \right) \Big) \\
	&= 1 - \sup_{\Y(1),\Y(0)} \Prob_\text{design} \left( | \overline{Y}_1^\text{new}(x) - \overline{Y}_0^\text{new}(x) | > K \text{ for $x=0$ or $x=1$} \right).
\end{align*}
So it suffices to show that the supremum goes to zero as $N \rightarrow \infty$, for all $K > 0$. Since $\Prob(A \cup B) \leq \Prob(A) + \Prob(B)$ for any events $A$ and $B$,
\begin{align*}
	&\sup_{\Y(1),\Y(0)} \Prob_\text{design} \left( | \overline{Y}_1^\text{new}(x) - \overline{Y}_0^\text{new}(x) | > K \text{ for $x=0$ or $x=1$} \right) \\
	&\leq 
\resizebox{0.98\linewidth}{!}{$\displaystyle
	\sup_{\Y(1),\Y(0)} \Prob_\text{design} \left( | \overline{Y}_1^\text{new}(1) - \overline{Y}_0^\text{new}(1) | > K \right) + \sup_{\Y(1),\Y(0)} \Prob_\text{design} \left( | \overline{Y}_1^\text{new}(0) - \overline{Y}_0^\text{new}(0) | > K \right).
$}
\end{align*}
Hence it suffices to show the result for each $x \in \{0,1\}$ at a time. Since
\begin{align*}
	| \overline{Y}_1^\text{new}(x) - \overline{Y}_0^\text{new}(x) |
	&= | \overline{Y}_1^\text{new}(x) - \overline{Y}_0^\text{new}(x) - \overline{Y(x)} + \overline{Y(x)} | \\
	&\leq | \overline{Y}_1^\text{new}(x) - \overline{Y(x)} | + | \overline{Y}_0^\text{new}(x) - \overline{Y(x)} |
\end{align*}
we have
\begin{align*}
	&\Prob_\text{design}(| \overline{Y}_1^\text{new}(x) - \overline{Y}_0^\text{new}(x) | > K) \\
	&\leq \Prob_\text{design}(| \overline{Y}_1^\text{new}(x) - \overline{Y(x)} | > K/2) + \Prob_\text{design}(| \overline{Y}_0^\text{new}(x) - \overline{Y(x)} | > K/2) \\
	&\leq 2 \exp \left[ \frac{- 2 (K/2)^2}{N_1 (y_\text{max} - y_\text{min})^2} \right] + 2 \exp \left[ \frac{- 2 (K/2)^2}{N_0 (y_\text{max} - y_\text{min})^2} \right] \\
	&= 2 \exp \left[ \frac{-2 (K/2)^2}{N (N_1/N) (y_\text{max} - y_\text{min})^2} \right] + 2 \exp \left[ \frac{- 2 (K/2)^2}{N (N_0/N) (y_\text{max} - y_\text{min})^2} \right].
\end{align*}
The second inequality follows by lemma \ref{lemma:Horowitz1990}. Thus
\begin{align*}
	&\sup_{\Y(1), \Y(0)} \Prob_\text{design}( | \overline{Y}_1^\text{new}(x) - \overline{Y}_0^\text{new}(x)| > K) \\
	&\leq 2 \exp \left[ \frac{-2 (K/2)^2}{N (N_1/N) (y_\text{max} - y_\text{min})^2} \right] + 2 \exp \left[ \frac{- 2 (K/2)^2}{N (N_0/N) (y_\text{max} - y_\text{min})^2} \right]
\end{align*}
since the right hand side does not depend on the exact values of the dgp $\Y(1)$ and $\Y(0)$. Moreover, this inequality holds regardless of the values of the realized data $\X$ and $\Y$. Finally, the right hand side converges to zero as $N \rightarrow \infty$. Here we also use that $N_1/N \rightarrow \rho \in (0,1)$.
\end{proof}

\bibliographystyle{ecta-fullname}
\bibliography{finite_pop_ident}

\makeatletter\@input{KM1aux.tex}\makeatother